\documentclass{aa}

\usepackage{amssymb}
\usepackage{epsfig} 
\usepackage{graphicx} 
\usepackage{color} 
\usepackage{times} 
\usepackage{amsmath} 
\usepackage{amssymb} 
\usepackage{xspace} 
\usepackage[varg]{txfonts} 
\usepackage{enumitem} 
\usepackage{natbib} 
\usepackage{algorithm} 
\usepackage{algorithmic} 
\usepackage{subfloat} 
\usepackage{subfig} 
\usepackage[normalem]{ulem}
\usepackage{booktabs}
\usepackage{lscape}
\usepackage{longtable}
\usepackage{hyperref}

\def\d3k{{\displaystyle {\rm d}{\bf k} \over \displaystyle (2\pi)^3}}

\newcommand{\Excursion}  {\mm{{\mathbb E}}}
\newcommand{\Mask}       {\mm{{\mathsf M}}}
\newcommand{\Mspace}       {\mm{{\mathbb M}}}
\newcommand{\Rspace}     {\mm{{\mathbb R}}} 
\newcommand{\Sspace}     {\mm{{\mathbb S}}} 
 
\newcommand{\Homology}[1]{\mm{{\sf H}_{#1}}}
\newcommand{\Rank}[1]    {\mm{{\rm rank\,}{#1}}}
\newcommand{\Betti}[1]   {\mm{{\beta}_{#1}}}
\newcommand{\relBetti}[1]{\mm{{b}_{#1}}}
\newcommand{\Euler}      {\mm{\sf EC}} 
\newcommand{\relEuler}      {\mm{\sf EC_{\rm rel}}} 
 
\newcommand{\dime}[1]    {\mm{\rm dim\,}{#1}}
\newcommand{\ssx}        {\mm{\sigma}}
\newcommand{\tsx}        {\mm{\tau}}

\newcommand{\lips}{{\cal L}}

\newcommand{\R}{\mathbb{R}}

\newcommand{\x}{\mathbf{x}}

\newcommand{\y}{\mathbf{y}}

\newcommand {\mm}[1] {\ifmmode{#1}\else{\mbox{\(#1\)}}\fi}

\newcommand{\Res}            {\mm{N}}

\newcommand{\Skip}[1]        {}

\def\LKC{Lipschitz-Killing curvature}

\def\npipe{\texttt{NPIPE }}
\def\ffp{\texttt{FFP10 }}

\newcommand{\pp}[1]{{\color{red}  #1}}

\newcommand{\reltabNpipe}{
        
        \begin{tabular}{|r|r||r|r|r||r|r|r||r|r|r|} \hline
                
                \multicolumn{11}{|c|}{Relative homology (\texttt{NPIPE})} \\
                \hline
                
                &       & \multicolumn{3}{|c|}{$\chi^2$ (theoretical)}  & \multicolumn{3}{|c|}{$\chi^2$ (empirical)} & \multicolumn{3}{|c|}{Tukey Depth} \\
                
                Res & \multicolumn{1}{|c||}{FWHM} & \multicolumn{1}{c}{$\relBetti{0}$} & \multicolumn{1}{c}{$\relBetti{1}$}
                
                & \multicolumn{1}{c}{$\relEuler$} & \multicolumn{1}{|c}{$\relBetti{0}$} & \multicolumn{1}{c}{$\relBetti{1}$}
                
                & \multicolumn{1}{c|}{$\relEuler$}
                
                & \multicolumn{1}{|c}{$\relBetti{0}$} & \multicolumn{1}{c}{$\relBetti{1}$}
                
                & \multicolumn{1}{c|}{$\relEuler$} \\ \hline \hline
                
                2048 &  5 &     0.709   &0.569& 0.858   &       0.677&  0.548&  0.852   &       0.545&  0.505&  0.912           \\
                \hline
                1024&   10&     0.649&  0.476&  0.383   &       0.628&  0.477&  0.363   &       0.513&  0.345&  \bf{0.000}      \\
                \hline
                512      &20&   0.157&  0.417&  0.203   &       0.162&  0.408&  0.183   &       \bf{0.000}&     \bf{0.000}&     \bf{0.000}              \\
                \hline
                256      &40&   0.398&  0.597&  0.772   &       0.403&  0.600&  0.757   &       0.253&  0.498&  0.642           \\
                \hline
                128     & 80&   0.356&  0.205&  0.563   &       0.383&  0.187&  0.555   &       0.267   &\bf{0.000}     &0.730          \\
                \hline
                64      &160&   0.389   &0.494  &0.768  &       0.402&  0.468   &0.747  &       0.443&  0.385&  0.755           \\
                \hline
                32&     320&    \bf{0.022}&     \bf{0.003} &    \bf{0.001}      &       \bf{0.028}&     \bf{0.013}&     \bf{0.002}      &       \bf{0.000}&     \bf{0.000}&     \bf{0.000}      \\      
                \hline
                16      &640&   0.975&  \bf{0.002} &    0.203   &       0.982&  \bf{0.035}&     0.227   &       0.973&  \bf{0.000}&     0.440   \\
                \hline

        \end{tabular}
}

\newcommand{\reltabffp}{
        
        \begin{tabular}{|r|r||r|r|r||r|r|r||r|r|r|} \hline
        
        \multicolumn{11}{|c|}{Relative homology (\texttt{FFP10})} \\
        \hline
        
        &       & \multicolumn{3}{|c|}{$\chi^2$ (theoretical)}  & \multicolumn{3}{|c|}{$\chi^2$ (empirical)} & \multicolumn{3}{|c|}{Tukey Depth} \\
        
        Res & \multicolumn{1}{|c||}{FWHM} & \multicolumn{1}{c}{$\relBetti{0}$} & \multicolumn{1}{c}{$\relBetti{1}$}
        
        & \multicolumn{1}{c}{$\relEuler$} & \multicolumn{1}{|c}{$\relBetti{0}$} & \multicolumn{1}{c}{$\relBetti{1}$}
        
        & \multicolumn{1}{c|}{$\relEuler$}
        
        & \multicolumn{1}{|c}{$\relBetti{0}$} & \multicolumn{1}{c}{$\relBetti{1}$}
        
        & \multicolumn{1}{c|}{$\relEuler$} \\ \hline \hline
        
        \multicolumn{11}{|c|}{threshold = 0.90} \\ \hline
                
                2048 & 5        &       0.431   &0.566  &0.890  &       0.397   &0.537  &0.900  &       \bf{0.000} &       0.360 & \bf{0.000}      \\      
                \hline
                1024 & 10       &       0.728   &0.525& 0.712   &       0.743&  0.533&  0.727   &       0.547&  0.477&  0.647           \\
                \hline
                512      & 20   &       0.146 & 0.688&  0.408   &       0.137&  0.687&  0.397   &       \bf{0.000}&     0.490&  \bf{0.000}              \\
                \hline
                256     & 40  & 0.338 & 0.498 & 0.636   &       0.313 & 0.490 &       0.650   &       \bf{0.000} &    0.370 & \bf{0.000}      \\      
                \hline
                128     & 80    & 0.080 & 0.304  &0.160 &       0.067&  0.327&  0.190   &       \bf{0.000}&     \bf{0.000}      & \bf{0.000}              \\
                \hline
                64       & 160   &      0.424 & 0.367   & 0.853 &       0.407   &0.353  &0.880  &       \bf{0.000} & \bf{0.000}    & \bf{0.000}    \\      
                \hline
                32      & 320   & 0.116 &       0.159 & \bf{0.050}      &       0.140   & 0.170    & 0.057        &       \bf{0.000} &    0.327 & \bf{0.000}      \\      
                \hline
                16      &       640 & 0.935 &   \bf{0.000}      & \bf{0.014}     &       0.933 & \bf{0.003} &    \bf{0.023}      &       0.910 & \bf{0.000} &       \bf{0.000}      \\      
                \hline

        \end{tabular}
}

\begin{document}

\title{Anomalies in the topology of the temperature fluctuations in the cosmic microwave background: An analysis of the \npipe and \ffp data releases}
\titlerunning{Anomalies in the topology of the temperature fluctuations in the CMB}

\author{Pratyush Pranav}

\authorrunning{Pranav}

\institute{Univ Lyon, ENS de Lyon, Univ Lyon1, CNRS, Centre de Recherche Astrophysique de Lyon UMR5574, F--69007, Lyon, France
}


\abstract{
        We present a topological analysis of the temperature fluctuation maps from the \textup{Planck 2020} Data Release 4 (DR4) \texttt{NPIPE} dataset and the \textup{Planck 2018} Data Release 3 (DR3) \texttt{FFP10} dataset. We performed a multiscale analysis in terms of the homology characteristics of the maps, invoking relative homology to account for the analysis in the presence of masks. We performed our analysis for a range of smoothing scales spanning sub- and super-horizon scales corresponding to a full width at half maximum $(FWHM) \text{ of } 5', 10', 20', 40', 80', 160', 320', \text{and } 640'$, and employed simulations based on the standard model for comparison, which assumes the initial fluctuation field to be an isotropic and homogeneous Gaussian random field. 
        
        Examining the behavior of topological components, represented by the 0D homology group, we find the observations to be approximately $2\sigma$ or less deviant from the simulations for all resolutions and scales for the \npipe dataset. For the \ffp dataset, we detect a $2.96\sigma$ deviation between the observations and simulations at $\Res = 128, FWHM = 80'$. For the topological loops, represented by the first homology group, the simulations and observations are consistent within $2\sigma$ for most resolutions and scales for both the datasets. However, for the \npipe dataset, we observe a high deviation between the observation and simulations in the number of loops at $FWHM = 320'$, but at a low dimensionless threshold $\nu = -2.5$. Under a Gaussian assumption, this would amount to a deviation of $\sim 4\sigma$ . However, the distribution in this bin is manifestly non-Gaussian and does not obey Poisson statistics either. In the absence of a true theoretical understanding, we simply note that the significance is higher than what may be resolved by $600$ simulations, yielding an empirical $p$-value of at most $0.0016$. Specifically in this case, our tests indicate that the numbers arise from a  statistically stable regime, despite being based on small numbers. For the \ffp dataset, the differences are not as strong as for the \npipe dataset, indicating a $2.77\sigma$ deviation at this resolution and threshold. The Euler characteristic, which is the alternating sum of the ranks of relative homology groups, reflects the deviations in the components and loops.  To assess the significance of combined levels for a given scale, we employed the empirical and theoretical versions of the $\chi^2$ test as well as the nonparametric Tukey depth test. Although all statistics exhibit a stable distribution, we favor the empirical version of the $\chi^2$ test in the final interpretation, as it indicates the most conservative differences. For the \npipe dataset, we find that the components and loops differ at more than $95\%$, but agree within the $99\%$ confidence level with respect to the base model at $\Res = 32, FWHM = 320'$. The Euler characteristic at this resolution displays a per mil deviation. In contrast, the \ffp dataset shows that the observations are consistent with the base model within the $95\%$ confidence level, at this and smaller scales. This is consistent with the observations of the Planck analysis pipeline via Minkowski functionals. For the largest smoothing scale, $\Res = 16, FWHM = 640'$, both datasets exhibit an anomalous behavior of the loops, where \ffp data exhibit a deviation that is larger by an order of magnitude than that of the \npipe dataset. In contrast, the values for the topological components and the Euler characteristic agree between observations and model to within a confidence level of $99\%$ . However, for the largest scales, the statistics are based on low numbers and may have to be regarded with caution. Even though both datasets exhibit mild to significant discrepancies, they also exhibit contrasting behaviors at various instances. Therefore, we do not find it feasible to convincingly accept or reject the null hypothesis. Disregarding the large-scale anomalies that persist at similar scales in WMAP and Planck,  observations of the cosmic microwave background are largely consistent with the standard cosmological model within  $2\sigma$.
}


\maketitle

\section{Introduction}
\label{sec:intro}

At the epoch of recombination, matter and radiation separate, allowing radiation to stream freely in the Universe. This free-streaming radiation permeating the Universe, which we observe as the \textup{cosmic microwave background} (CMB) radiation, encodes a treasure trove of information about the initial conditions in the Universe \citep{ryden2003,jones2017precision}. Although it has a remarkably consistent average temperature, the CMB still exhibits tiny deviations of about $10^{-5}$ from the background average. The temperature fluctuations in the CMB trace the fluctuations in the underlying matter distribution in the early Universe that are linked to the spontaneous quantum fluctuations generated in an otherwise homogeneous medium  \citep{harrison1970,peebles1970}. Studying the properties of the temperature fluctuations in the CMB is therefore essential for understanding the properties of the primordial matter field. 

The lambda cold dark matter (LCDM) paradigm is the standard paradigm of cosmology. Together with the inflationary models in their simplest form \citep{starobinsky1982,guthpi1982}, the standard model of cosmology predicts the nature of the primordial stochastic matter distribution field to be that of an isotropic and homogeneous Gaussian random field \citep{harrison1970,guth1981}. This prediction finds allies theoretically in the central limit theorem, and observationally in the various measurements of the CMB temperature anisotropy field via ground- and space-based probes such as the \texttt{BOOMERanG} balloon-based experiment \citep{boomerang} and the Wilkinson Microwave Anisotropy Probe (WMAP) satellite \citep{wmap9}. The latest endeavor of measuring the CMB temperature anisotropies was the launch of the \textup{Planck} satellite, which boasts of the highest resolution in measurements to date. The resolution is at scales of a few arcminutes \citep{planckOverview2018}. Despite the general consensus that the CMB exhibits the characteristics of an isotropic and homogeneous Gaussian random field, a growing body of evidence shows anomalies in the observed CMB field with respect to the base model.  They include in particular the observed hemispherical asymmetry in the CMB power spectrum \citep{eriksen2004} as well as the alignment of low multipoles \citep{multipoles}; see \cite{cmbanomaliesstarkman} for a review. These observed anomalies raise doubts about the assumption of statistical isotropy and homogeneity, respectively. 

Testing the assumption of Gaussianity requires tools that encode information about higher orders. Traditional endeavor in this direction has focused on  higher-order correlation functions \citep{durrer1996}, which are generally extremely resource-intensive computationally \citep{planckcollaboration2016a}. Recently, attention has turned toward developing alternative tools beyond the correlation functions and multispectra, which may potentially encode information of all orders. The principal tools in this regard have arisen from integral geometry and involve computing the \textup{Minkowski functionals} or the \textup{Lifshitz-Killing curvatures} \citep{adler1981,mecke94,schmalzing1996,schmalzinggorski,sahni1998,codis2013,ducout2013,matsubara2010,chingangbam2017,pranav2019a,appleby2021minkowskiSDSS}. The $j$-th Minkowski functional and $(D-j)$-th Lifshitz-Killing curvature of a $D$-dimensional manifold $\Mspace$ are related by $Q_j(\Mspace) \ =\  j! \omega_j  \lips_{D-j}(\Mspace)$,   where $ j=0,\dots,D,$ and $\omega_j$ is the volume of the $j$-dimensional unit ball. There are $d$ such quantifiers for a $D$-dimensional set, where $d = 0, \ldots, D$. All but one are purely geometrical quantities that are related to the $d$-dimensional volume of the manifold. The exception is the $0$-th Lifshitz-Killing curvature, or equivalently, the $D$-th Minkowski functional, which is related to a purely topological quantity, the \textup{Euler characteristic} \citep{euler1758,adler1981,pranav2019a,eecestimate}, via Gauss's \textup{Theorema Egrerium} \citep{gauss1900,adler1981,pranav2019a}. The Minkowski functional computations of the CMB have consistently shown the observations to be congruent with the standard model \citep{planckIsotropy2015}. 

More recently, developments in computational topology have paved the way for extracting topological information from datasets at the level of \textup{homology} \citep{munkres1984,edelsbrunnerharer10,isvd10,pranavthesis,pranav2017,moraleda2019,pranavReview2021} and its hierarchical extension, \textup{persistent homology} \citep{edelsbrunnerharer10,pranav2017,shivashankar2015,rst,pranav2021topology2,heydenreich2021}. \textup{Topological data analysis} (TDA) involving homology and persistent homology has recently started finding application in astrophysical disciplines, for example, in the context of structure identification \citep{shivashankar2015,xu2019} and quantification of large-scale structures \citep{kono2020,wilding2020}, including detection and quantification of non-Gaussianities \citep{feldbrugge2019,biagetti2020}. Homology describes the topology of a space by identifying the holes and the topological cycles that bound them. A $d$-dimensional space may contain topological cycles of $0$ up to $d$ dimensions. The cycles and holes are associated with the \textup{homology groups} of the space. The $p$-th \textup{Betti number}, $\Betti{p}$, is the rank of the $p$-th homology group, $\mathbb{H}_{p, p = 0 \ldots d}$. While itself a purely topological quantity, the Euler characteristic is also the alternating sum of the Betti numbers of all ambient dimensions of a manifold, as denoted by the Euler-Poincar\'{e} formula \citep{adler1981,pranav2017,pranav2019a}. The Euler characteristic has a long history in the analysis of cosmological fields \citep{gdm86,pogosyan2009,ppc13,appleby2020}. 

Building on and refining existing tools from computational topology in the context of analyzing the CMB field, this paper presents the homology characteristics of the temperature fluctuation maps of the cosmic microwave background obtained by the \textup{Planck} satellite \citep{planckOverview2018}. We perform our experiments on the fourth and final data release, \textup{Planck 2020} Data Release 4 (DR4), which is based on the \texttt{NPIPE} data-processing pipeline \citep{npipe}. The \texttt{NPIPE} dataset represents a natural evolution of the Planck data-processing pipeline, integrating the best practices from the LFI and HFI pipelines separately. The result is an overall amplification of signal and reduction in the associated systematic, noise, and residuals at almost all angular scales \citep{npipe}. For comparison, we also present results for the \textup{Planck 2018} Data Release 3 (DR3) \citep{planckOverview2018}, which is based on the \textup{Full Focal Plane} (FFP) data processing pipeline \citep{plancksims}, resulting in the \texttt{FFP10} simulations \citep{ffp10}. The paper follows the spirit of \cite{pranav2019b}  in methods and analysis, where we present results for the intermediate \textup{Planck 2015} Data Release 2 (DR2) \citep{planckIsotropy2015}; also see \cite{rst}. The novel aspect of the methods presented here and in \cite{pranav2019b} is an analysis pipeline that takes regions with unreliable data on $\Sspace^2$ into account. In the case of CMB, this is reflected in the obfuscation effects of the measurements that are due to foreground objects such as our own galaxy, as well as other extra- and intragalactic foreground sources. We masked these regions and computed the homology of the excursion sets relative to the mask. 

We present a brief description of the topological background in Section~\ref{sec:topology}, followed by the results in Section~\ref{sec:result}. We discuss the ramifications of the results and conclude the main body of the paper in Section~\ref{sec:discussion}. The appendices present a brief description of the datasets and the computational pipeline as well as a validation for the statistical tests.

\begin{figure}
        \centering
        \subfloat{\includegraphics[width=0.5\textwidth]{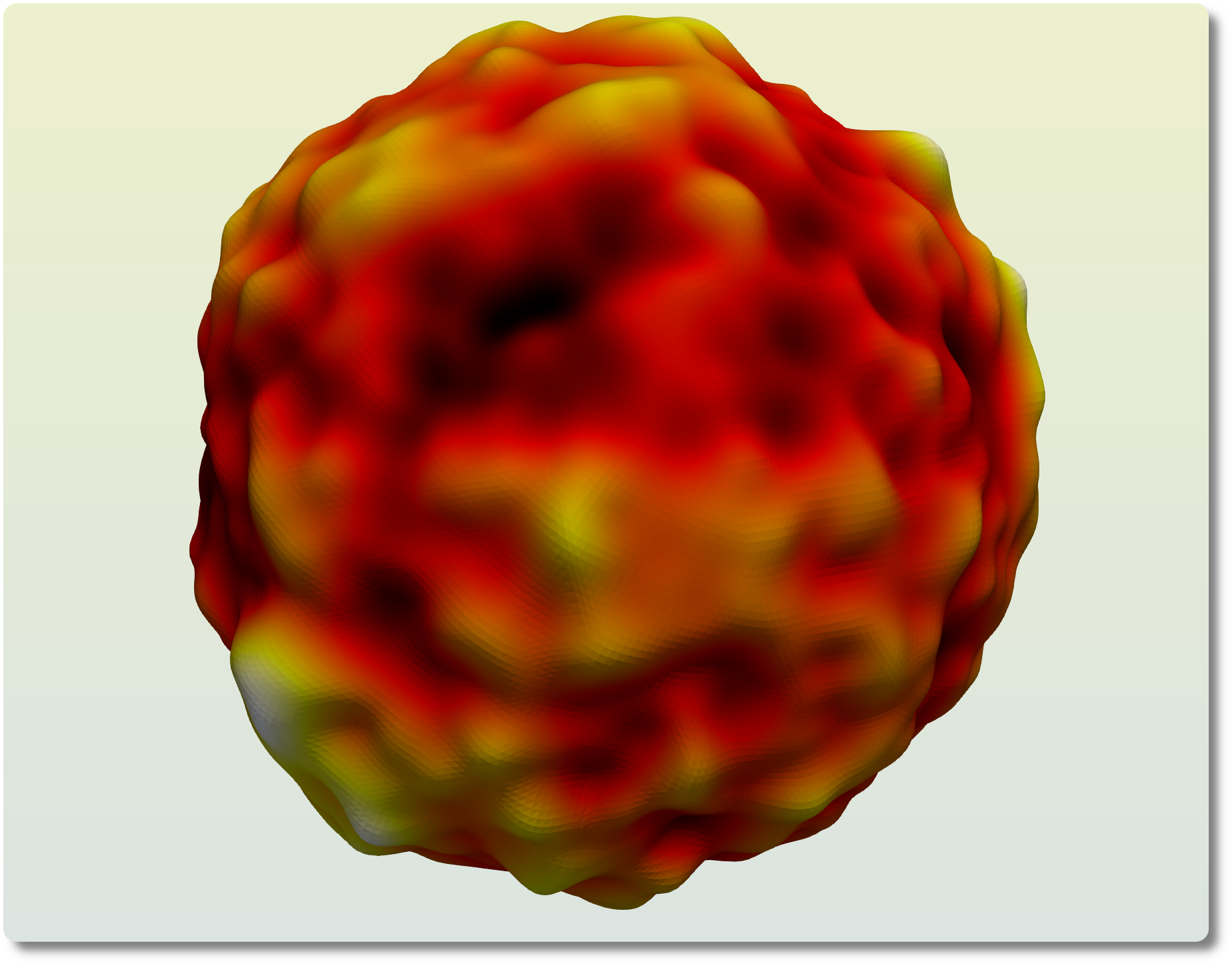} }
        \caption{Visualization of the temperature fluctuations in the CMB sky. The survey surface $\Sspace^2$ is distorted at each point in the direction of the surface normal. The distortion is proportional to the fluctuation in direction and magnitude. The visualization is based on the observed CMB sky cleaned by the \texttt{NPIPE} pipeline and smoothed at $5$ degrees.}
        \label{fig:cmbSky}
\end{figure}

\begin{figure*}
        \centering   
        \subfloat[]{\includegraphics[width=0.49\textwidth]{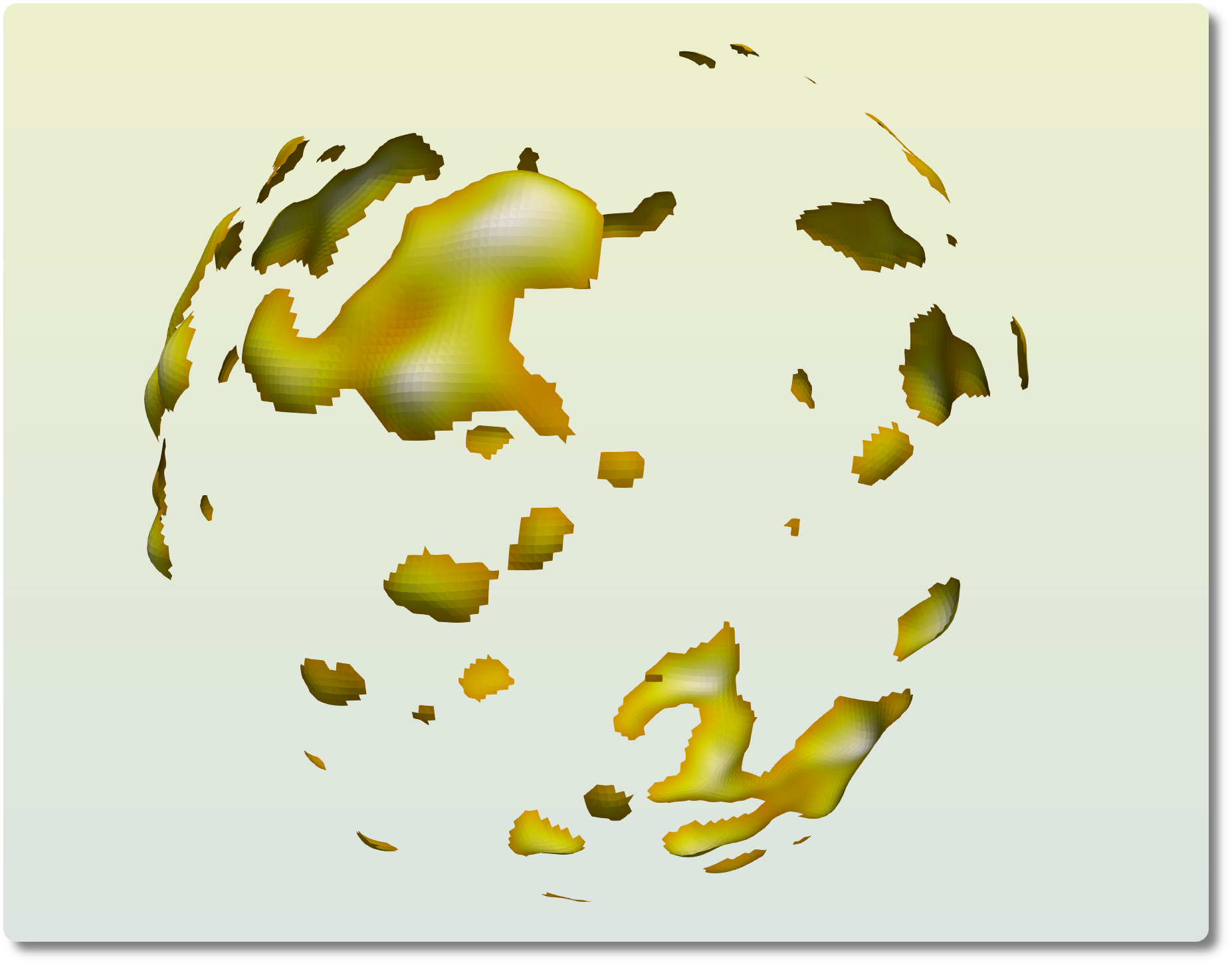}}
        \subfloat[]{\includegraphics[width=0.49\textwidth]{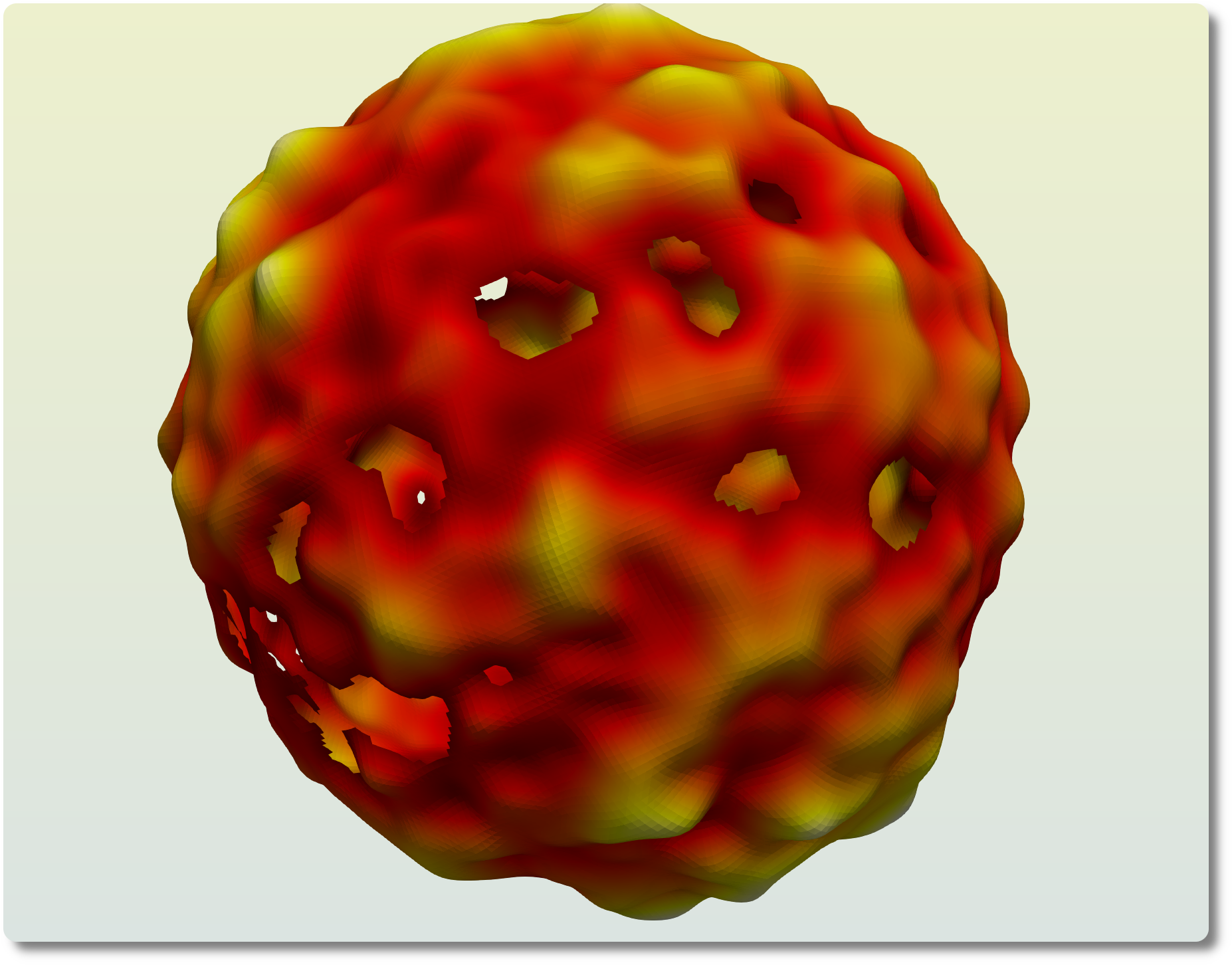}}
        
        \caption{Cosmic microwave background sky thresholded at moderately positive (left) and negative (right) levels. For high thresholds, the excursion set is dominated by isolated components, while at low thresholds, it gives the appearance of a single connected surface indented by numerous holes. For sufficiently low thresholds, the holes fill up, and the excursion set covers the entire sphere, which is composed of a connected surface without a boundary that encloses a single void (cf. Figure~\ref{fig:cmbSky}).}
        \label{fig:cmbThld}
\end{figure*}

\begin{figure}
        \centering
        \subfloat{\includegraphics[width=0.5\textwidth]{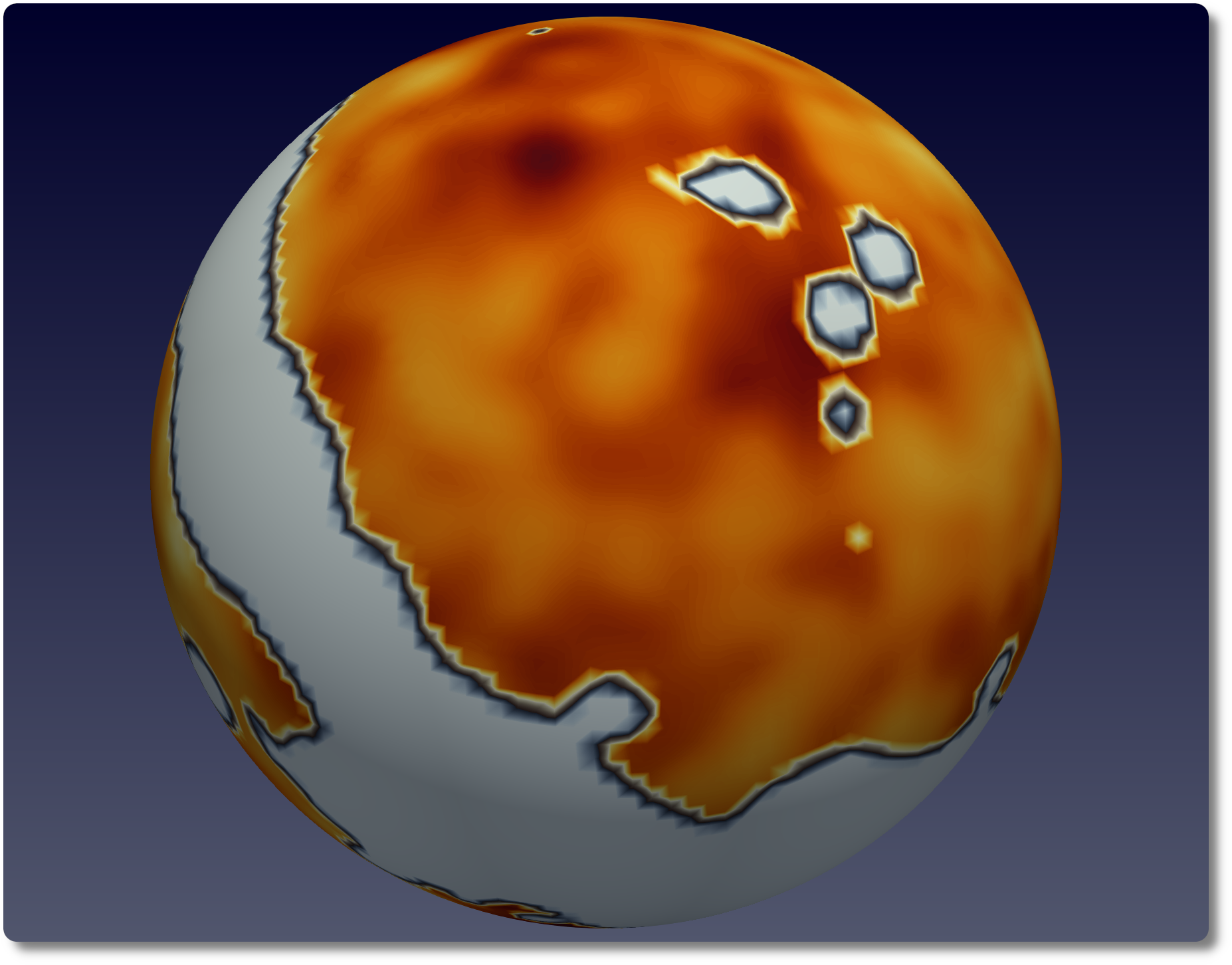} }
        \caption{Visualization of the temperature fluctuations in the CMB sky in the presence of masks (plotted in gray). It consists of an equatorial belt and numerous patches on the northern and southern cap, which correspond to our galaxy and other bright foreground objects. The visualization is based on the observed CMB sky cleaned by the \texttt{NPIPE} pipeline and smoothed at $5$ degrees.}
        \label{fig:maskedSky}
\end{figure}

\section{Topological background}
\label{sec:topology}

Commensurate with our intention of analyzing the topology of the CMB temperature fluctuations, we restricted ourselves to topological definitions on $\Sspace^2$ \citep{pranav2019b} and invoked \textup{relative homology} to account for analysis in the presence of masked regions. The standard reference for this section is \cite{edelsbrunnerharer10}; also see \cite{pranav2019b} for discussion in the context of CMB, as well as \cite{heydenreich2021} in the context of cosmic shear fields.

\subsection{Homology characteristics of excursion sets of $\Sspace^2$}

Denoting the CMB temperature fluctuations on $\Sspace^2$ as $f \colon \Sspace^2 \to \Rspace$, we define the \textup{excursion set} \footnote{Excursion sets are also known as \textup{superlevel sets} and their boundary thresholds as \textup{levelsets}, or simply \textup{levels}.} at a temperature $\nu$ as the subset of $\Sspace^2$ where the temperature is higher than or equal to $\nu$,
\begin{equation}
\Excursion (\nu)  =  \{ x \in \Sspace^2  \mid  f(x) \geq \nu \}.
\end{equation}

In the cosmological setting, the usual practice is to examine the dimensionless threshold, which is the mean-subtracted and variance-scaled field derived from the original field, in which case, $\nu = (f - \mu_f)/\sigma_f$, and $\mu_f$  and $\sigma_f$ are the mean and the standard deviation of the field $f$. If $\Excursion (\nu)$ does not cover the entire $\Sspace^2$, it may be composed of isolated components and holes. Figure~\ref{fig:cmbThld} presents excursion sets corresponding to two different thresholds. For high thresholds, presented in the left panel, the excursion set is dominated by components, while for low thresholds, presented in the right panel, the excursion set is dominated by a few large connected objects indented with holes, which are bounded by loops. The \textup{Betti numbers} $\Betti{0}$ and $\Betti{1}$ count the number of independent components and loops of the excursion set, respectively. In general, for a $d$-dimensional topological space, $\Betti{p}$ is the rank of the $p$-th \textup{homology group}, $\Homology{p};p = 0, \ldots, d$, and counts the number of independent $p$-dimensional cycles \citep{munkres1984,edelsbrunnerharer10,pranav2017}. If $\Excursion (\nu)$ does not cover the entire $\Sspace^2$, the number of independent loops is one less than the total number of loops. If $\Excursion (\nu)$ covers the entire $\Sspace^2$, there are no loops, and $\Betti{2} = 1$, because of the void that is enclosed by the boundary-less surface of the sphere. A related quantity that has a long history of usage in cosmological analyses is the \textup{Euler characteristic}, or alternatively, the \textup{genus} \citep{gdm86,ppc13}, which is the alternating sum of the Betti numbers of the excursion set,
\begin{equation}
\Euler (\nu) = \Betti{0} (\nu) - \Betti{1} (\nu) + \Betti{2} (\nu).
\end{equation}

The Euler characteristic also has a geometric interpretation as one of the \textup{Lifshitz-Killing} curvatures of the manifold \citep{adler1981,pranav2019b}.

\subsection{Masks and relative homology}

The measurement of the CMB signal is unreliable in certain parts of the sky due to interference from bright foreground objects. These include extended objects such as our galaxy, as well as bright point sources. We masked these regions and computed the homology characteristics of the excursion set relative to the mask. Figure~\ref{fig:maskedSky} presents a visualization of the masked CMB sky. Letting $\Mask \subseteq \Sspace^2$ be the mask and $\Excursion (\nu)$ the excursion set,  we considered the \textup{relative homology} of the pair of closed spaces, $(E, M)$, where $E = \Excursion (\nu)$ and $M = \Mask \cap \Excursion (\nu)$. We note that $M$ is contained in $E$ and is an open set by definition in this context. We denote the rank of the \textup{relative homology groups} of the pair $(E, M)$ by $\relBetti{p} = \Rank{\Homology{p} (E, M)}; p = 0, 1, 2$. The Betti numbers computed considering the pair $(E, M)$ are different from the Betti numbers of the excursion set without a mask.  For a more detailed discussion about relative homology in the context of masked CMB sky, see \cite{pranav2019b}. The \textup{relative Euler characteristic}, as in the case of absolute homology, is the alternating sum of the rank of relative homology groups,
\begin{equation}
\relEuler (\nu) = \relBetti{0} (\nu) - \relBetti{1} (\nu) + \relBetti{2} (\nu).
\end{equation}

\begin{figure*}
        \centering
        \subfloat[][]{\includegraphics[width=0.8\textwidth]{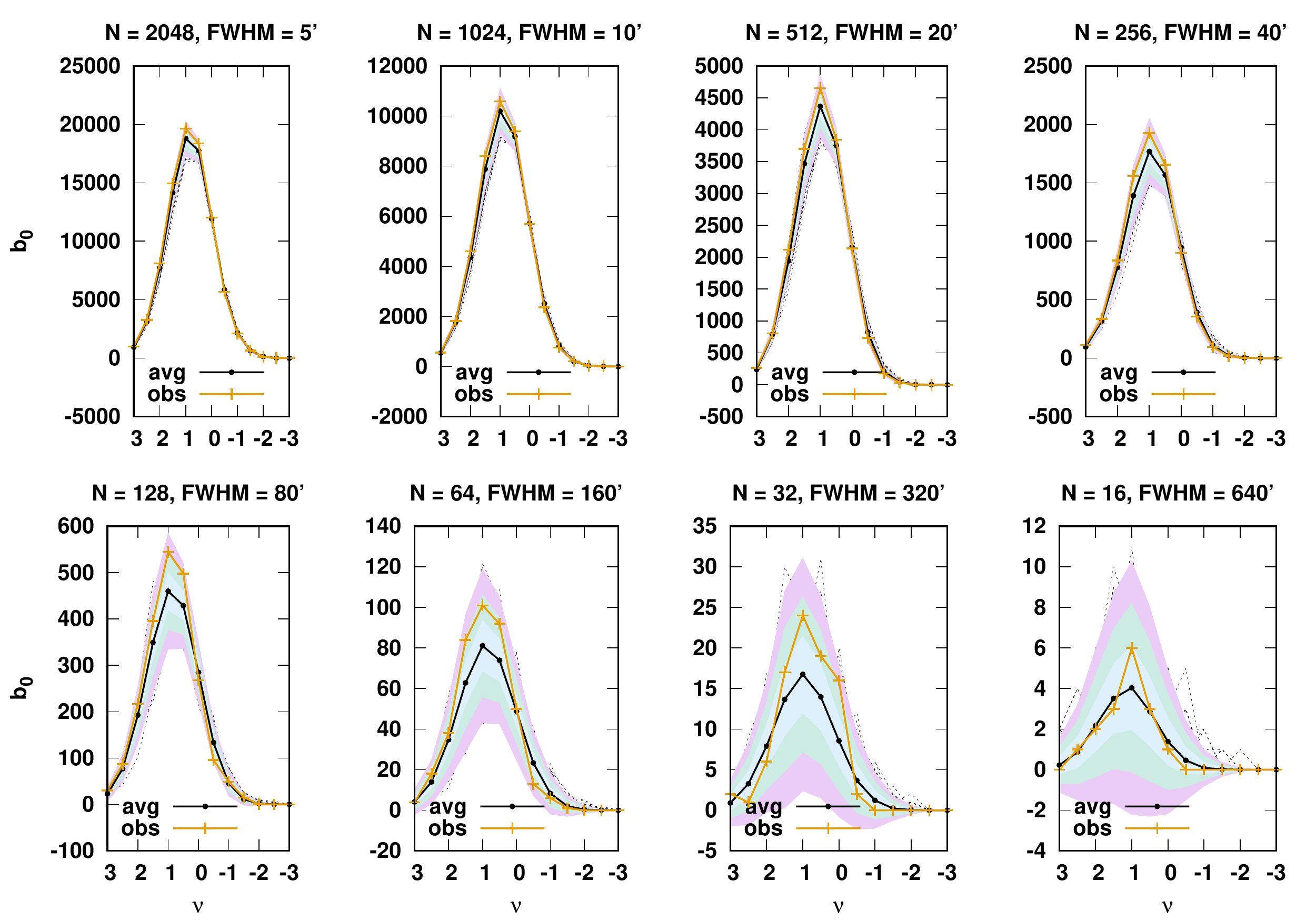}}\\
        \subfloat[][]{\includegraphics[width=0.6\textwidth]{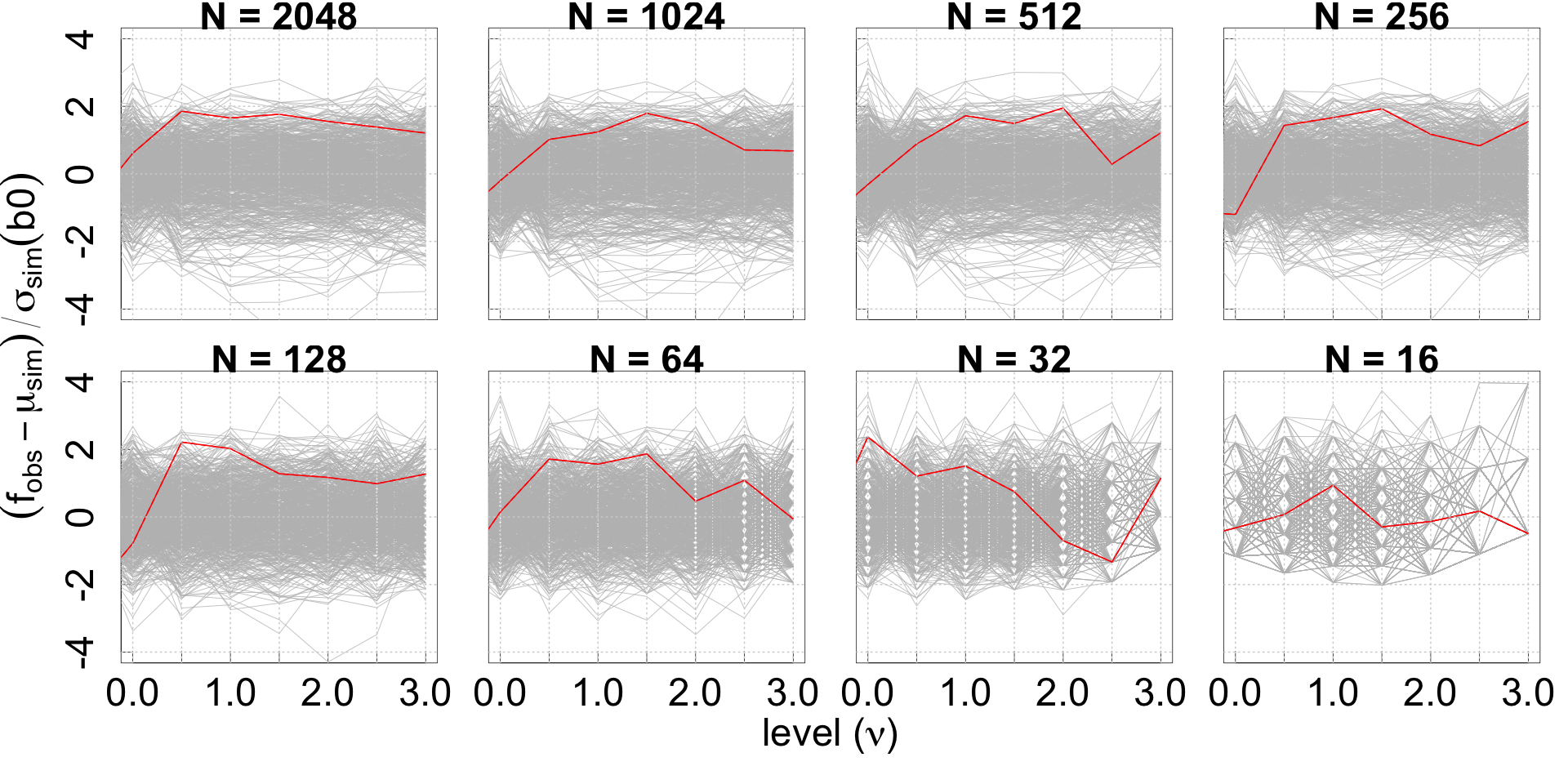} }\\
        
        \caption{Graphs of $\relBetti{0}$ for the \texttt{NPIPE} dataset for various degraded and smoothing scales. Panel (a) presents the observational curve in yellow, and the curves corresponding to the average of simulations are presented in gray. Error bands corresponding to $(1\sigma:3\sigma)$ are also drawn.  Panel (b) presents the curve for the significance of the differences, where the observed curves are presented in red, and the values for the simulations are presented as dotted gray lines.}
        \label{fig:betti0_graph_npipe}
\end{figure*}

\begin{figure*}
        \centering
        
        \subfloat[][]{\includegraphics[width=0.8\textwidth]{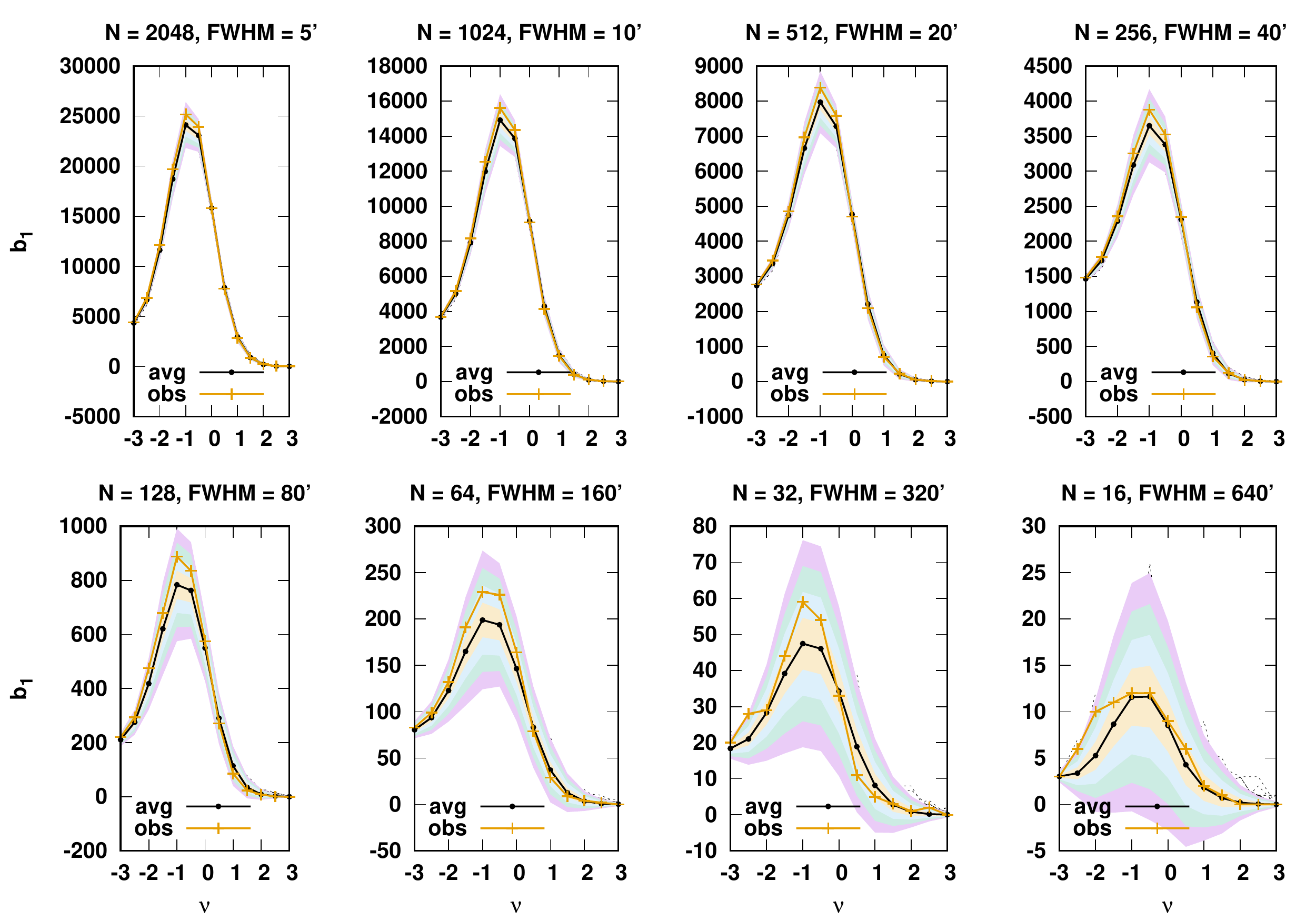}}\\
        \subfloat[][]{\includegraphics[width=0.6\textwidth]{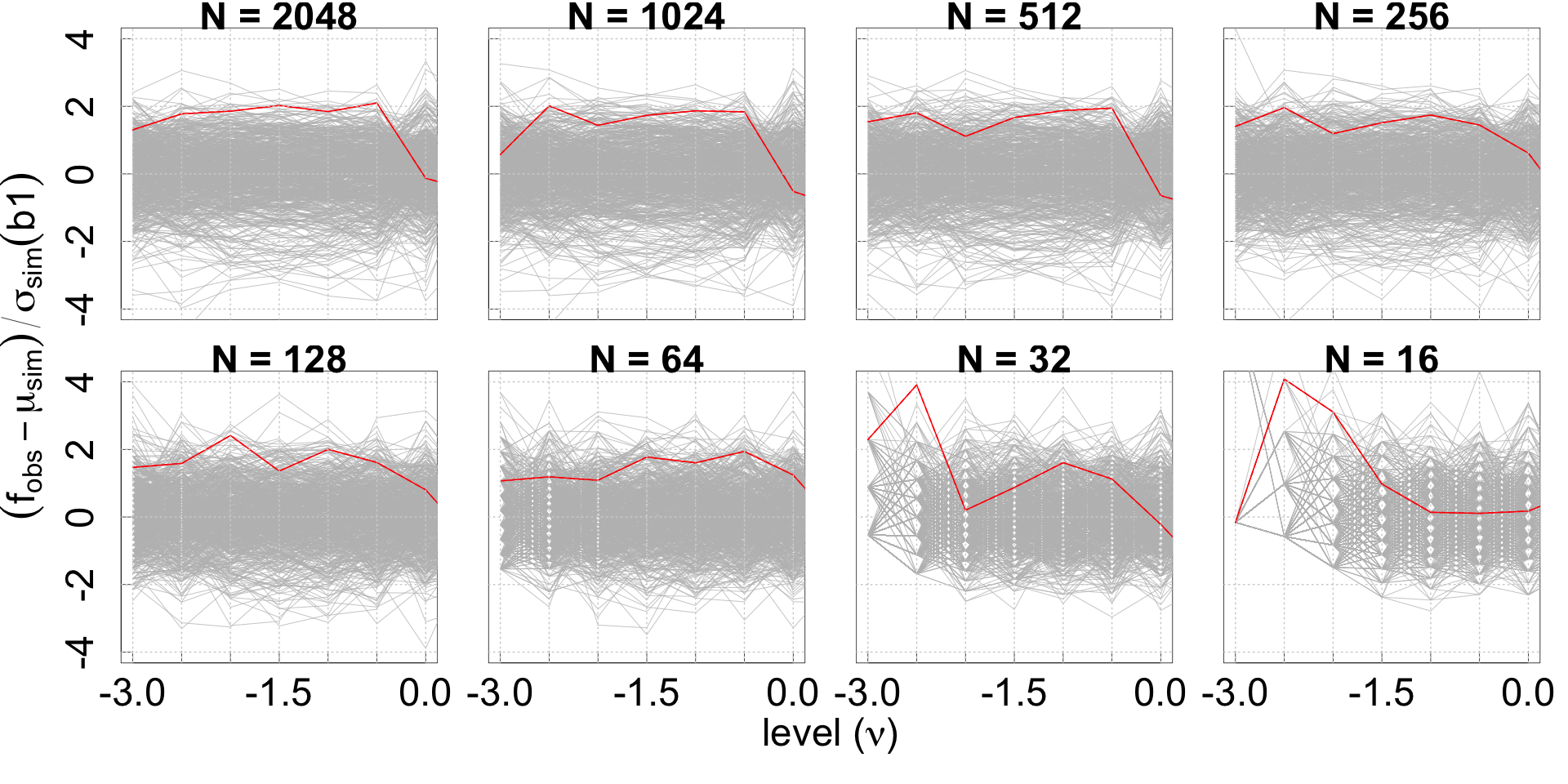} }\\
        
        \caption{Graphs of $\relBetti{1}$ for the \texttt{NPIPE} dataset for various resolutions and smoothing scales. In panel (a), the observational curves are presented in yellow, and the curves corresponding to the average of simulations are presented in gray, while panel (b) presents the significance of the differences. Error bands corresponding $(1\sigma:4\sigma)$ are also presented in various colors. }
        \label{fig:betti1_graph_npipe}
\end{figure*}

\begin{figure*}
        \centering
        \subfloat{\rotatebox{-90}{\includegraphics[height=0.8\textwidth]{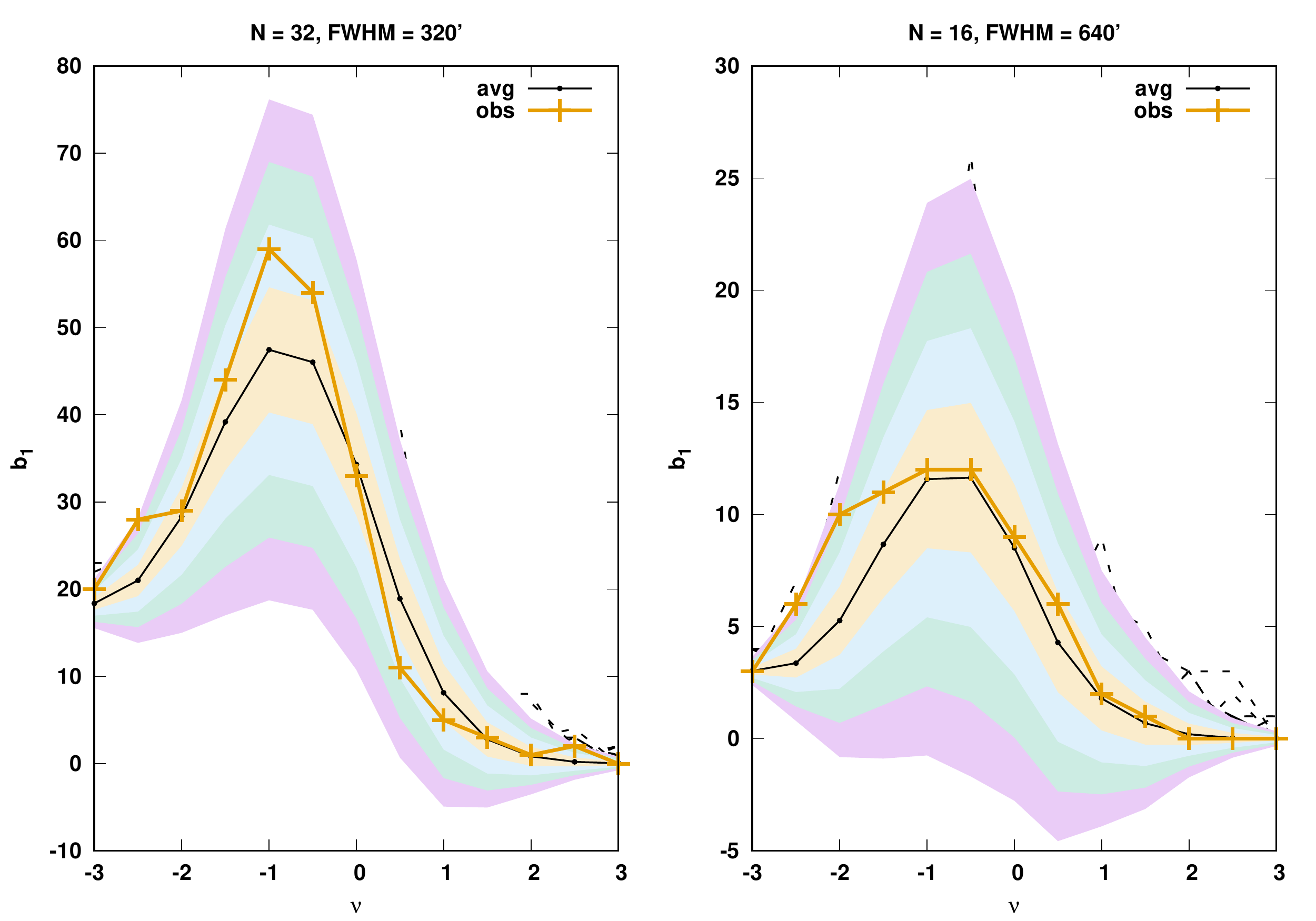}}}\\
        \subfloat{\includegraphics[width=0.43\textwidth]{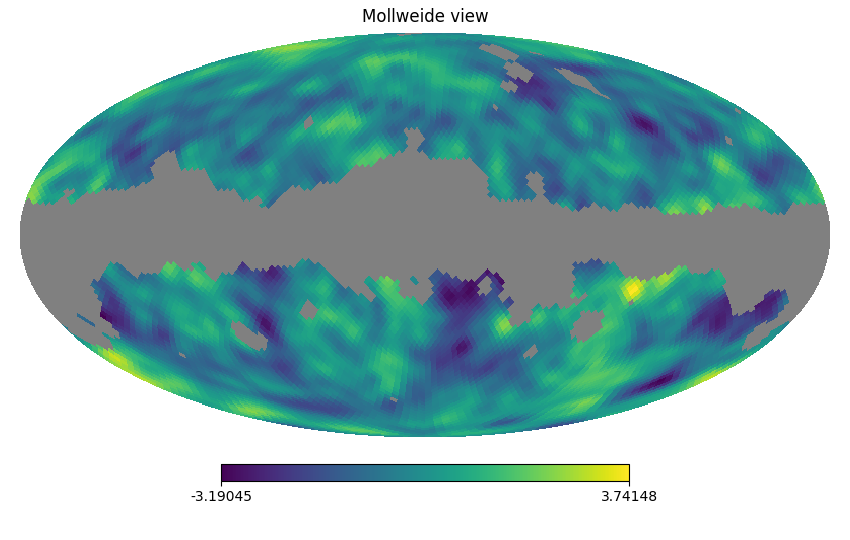} }
        \subfloat{\includegraphics[width=0.43\textwidth]{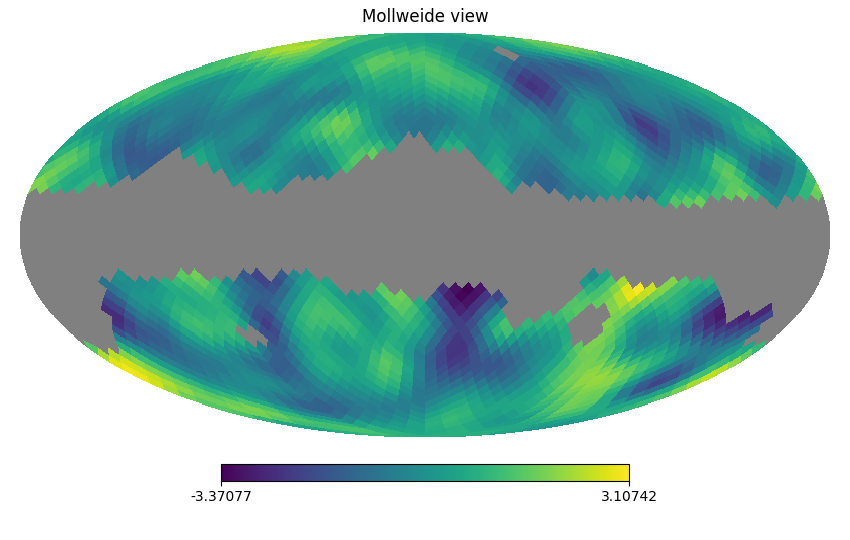} }\\
        \caption{Graph of $\relBetti{1}$, selectively for $\Res= 32$ and $\Res= 16$, corresponding to $FWHM = 320'$ and $FWHM = 640'$, presenting an enlarged view of the concerned resolutions. The observational curve is presented in yellow, and the curves corresponding to the average of simulations are presented in gray. Error bands corresponding to  $(1\sigma:4\sigma)$ are also drawn. The bottom two panels present the visualization of the scalar temperature field in order to facilitate an appreciation of the structure of the field.}
        \label{fig:betti1_graph_32_16}
\end{figure*}

\begin{figure}
        \centering
        \includegraphics[width=0.5\textwidth]{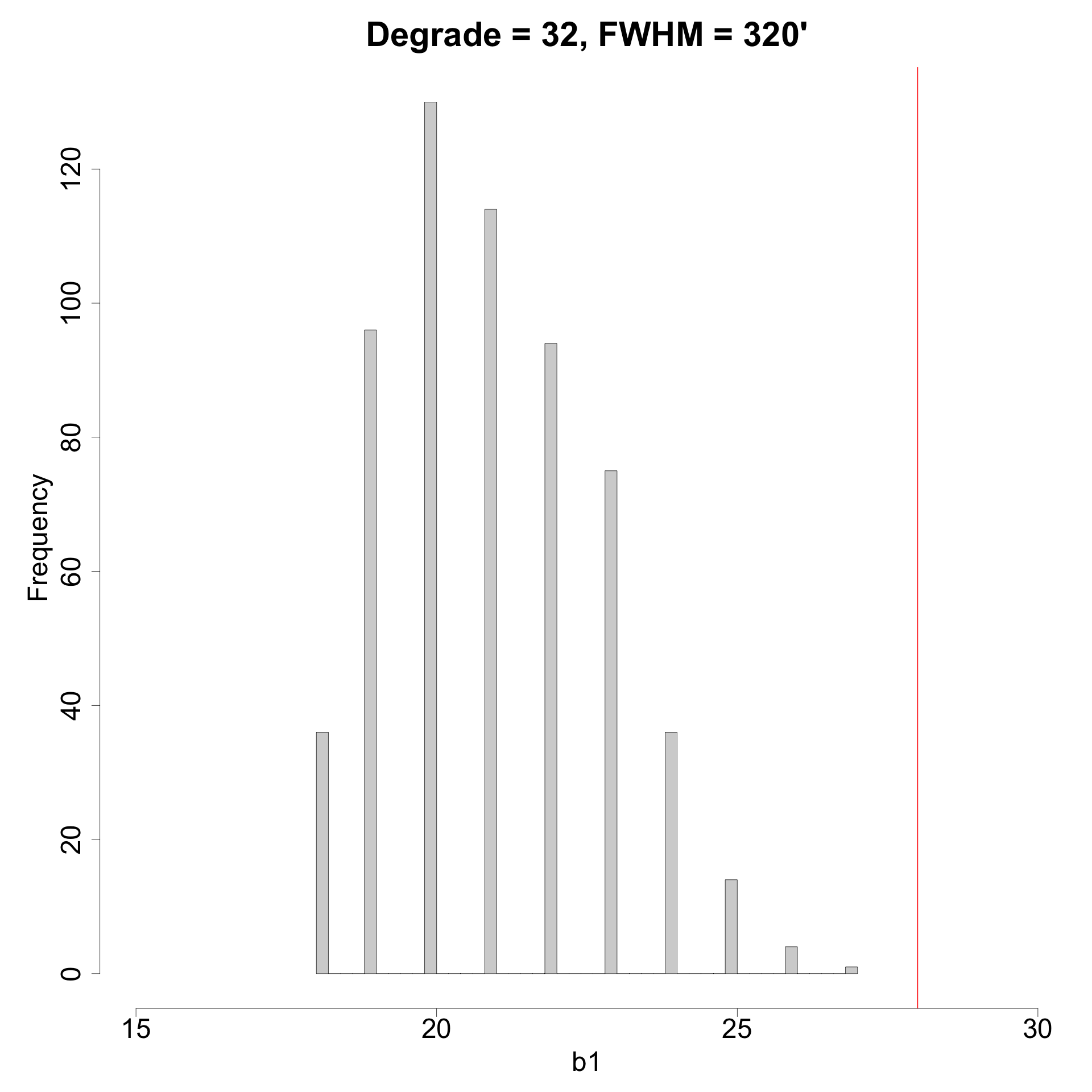}
        \caption{Histogram of  $b_1$ at $\nu = -2.5$, where the $\sim4\sigma$ deviation occurs. The minimum value attained across $600$ simulations is $18$, with a mean at $\sim21$ and a standard deviation of $\sim1.7$. The observed map exhibits $b_1 = 28$ and is well outside the distribution. }
        \label{fig:b1_level-2.5_distr}
\end{figure}

\begin{figure*}
        \centering
        
        \subfloat[][]{\includegraphics[width=0.8\textwidth]{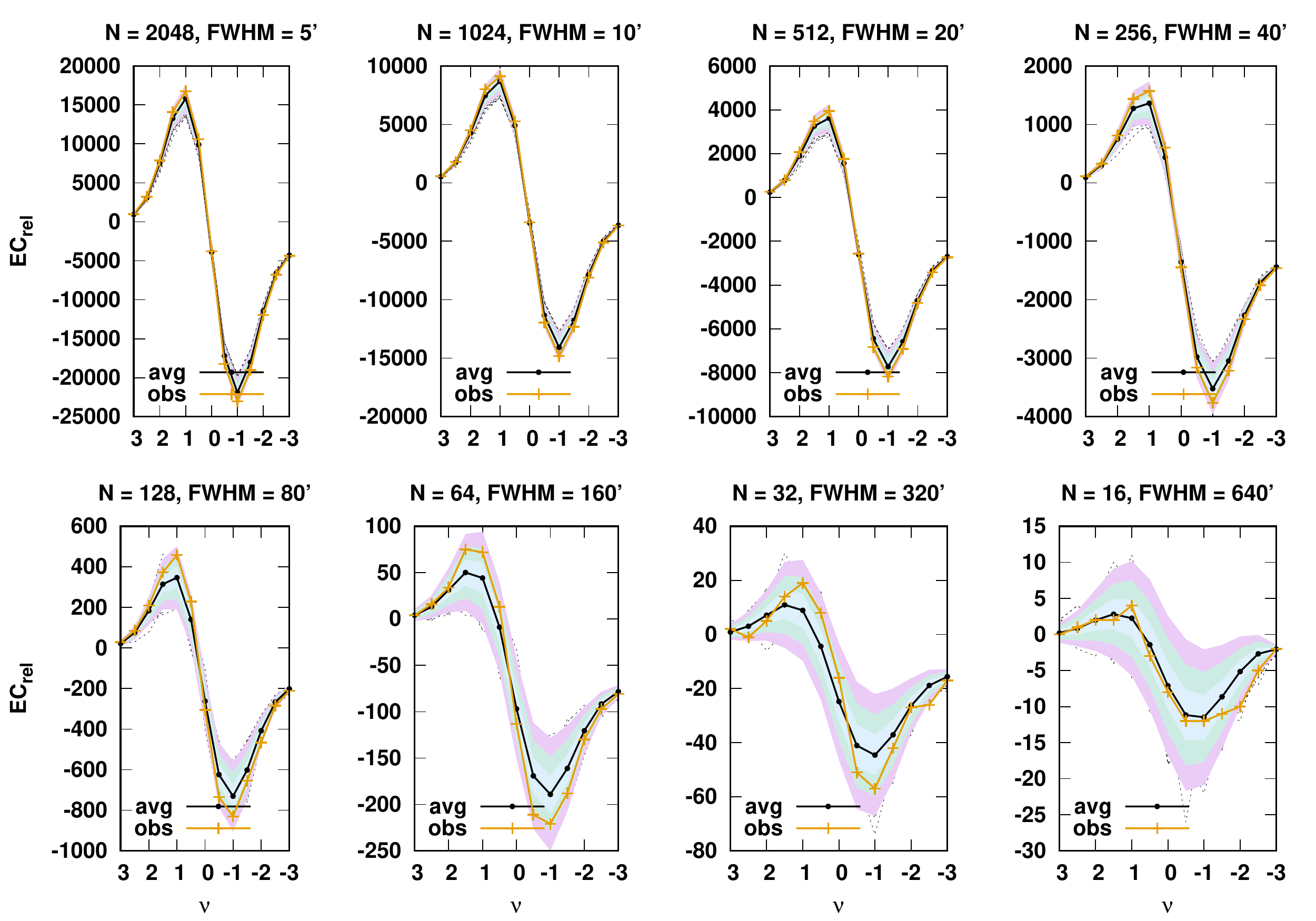}}\\
        \subfloat[][]{\includegraphics[width=0.6\textwidth]{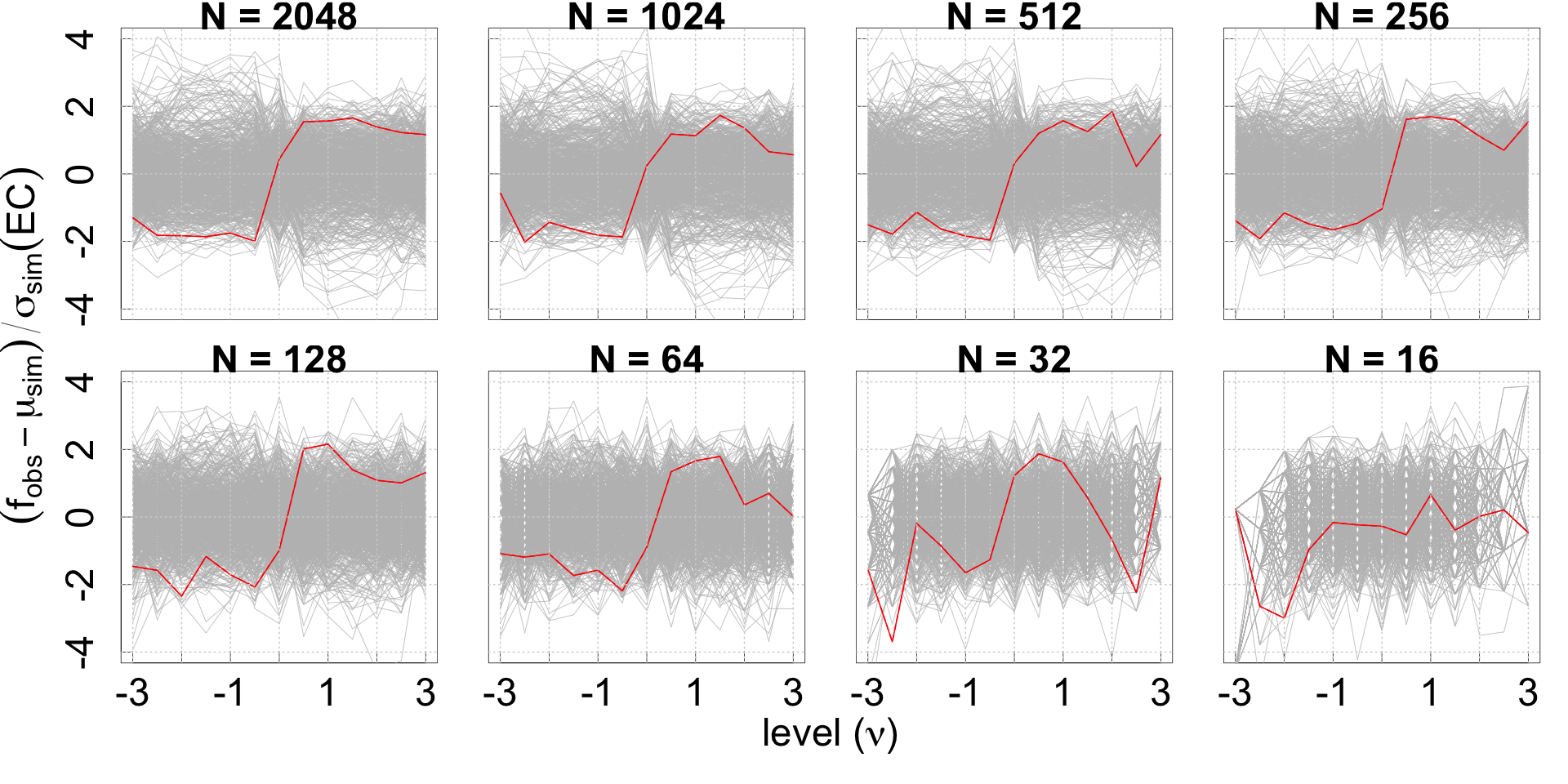}}\\
        
        \caption{Graphs of $\relEuler$ for the \texttt{NPIPE} dataset. Panels (a) and (b) present similar information as Figures~\ref{fig:betti0_graph_npipe} and~\ref{fig:betti1_graph_npipe}. $\relEuler$ exhibits deviations commensurate with those in $\relBetti{0}$ and $\relBetti{1}$.}
        \label{fig:ec_graph_npipe}
\end{figure*}

\begin{figure*}
        \centering
        \subfloat[][]{\includegraphics[width=0.8\textwidth]{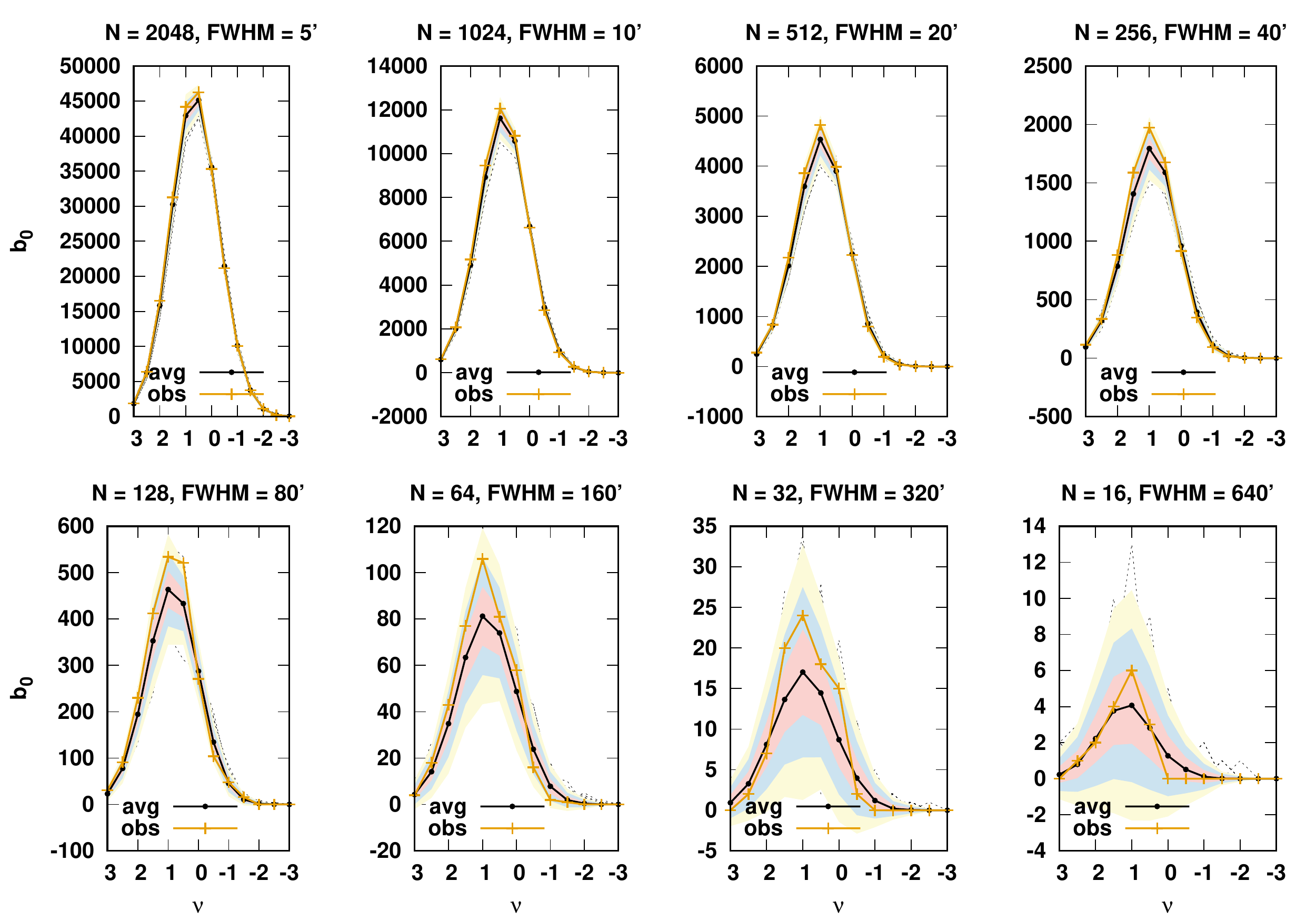}}\\
        \subfloat[][]{\includegraphics[width=0.6\textwidth]{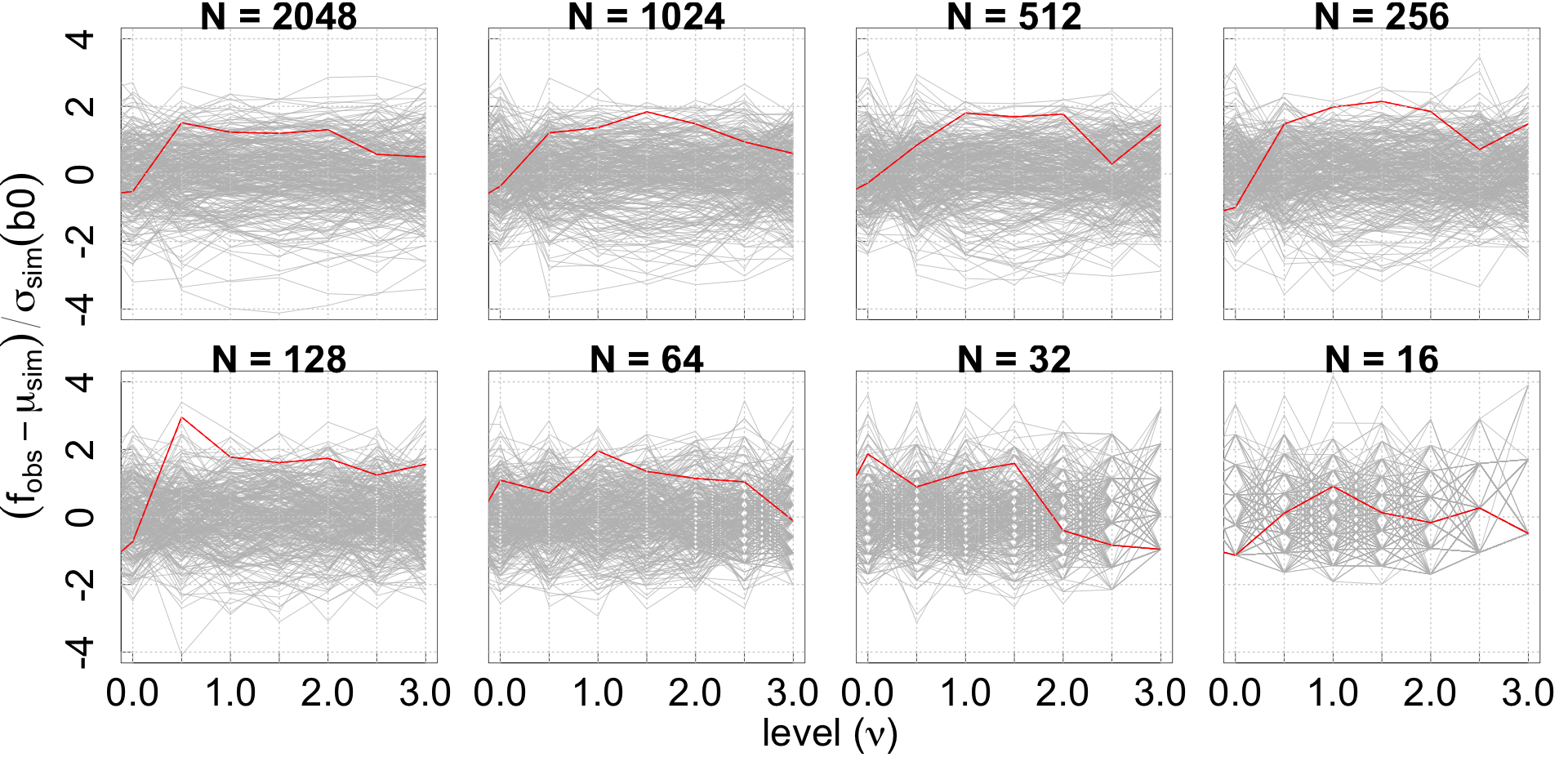} }
        
        \caption{Graph of $\relBetti{0}$ for the \texttt{FFP10} dataset. In panel (a), the observational curves are presented in yellow, and the curves corresponding to the average of simulations are presented in gray. Error bands corresponding to $(1\sigma:3\sigma) $ are also drawn. Panel (b) presents the significance of the differences. The dataset exhibits milder deviations than the \npipe dataset in general. However, we note a $2.96\sigma$ deviation in the number of components between the simulations and observation at $\Res = 128, FWHM = 80'$.}
        \label{fig:betti0_graph_ffp10}
\end{figure*}

\begin{figure*}
        \centering
        
        \subfloat[][]{\includegraphics[width=0.8\textwidth]{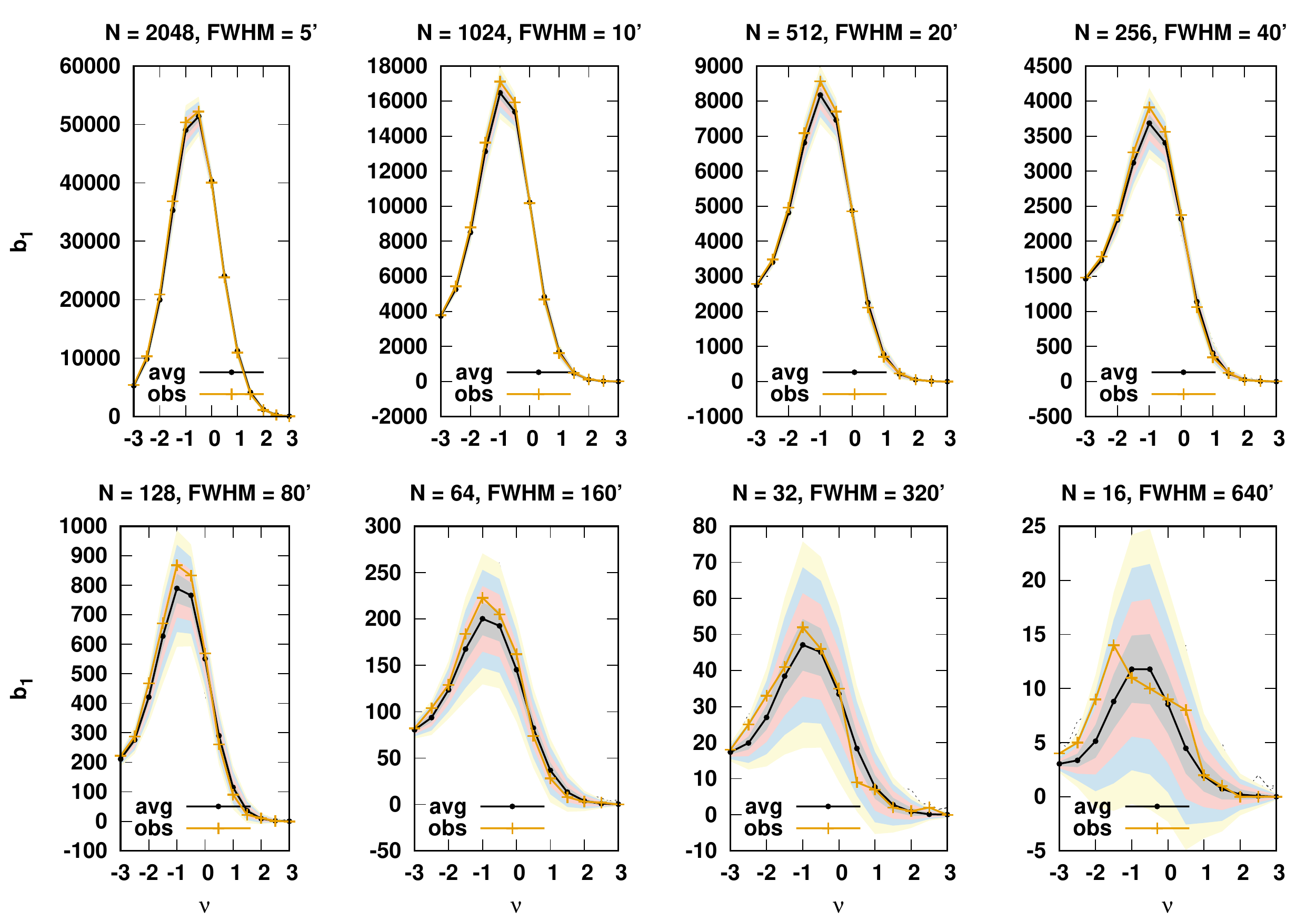}}\\
        \subfloat[][]{\includegraphics[width=0.6\textwidth]{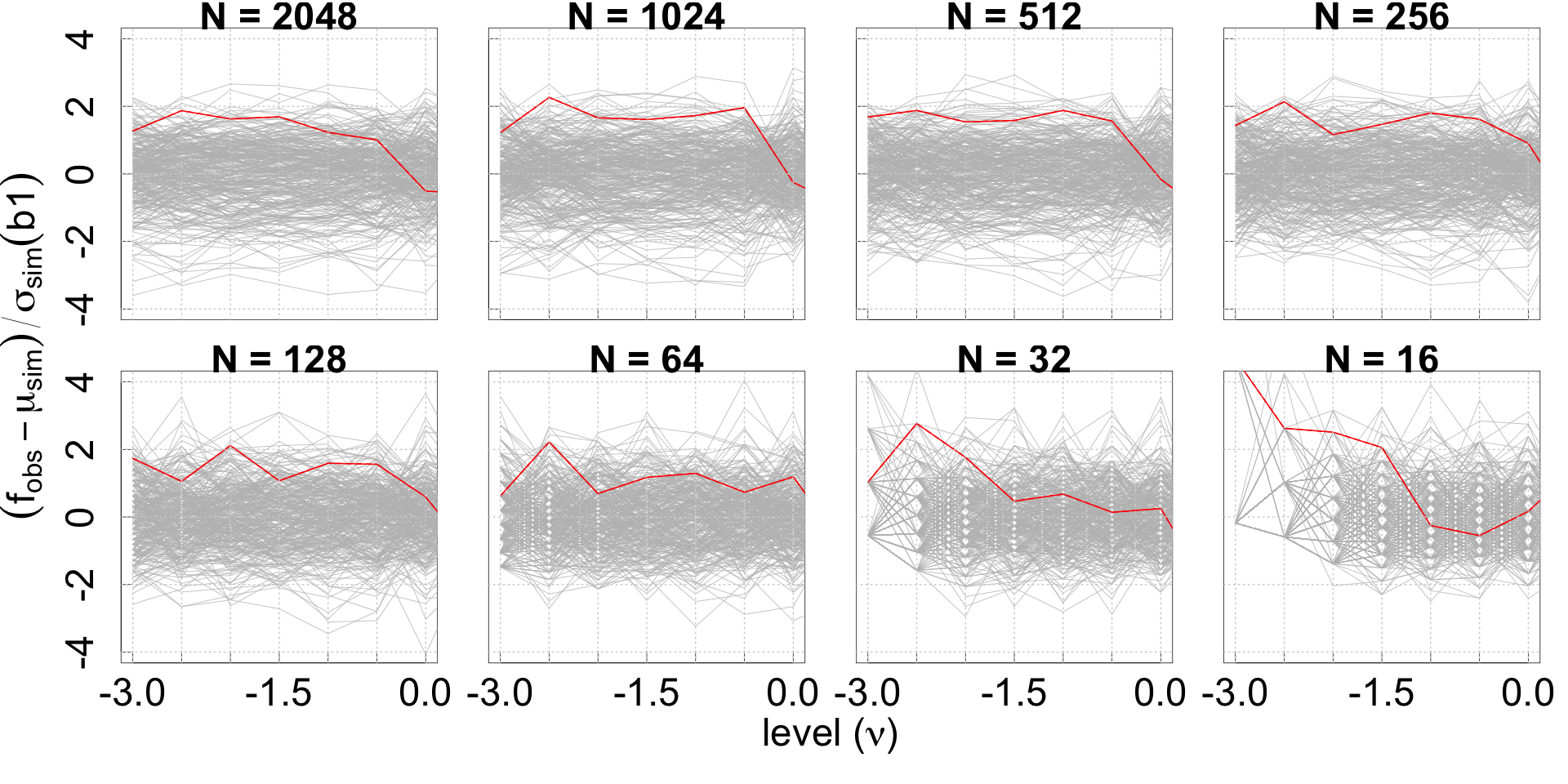}}
        
        \caption{Graph of $\relBetti{1}$ for the \texttt{FFP10} dataset. In panel (a),  the observational curves are presented in yellow, and the curves corresponding to the average of simulations are presented in gray. Error bands corresponding to $(1\sigma:4\sigma)$ are also drawn. Panel (b) presents the significance of the differences. The dataset exhibits milder deviations than the \npipe dataset in general, but there are mildly significant deviations at  $2.77\sigma$, at scales and thresholds where the \npipe dataset shows strong deviations as well.}
        \label{fig:betti1_graph_ffp10}
\end{figure*}

\begin{figure*}
        \centering
        
        \subfloat[][]{\includegraphics[width=0.8\textwidth]{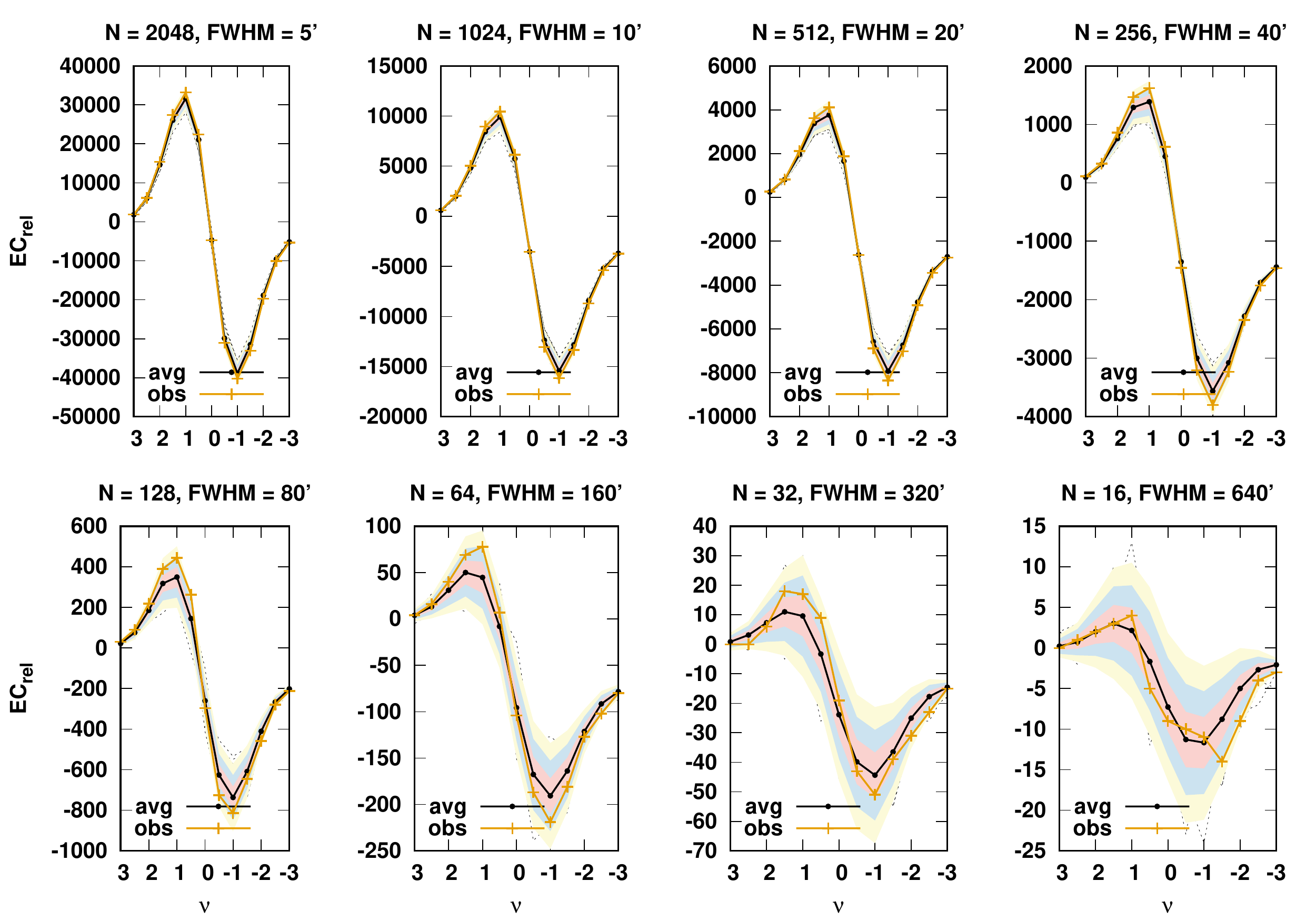}}\\
        \subfloat[][]{\includegraphics[width=0.6\textwidth]{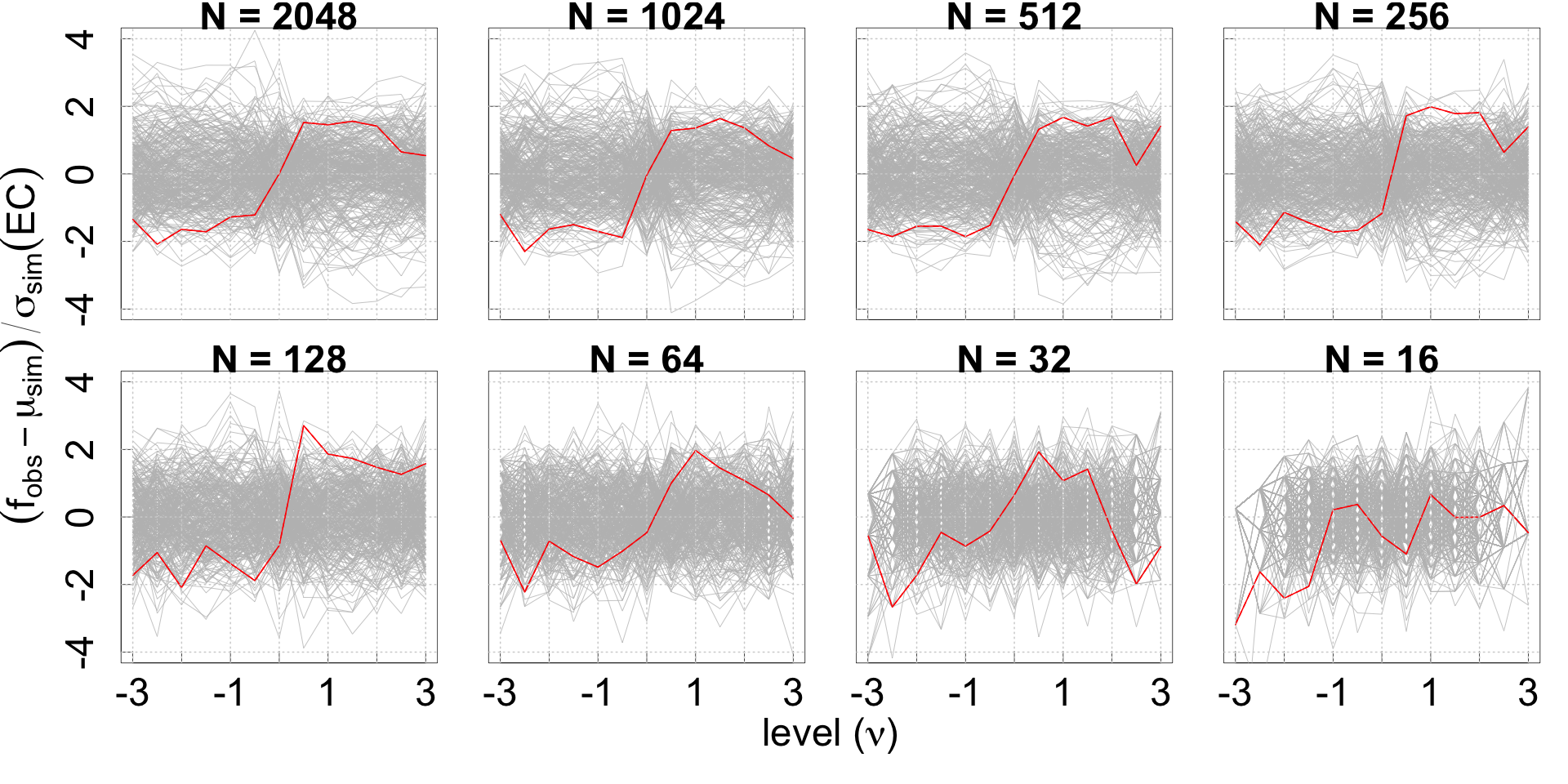} }
        
        \caption{Graph of $\relEuler$ for the \texttt{FFP10} dataset. In panel (a), the observational curve is presented in yellow, and the curves corresponding to the average of simulations are presented in gray. Error bands corresponding to $(1\sigma:3\sigma)$ are also drawn. Panel (b) presents the significance of the differences. The Euler characteristic exhibits deviations commensurate with $\relBetti{0}$ and $\relBetti{1}$.}
        \label{fig:ec_graph_ffp10}
\end{figure*}

\section{Results}
\label{sec:result}

In this section, we present our results in terms of the ranks of homology groups, $\relBetti{p}$, $p = 0, 1$, as well as the Euler characteristic, $\relEuler$, relative to the mask. Section~\ref{sec:npipe_result} presents results from the \texttt{DR4 NPIPE} dataset based on $600$ simulations, obtained using the \texttt{SEVEM} component separation pipeline. Similar results for the \texttt{DR3 FFP10} dataset based on $300$ simulations,  obtained using the \texttt{SMICA} component separation pipeline, are presented in Section~\ref{sec:ffp10_result} \footnote{The Planck CMB maps are hosted at {\url{https://pla.esac.esa.int/}}. The secondary datasets encapsulating topological information of CMB maps are hosted at {\url{https://www.pratyushpranav.org/cmb_data/cmb_data_archive.tar.gz}}. The analysis codes are available at {\url{https://www.pratyushpranav.org/codes/topos2.tar.gz}}.}. The latter dataset facilitates a direct  comparison with the topo-geometrical studies of the CMB performed by the Planck team in \cite{planckIsotropy2018}. For both datasets, we begin by examining the graphs of $\relBetti{0}$, $\relBetti{1}$, and of the (relative) Euler characteristic, $\relEuler$. This is followed by statistical tests that estimate the significance of results, taking into account all of the levels together for a given resolution and scale. We employed the standard  $\chi^2$-test in the empirical and theoretical settings, as well as the model-independent \textup{Tukey depth} test for this purpose.  We chose a-priori levels $\ell_{k}$, where $\ell_k = k/2$, setting $k_{\relBetti{0}} [0:6], k_{\relBetti{1}} = [-6:0], k_{\relEuler} = [-6:6]$.  We did this to restrict ourselves to analyzing $\relBetti{0}$ for positive thresholds, $\relBetti{1}$ for negative thresholds, and $\relEuler$ across the full threshold range. The choice of regions is determined by the fact that $\relBetti{0}(\nu)$  tends to be  small and carries little information for $\nu<0$, $\relBetti{1}(\nu)$ tends to be small for $\nu>0$, while the Euler characteristic is informative over the full range of levels.  These levels  are consistent with the levels investigated in \cite{planckIsotropy2015} and \cite{planckIsotropy2018}.

\subsection{Planck 2020 Data Release 4: \texttt{NPIPE} dataset}
\label{sec:npipe_result}

In this section, we analyze the graphs of the topological descriptors for the \texttt{NPIPE} dataset. Treating the topological descriptors as discrete random variables,  we compute the mean, $\mu_{sim}$, and  the standard deviation, $\sigma_{sim}$, from the simulations in the usual sense for each of the levels.  We use them  to assess the model-independent significance of the difference between the simulations and observations.

\subsubsection{Graph of $\relBetti{0}$} 

 Figure~\ref{fig:betti0_graph_npipe} presents the graphs of $\relBetti{0}$ for the various degraded resolutions and the associated smoothing scales. The mean and the standard deviation are computed from the simulations. In general, we find  that the significance across resolutions and levels is approximately $2\sigma$ or lower. 

\subsubsection{Graph of $\relBetti{1}$} 

Figure~\ref{fig:betti1_graph_npipe} presents the graphs of $\relBetti{1}$ for the various resolutions and corresponding scales. For resolutions between $\Res = 2048,\ldots,64$, corresponding to $FWHM = 5',\ldots,160'$, we observe the significance to be approximately $2\sigma$ or lower. However, for the next higher smoothing scales corresponding to $FWHM = 320'$ (approximately $5$ degrees) and $FWHM = 640'$ (approximately $10$ degrees), we note interesting deviations that we examine next. These cases are also presented in Figure~\ref{fig:betti1_graph_32_16} in an enlarged view for clarity.

Concentrating on the top left panel of Figure~\ref{fig:betti1_graph_32_16}, at a Gaussian smoothing scale of $FWHM = 320'$, approximately 5 degrees, and at a moderately low dimensionless density threshold $\nu = -2.5$, we find a $3.91\sigma$ deviation between the simulations and the observation. We examine the behavior of $b_1$ at this level further in Figure~\ref{fig:b1_level-2.5_distr}, where we present the histogram of the distribution of $b_1$ from the simulations in gray boxes. The observed value is indicated by a red vertical line. Evidently, the distribution is nonsymmetric and hence non-Gaussian. Our tests also indicate that the Poisson distribution is a poor fit to the data as well. We detect $28$ loops in the observational curves, compared to the average of $\sim 21$ loops in the simulations, with a standard deviation of $\sim1.78$. Within the Gaussian context, this would yield the significance of the difference at approximately $4\sigma$. However, since the distribution in this bin is distinctly non-Gaussian and does not obey Poisson statistics, ascribing a $\sigma$ significance in the usual sense is not viable in this case. As a result, we simply note that the significance is higher than what may be resolvable by $600$ simulations, yielding an empirical $p$-value of at most $0.0016$. While the generally low number of loops in both simulations and observations, owing to the large scale of probing, push us to the regime of small numbers, the behavior of the statistics indicates a stable regime.

\begin{figure}
        \centering
        \subfloat{\includegraphics[width=0.5\textwidth]{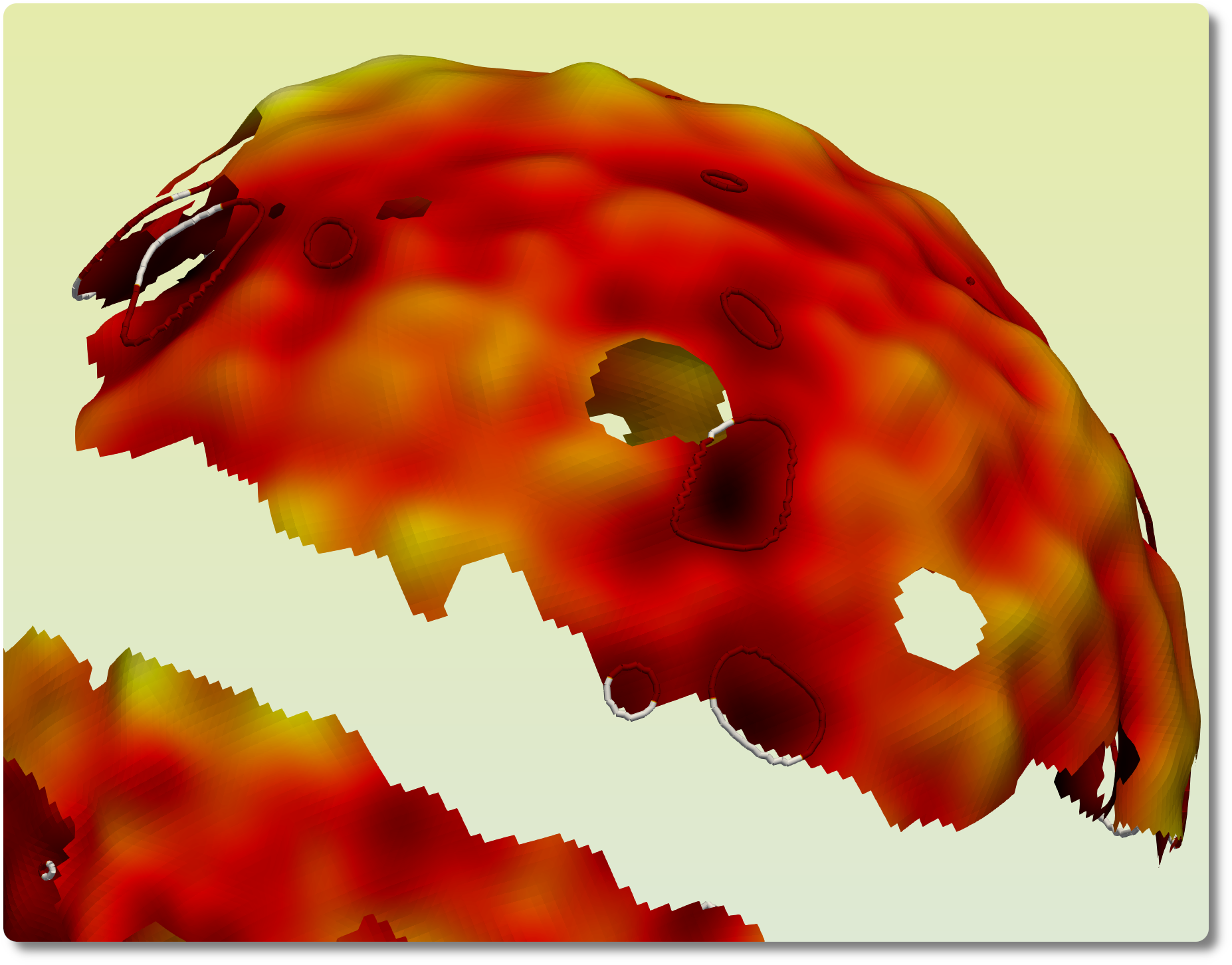} }\\
        \caption{Visualization of the loops surrounding the low-density regions at a moderately negative threshold. The gaps in the manifold correspond to the mask that has been removed. We clearly note the equatorial belt corresponding to our galaxy, and more patches in the visible cap. Some loops live fully in the excursion region that is not influenced by the mask. We show them in red. We also depict some representative relative loops whose end points are masked. We draw closed loops for this category as well and show the portions in which they overlap the masked regions in white. The visualization is based on the observed CMB map from the \texttt{NPIPE} data set smoothed at $5$ degrees.}
        \label{fig:loops}
\end{figure}

Further evidence in support of this argument is presented in Appendix~\ref{sec:stat_validity}, where we examine the distribution of $\chi^2$ and Tukey depth from the simulations, and compare them with the observations. In general, for the $\chi^2$-test, we find good agreement between the histogram of simulations and the theoretical curve. Therefore, the highly significant deviation merits consideration.  The deviant behavior of loops in the earlier analysis of \textup{Planck 2015} Data release 2 (DR2) presented in \cite{pranav2019b} at similar scales is also noteworthy in this context, as is the deviant Euler characteristic from WMAP observational data reported by \cite{eriksen04ng} and \cite{park2004}. Figure~\ref{fig:loops} presents a visualization of some of these loops at moderately negative thresholds for observational maps smoothed at $FWHM = 320'$.

At the next higher smoothing scale of $FWHM = 640'$, approximately $10$ degrees, presented in the right panel of Figure~\ref{fig:betti1_graph_32_16}, the shapes of the observational curve and the mean of the simulated curves for the negative thresholds, where topological loops are the dominant entities, are widely different. As differences in shapes may be a stronger indicator of inherent differences in the models than simply differences in the amplitudes, our assessment is that this case merits scrutiny as well. However, since the numbers involved are small, about $5$ and fewer in some bins \footnote{The standard requirement for the validity of $\chi^2$ test is a minimum of five samples in each bin considered for computing the statistic.}, we consider the statistics emerging from this resolution as not stable and reject them as possible statistical fluke.

\subsubsection{Graph of $\relEuler$ } 

Figure~\ref{fig:ec_graph_npipe} presents the graphs of $\relEuler$ for the various resolutions and corresponding scales. The relative Euler characteristic also deviates for the resolutions where $b_0/b_1$, or both, deviate.  However, the significance of the difference shows milder characteristics owing to cancellation effects \citep{pranav2019b}.  This is because the Euler characteristic is not strictly an independent quantity because it is an alternating sum of the ranks of the relative homology groups. It therefore merely reflects the deviations in the contributing Betti numbers.

\subsection{Planck 2018 Data Release 3: \texttt{FFP10} dataset}
\label{sec:ffp10_result}

In this section, we analyze the topological characteristics of the \texttt{FFP10} dataset. We also compare our results for the Euler characteristic with those presented in \cite{planckIsotropy2018}.

\subsubsection{Graph of $\relBetti{0}$} 

 Figure~\ref{fig:betti0_graph_ffp10} presents the graphs for the topological components in panels (a) and (b), similar to Figure~\ref{fig:betti0_graph_npipe}. In general, the observational values are within the $2\sigma$ band when compared with the simulations for almost all resolutions and levels. The exception is the number of components at $\Res = 128, FWHM = 80'$, where the observations deviate from the simulations at $2.96\sigma$ at the dimensionless density threshold $\nu = 0.5$. The smoothing scale, approximately a degree, and the moderate threshold at which the deviation occurs ensure that the statistics are away from the low-number regime.

\subsubsection{Graph of $\relBetti{1}$} 

Figure~\ref{fig:betti1_graph_ffp10} presents the graphs for the topological loops. The deviations between the observations and simulations are within approximately $2\sigma$ for most of the smoothing scales and thresholds. However, for the dimensionless threshold, $\nu = -2.5$, we note a more than $2\sigma$ deviation for $\Res = 64, FWHM = 160'$ onward toward higher smoothing scales. For $\Res = 32, FWHM = 320'$, the deviation is $2.77\sigma$, and similar for $\Res = 16, FWHM = 640'$. These deviations, while not as significant as exhibited in the \texttt{NPIPE} dataset, occur at similar scales and thresholds. In addition, we also note a more than $4\sigma$ deviation  for $\Res = 16, FWHM = 640'$ at $\nu = -3$. However, we reject this as a possible statistical fluke because of the extremely low numbers involved, which are about $5$ or fewer.

\subsubsection{Graph of $\relEuler$} 

Figure~\ref{fig:ec_graph_ffp10} presents the graphs of $\relEuler$ for the various resolutions and corresponding scales. The Euler characteristic shows commensurate deviations with respect to $\relBetti{0} (\relBetti{1})$ because of their contributions in the alternating sum. 
 
\subsection{Statistical significance from combined thresholds}

\begin{table*}
	\caption{Table displaying the two-tailed $p$-values for relative homology for the datasets. } 
	\tabcolsep=0.09cm
	\centering
	\subfloat[][]{\reltabNpipe}\\
	\subfloat[][]{\reltabffp}\\
	\tablefoot{The $p$-values are obtained via the  parametric (Mahalanobis distance) and non-parametric (Tukey depth) tests, for different resolutions and smoothing scales for the \texttt{NPIPE} (panel (a)) and the \texttt{FFP10} (panel (b)) dataset. Marked in boldface are $p$-values $0.05$ or smaller.}
	
	\label{tab:npipe_degrade-pvalues}
\end{table*}

We considered the two methods detailed in Appendix~\ref{sec:stat} to compute the statistics of the combined thresholds and present $p$-values of the observed maps for them. The first method is the Mahalanobis distance \citep{mahalanobis}, also known as the $\chi^2$-test, which works in model-dependent and in empirical settings. The second method is the nonparametric Tukey depth test. In Appendix~\ref{sec:stat_validity} we examine the distribution characteristics of these statistics from the simulations. The simulation and theoretical curve agree well for the $\chi^2$ statistic. Despite this, we computed the $p$-values both theoretically and from the empirical distribution, and found that the empirical distributions yield a milder significance.  We also find that the depth histograms and distributions are well behaved and represent a meaningful quantification. Considering all the three topological quantities $\relBetti{0}$, $\relBetti{1}$ and $\relEuler$, we computed the statistics for all resolutions separately to ascribe a scale dependence to the signals.


\subsubsection{\texttt{NPIPE} dataset}
All entries before the last in Table~\ref{tab:npipe_degrade-pvalues}, panel (a), present $p$-values for the variables, computed from maps degraded at specific resolutions, with additional Gaussian smoothing applied. The $\chi^2$ results are presented for the theoretical and empirical distributions in the left and middle blocks, respectively.  The results for the Tukey depth are presented in the rightmost block. When we consider all the three tests, the $p$-values computed from the empirical distribution using the $\chi^2$ statistic yield the most conservative estimate of the significance,  and we favor it in order to be conservative in our interpretation. $\relBetti{1}$ shows a significant difference between the observational maps and the simulations for $\Res = 32$ and $\Res = 16$. Additionally, $\relBetti{0}$ also shows a significant difference at $\Res = 32$. However, we note in this context that   the distributions in this case are manifestly non-Gaussian, and their form is poorly understood theoretically. As a result, the $\chi^2$ values may be regarded with caution. This also presents the case for admitting the nonparametric Tukey depth test, which detects an outlier event in this case, yielding a $p$-value of $0.0$. The Euler characteristic shows highly significant difference at $\Res = 32$, which is an order of magnitude larger than either $b_0$ or $b_1$. This is due to the significant difference shown by the contributing $\relBetti{0}$ and $\relBetti{1}$ at this resolution, in combination with the fact that these deviations occur near the tail of the distribution. In general, because the Euler characteristic is an alternating sum of the Betti numbers, its behavior is influenced by both the Betti numbers. Cancellation effects are dominant if the deviations in the contributing Betti numbers occur in the zone of overlap, which is more toward median thresholds. If the deviations occur toward the tail, where either one of the Betti numbers is dominant, the signals of the contributing Betti numbers are amplified in the Euler characteristic signal. As an example, for the next lower resolution, $\Res = 16$, the Euler characteristic shows no significant difference even though $\relBetti{1}$ exhibits significant difference. This is because of the highly nonsignificant behavior of $\relBetti{0}$, whose contribution cancels the contribution of $\relBetti{1}$  toward the Euler characteristic.  In many instances, the Tukey depth test yields $p$-values of $0.0$ for all the descriptors, indicating that the observation is a true outlier compared to the simulations. We note the trend that these instances occur when the $p$-values computed from the Mahalanobis distance also exhibit generally low values. We help interpret the Tukey depth values in Appendix~\ref{sec:stat_validity}, where we examine the trends in their distribution.

\subsubsection{\texttt{FFP10} dataset}
Results for the \texttt{FFP10} dataset are presented in panel (b) of Table~\ref{tab:npipe_degrade-pvalues}. We concentrate our analysis on the middle block, which presents the $p$-values from the $\chi^2$ statistic using the empirical distribution.   For $\Res = 32, FWHM = 320'$, we note that all the topological descriptors show a nonsignificant deviation between the simulations and the observation, which is in contrast with the behavior at the same resolution in the \texttt{NPIPE} dataset, where the values are an order of magnitude lower. For the next larger scale of probing, $\Res = 16, FWHM = 640'$, we note a per mil significance of the difference between simulations and observations for the number of loops, which is an order of magnitude smaller than in the \texttt{NPIPE} dataset. Additionally, we also note the low $p$-value for $b_0$ at $N =128, FWHM = 80'$, which is the scale at which an approximately $3\sigma$ deviation occurs for the number of components between the simulations and the observation. When examining the Euler characteristic at this scale, we note that the values are mildly significant, in contrast with the \texttt{NPIPE} dataset, which exhibits nonsignificant behavior. The Tukey depth test shows the observations to be true outliers in instances  at different resolutions for all the three variables, yielding $p$-values of $0.0$. As in the case with the \texttt{NPIPE} dataset, these instances occur when the $\chi^2$ test also exhibits small $p$-values.

\subsection{Comparison with earlier topo-geometrical results in the literature} 

Topo-geometrical studies aimed at testing isotropy, homogeneity, and Gaussianity of the cosmic fields, such as the CMB and the 3D density distribution, are standard in the cosmological literature. The principle tools employed for this purpose are the geometrical Minkowski functionals and the topological Euler characteristic. The bridge between topology and geometry is provided by the \textup{Theorema Egrerium} of Gauss, which establishes the equivalence between the geometric  $0$-th  \LKC~(equivalently the $D$-th Minkowski functional), and the purely topological notion of Euler characteristic of a $D$-dimensional space. Our results, on the other hand, involve novel and purely topological notions arising from homology theory quantified by the Betti numbers. The Euler characteristic has connections to these topological measures due to the Euler-Poincar\'{e} formula, which expresses the Euler characteristic as the alternating sum of Betti numbers.

The pioneering works that examined the Minkowski functionals and Euler characteristic of the CMB are by \cite{eriksen04ng} and \cite{park2004}, performed on the WMAP data \citep{wmap9}. While \cite{eriksen04ng} examined the whole set of Minkowski functionals as well as the Morse-theoretical concept of a skeleton length, \cite{park2004} restricted themselves to genus measurement, which is linearly related to the Euler characteristic. The purely geometric Minkowski functionals, namely the area functional and the contour length functional, show consistency with the standard model. In contrast, both works reported anomalous measurements of the Euler characteristic or genus. \cite{eriksen04ng} reported that the Euler characteristic at negative density thresholds is anomalous with the base model simulations at more than $3\sigma$ for the northern hemisphere at smoothing scales of $FWHM = 3.40^\circ$. The corresponding $\chi^2$ value reported for this scale is at  the $95\%$ confidence level, which translates into a $p$-value of $0.05$. \cite{park2004} reported an anomalous genus at more than $2.5\sigma$. The anomalous behavior of genus or the Euler characteristic, specifically at negative levels, is linked to and generated by the anomalous behavior of the $\text{fir}$st homology group, represented by topological loops, and establishes a connection between our results and previous results, showing a consistency across datasets.

The analysis of the Minkowski functionals and Euler characteristic on the Planck datasets was performed by the Planck collaboration itself and was reported in \cite{planckIsotropy2015} for DR2 (FFP8)  and in \cite{planckIsotropy2018} for DR3 (FFP10). \cite{planckIsotropy2018} examined the Minkowski functionals, and the graphs for the Euler characteristic were presented for $\Res = 1024, 256, 32$. The general indication is that the observational values are within the $2.5\sigma$ band computed from the simulations. Our results are largely consistent with this observation from the Planck analysis for the given resolutions, which is also exhibited by the nonanomalous $p$-values obtained from the empirical $\chi^2$ test. However, we also note that our results show stronger differences than the Planck results. In this context, it is important to note that in the Planck analysis pipeline, it has been the practice to combine all the Minkowski functionals to perform the statistical tests and subsequently present the combined $p$-values. Beyond the fact that this combination mixes signals from topo-geometrical descriptors that are a priori known to represent independent properties, it also prevents a direct comparison with our results of Euler characteristic computations.

In this context, another subtle but important point must be considered. In a  detailed treatment, \cite{pranav2019a} showed that the standard equations for Minkowski functionals used in cosmology are volume-normalized representations and represent exact computations only under specific conditions. The simplest case is a compact manifold without a boundary, for example, the complete $2$-sphere in 2D, or a periodic 3D Euclidean grid. When the manifold has boundaries, for example, the masked CMB sphere, additional boundary terms are involved in the exact computation of the Minkowski functionals. For purely geometric quantities, due to their localized nature, these boundary terms may be ignored when the manifold is large compared to the boundary. For topological quantities, which are nonlocal by nature, there is an interplay between the size of the topological objects, represented by the smoothing scale, and the size of the manifold. As a result, in case of topological descriptors such as the Euler characteristic and the Betti numbers, for large scales, boundary effects become increasingly more important. A full exact computation that takes the boundary effect generated by the complicated mask into account may therefore be a more accurate reflection of reality. Our computational methods for the topological descriptors perform exact computations in the presence of arbitrary masks. For the purely geometric components of the Minkowski functionals, this will involve a  computation via the full Gaussian Kinematic Formula \citep{adl10,pranav2019b}, which takes the boundary terms into account that were, for example, developed in \cite{fantaye2015}.

\section{Discussions and conclusion}
\label{sec:discussion}

We presented a topological analysis of the temperature fluctuations  in the CMB in terms of homology. To account for regions with unreliable data, we computed the homology of the excursion sets relative to the mask covering these areas. We performed a multiscale analysis by degrading and smoothing the maps for a range of pixel resolutions, and a subsequent convolution with a Gaussian filter for a range of scales. We performed our analysis on the fourth and final \texttt{NPIPE} data release from the Planck team. The pipeline represents a natural evolution of the data-processing pipeline, commensurate with better understanding of systematics, residuals, and noise over a period of time in successive data releases. It incorporates the best strategies from the LFI and HFI processing pipeline, so that the level of noise and residuals at all scales is reduced \citep{npipe}. We also investigated the \textup{Planck 2018} Data release 3 (DR3), accompanied by the  \texttt{FFP10} simulations \citep{ffp10}, for comparison and completeness. The present paper is a successor to \cite{pranav2019b}, where we investigated the topology of the temperature fluctuation maps  based on the intermediate  \textup{Planck 2015 } Data Release 2 (DR2), accompanied by \texttt{FFP8} simulations. These two investigated topological characteristics of the CMB temperature fluctuation maps for the latest three data releases by the Planck team. The overall trends in the results in the datasets show the consistency of the data processing pipeline and our own methods.

Examining the behavior of topological components or isolated objects represented by the $0$D homology group, we find that the observations deviate by approximately $2\sigma$ or less from the simulations for the \texttt{NPIPE} dataset. For the \texttt{FFP10} dataset, $\relBetti{0}$ exhibits  a deviation of $2.96\sigma$ at $\Res = 128, FWHM = 80'$. The $1$D homology group, representing and quantifying the topological loops, is also consistent with the base model at $2\sigma$ for most of the resolutions and scales. However, for $FWHM = 320'$ and at moderately low negative thresholds, we record a high deviation between the simulations and observations for the \texttt{NPIPE} dataset. In a Gaussian setting, this would amount to an  approximately $4\sigma$ significance. However, the distribution characteristics at this threshold are manifestly non-Gaussian, rendering the usual $\sigma$ significance nonviable. The distribution characteristics do not obey the Poisson statistics either. In view of this, and because we lack a true theoretical understanding of the distribution characteristics, we simply note that the significance is larger than what may be resolved by $600$ simulations, yielding an empirical $p$-value of at most $0.0016$.  The \texttt{FFP10} dataset exhibits a deviant behavior at $\sim 2.8\sigma$ at the same threshold and resolution for the loops. Although the deviation occurs in a regime in which the low threshold and large smoothing scale may indicate statistics with low numbers, we find the statistics to be well behaved and stable.  The high significance value, combined with the fact that the deviation is at a scale at which anomalies have been reported in several methods and datasets \citep{eriksen04ng,park2004,cmbanomaliesstarkman}, the case merits consideration. However, in this context, we also note that the statistical analysis is  based on merely $600$ and $300$ simulations for the \texttt{NPIPE} and \texttt{FFP10} datasets, respectively. For the Euler characteristic,  the \texttt{NPIPE} dataset exhibits a high significance due to the high significance exhibited by the loops. However, for the  \texttt{FFP10} data set, the Euler characteristic is within approximately the $2.5\sigma$ band computed from simulations, which is largely consistent with the observations in \cite{planckIsotropy2018}. However, we note that our results generally exhibit stronger differences than the Planck analysis pipeline. While ascertaining the source of this discrepancy is beyond the scope of this paper, a plausible source of the difference may be the different methods adopted in estimating the Euler characteristic and Minkowski functionals in general in \cite{planckIsotropy2018}, which are based on theoretical equations for boundary-less manifolds. At the next higher smoothing scale, $\Res = 16, FWHM =640'$, both datasets exhibit a significantly anomalous behavior with respect to the number of loops, but not the number of components. However, we disregard this  resolution because of the low numbers involved in computing the statistics, owing to the large smoothing scale and low thresholds.

In order to test for nonrandom discrepancies, we computed the $p$-values using the theoretical model-dependent $\chi^2$ test, as well as its nonparametric version computed from empirical distributions, by combining all relevant thresholds at a given resolution. We also presented the nonparametric and model-independent Tukey depth test, which indicated that the observations are true outliers with respect to the simulations in numerous instances. However, in the final interpretation, we focused on the most conservative $p$-values, estimated using the $\chi^2$ test from the empirical distributions. For the  \texttt{NPIPE} dataset, the observations are consistent with the simulations, with $p$-values higher than $0.05$ for most resolutions. However, the number of components and loops shows mildly significant deviations at $\Res = 32, FWHM = 320'$. The corresponding deviation for the Euler characteristic exhibits an order-of-magnitude lower value at per mil levels. For the next larger scale, we find that the components and the Euler characteristic are consistent with the base model, while the loops exhibit a mildly anomalous $p$-value of $0.03$. While emphasizing that the statistics are based on low numbers, we note the contrasting behavior of the different topological descriptors. For the \texttt{FFP10} dataset, the components and the Euler characteristic exhibit $p$-values higher than $0.01$ for all resolutions. This is consistent with the observation in \cite{planckIsotropy2018} that the $p$-values obtained from the combined Minkowski functionals are consistent with the base model within the $99\%$ confidence level. The the number of loops exhibits a significantly anomalous behavior, yielding per mil $p$-values, in contrast with the number of components and the Euler characteristic, at $\Res = 16, FWHM = 640$. However, we  note that this deviation occurs at large smoothing scales, where the numbers involved in computing the statistics are small. Disregarding this anomalous behavior of loops at large scales, which might be affected by low-number statistics, we find that the \texttt{FFP10} dataset is consistent with the standard model simulations at the $99\%$ confidence level.

In summary, while both datasets are largely consistent with the simulations based on the standard model, we find instances of interesting discrepancies in both datasets, which may be difficult to reject summarily as statistical flukes. Although most but not all  of the anomalies occur on large scales, which inherently indicates statistics based on small numbers, we find them to be in statistically stable regimes in most instances. Based on the evidence presented in this paper, our assessment is that it may be a difficult task to accept or reject the null hypothesis summarily. A primary but crucial requirement may be a significantly larger number of simulations in order to achieve a clear verdict.  In the absence of a true understanding of the origin of the possible anomalous behavior, any attempt to classify it will merely be speculative in nature, and we refrain from it. However, to facilitate a deeper understanding, we envision future research directed toward examining the fiducial frequency maps, as well as the polarization maps, among others, possibly with a larger suite of simulations. This will also involve testing smaller patches of the sky in order  to determine whether the effect is global in nature or restricted to a specific part of the sky.

\section*{Acknowledgements}

I am indebted to the anonymous referee, whose incisive, yet extremely insightful comments have helped bring this draft to a balanced place. I am greatly indebted to Robert Adler, Thomas Buchert, Herbert Edelsbrunner, Bernard Jones, Armin Schwarzman, Gert Vegter, and Rien van de Weygaert for encouraging this solo venture, and for extremely helpful discussions. My gratitude also to Julian Borrill and Reijo Keskitalo for their patience in clarifying doubts, and their constructive comments on the draft. I also thank Tal Eliezri for insightful comments on the artwork. This work is supported by the ERC advanced grant ARThUs (grant no: 740021; PI: Thomas Buchert), with contributing influence from ERC advanced grant URSAT (grant no: 320422; PI: Robert Adler). I gratefully acknowledge the support of PSMN (P\^ole Scientifique de Mod\'elisation Num\'erique) of the ENS de Lyon, and the Department of Energy’s National Energy Research Scientific Computing Center (NERSC) at Lawrence Berkeley National Laboratory, operated under Contract No. DE-AC02-05CH11231, for the use of computing resources.

\bibliographystyle{aa}
\bibliography{master_references.bib}

\begin{thebibliography}{65}
\expandafter\ifx\csname natexlab\endcsname\relax\def\natexlab#1{#1}\fi

\bibitem[{Adler(1981)}]{adler1981}
Adler, R. 1981, The Geometry of Random Fields, Classics in applied mathematics
  (Society for Industrial and Applied Mathematics (SIAM, 3600 Market Street,
  Floor 6, Philadelphia, PA 19104))

\bibitem[{Adler {et~al.}(2017)Adler, Agami, \& Pranav}]{rst}
Adler, R.~J., Agami, S., \& Pranav, P. 2017, Proceedings of the National
  Academy of Sciences, 114, 11878

\bibitem[{Adler \& Taylor(2010)}]{adl10}
Adler, R.~J. \& Taylor, J.~E. 2010, Random Fields and Geometry, Springer
  Monographs in Mathematics (Springer)

\bibitem[{{Appleby} {et~al.}(2020){Appleby}, {Park}, {Hong}, {Hwang}, \&
  {Kim}}]{appleby2020}
{Appleby}, S., {Park}, C., {Hong}, S.~E., {Hwang}, H.~S., \& {Kim}, J. 2020,
  \apj, 896, 145

\bibitem[{{Appleby} {et~al.}(2021){Appleby}, {Park}, {Pranav}, {Hong}, {Hwang},
  {Kim}, \& {Buchert}}]{appleby2021minkowskiSDSS}
{Appleby}, S., {Park}, C., {Pranav}, P., {et~al.} 2021, arXiv e-prints,
  arXiv:2110.06109

\bibitem[{Bauer {et~al.}(2014)Bauer, Kerber, Reininghaus, \& Wagner}]{phat}
Bauer, U., Kerber, M., Reininghaus, J., \& Wagner, H. 2014, in Mathematical
  Software {\textendash} {ICMS} 2014 (Springer Berlin Heidelberg), 137--143

\bibitem[{{Bennett} {et~al.}(2013){Bennett}, {Larson}, {Weiland}, {Jarosik},
  {Hinshaw}, {Odegard}, {Smith}, {Hill}, {Gold}, {Halpern}, {Komatsu}, {Nolta},
  {Page}, {Spergel}, {Wollack}, {Dunkley}, {Kogut}, {Limon}, {Meyer}, {Tucker},
  \& {Wright}}]{wmap9}
{Bennett}, C.~L., {Larson}, D., {Weiland}, J.~L., {et~al.} 2013, Astrophys. J.
  Suppl., 208, 20

\bibitem[{Biagetti {et~al.}(2021)Biagetti, Cole, \& Shiu}]{biagetti2020}
Biagetti, M., Cole, A., \& Shiu, G. 2021, Journal of Cosmology and
  Astroparticle Physics, 2021, 061

\bibitem[{{Chingangbam} {et~al.}(2017){Chingangbam}, {Yogendran}, {Joby},
  {Ganesan}, {Appleby}, \& {Park}}]{chingangbam2017}
{Chingangbam}, P., {Yogendran}, K.~P., {Joby}, P.~K., {et~al.} 2017, \jcap, 12,
  023

\bibitem[{{Codis} {et~al.}(2013){Codis}, {Pichon}, {Pogosyan}, {Bernardeau}, \&
  {Matsubara}}]{codis2013}
{Codis}, S., {Pichon}, C., {Pogosyan}, D., {Bernardeau}, F., \& {Matsubara}, T.
  2013, \mnras, 435, 531

\bibitem[{Copi {et~al.}(2015)Copi, Huterer, Schwarz, \& Starkman}]{multipoles}
Copi, C., Huterer, D., Schwarz, D., \& Starkman, G. 2015, Monthly Notices of
  the Royal Astronomical Society, 449, 3458

\bibitem[{{Ducout} {et~al.}(2013){Ducout}, {Bouchet}, {Colombi}, {Pogosyan}, \&
  {Prunet}}]{ducout2013}
{Ducout}, A., {Bouchet}, F.~R., {Colombi}, S., {Pogosyan}, D., \& {Prunet}, S.
  2013, MNRAS, 429, 2104

\bibitem[{{Durrer} {et~al.}(1996){Durrer}, {Gangui}, \&
  {Sakellariadou}}]{durrer1996}
{Durrer}, R., {Gangui}, A., \& {Sakellariadou}, M. 1996, Physical Review
  Letters, 76, 579

\bibitem[{Edelsbrunner \& Harer(2010)}]{edelsbrunnerharer10}
Edelsbrunner, H. \& Harer, J. 2010, Computational Topology - an Introduction
  (American Mathematical Society), I--XII, 1--241

\bibitem[{{Eriksen} {et~al.}(2004{\natexlab{a}}){Eriksen}, {Hansen}, {Banday},
  {G{\'o}rski}, \& {Lilje}}]{eriksen2004}
{Eriksen}, H.~K., {Hansen}, F.~K., {Banday}, A.~J., {G{\'o}rski}, K.~M., \&
  {Lilje}, P.~B. 2004{\natexlab{a}}, \apj, 605, 14

\bibitem[{{Eriksen} {et~al.}(2004{\natexlab{b}}){Eriksen}, {Novikov}, {Lilje},
  {Banday}, \& {G{\'o}rski}}]{eriksen04ng}
{Eriksen}, H.~K., {Novikov}, D.~I., {Lilje}, P.~B., {Banday}, A.~J., \&
  {G{\'o}rski}, K.~M. 2004{\natexlab{b}}, \apj, 612, 64

\bibitem[{Euler(1758)}]{euler1758}
Euler, L. 1758, Novi Commentarii academiae scientiarum Petropolitanae, 4, 140

\bibitem[{Fantaye {et~al.}(2015)Fantaye, Marinucci, Hansen, \&
  Maino}]{fantaye2015}
Fantaye, Y., Marinucci, D., Hansen, F., \& Maino, D. 2015, Phys. Rev. D, 91,
  063501

\bibitem[{Feldbrugge {et~al.}(2019)Feldbrugge, van Engelen, van~de Weygaert,
  Pranav, \& Vegter}]{feldbrugge2019}
Feldbrugge, J., van Engelen, M., van~de Weygaert, R., Pranav, P., \& Vegter, G.
  2019, Journal of Cosmology and Astroparticle Physics, 2019, 052–052

\bibitem[{{Gauss}(1900)}]{gauss1900}
{Gauss}, C.~F. 1900, K. Gesellschaft Wissenschaft, 8

\bibitem[{{G{\'o}rski} {et~al.}(2005){G{\'o}rski}, {Hivon}, {Banday},
  {Wandelt}, {Hansen}, {Reinecke}, \& {Bartelmann}}]{healpix1}
{G{\'o}rski}, K.~M., {Hivon}, E., {Banday}, A.~J., {et~al.} 2005, Astrophysical
  Journal, 622, 759

\bibitem[{{Gott} {et~al.}(1986){Gott}, {Dickinson}, \& {Melott}}]{gdm86}
{Gott}, III, J.~R., {Dickinson}, M., \& {Melott}, A.~L. 1986, Astrophysical
  Journal, 306, 341

\bibitem[{{Guth}(1981)}]{guth1981}
{Guth}, A.~H. 1981, Physical Review D, 23, 347

\bibitem[{{Guth} \& {Pi}(1982)}]{guthpi1982}
{Guth}, A.~H. \& {Pi}, S.-Y. 1982, Physical Review Letters, 49, 1110

\bibitem[{{Harrison}(1970)}]{harrison1970}
{Harrison}, E.~R. 1970, \prd, 1, 2726

\bibitem[{Heydenreich {et~al.}(2021)Heydenreich, Brück, \&
  Harnois-D\'{e}raps}]{heydenreich2021}
Heydenreich, S., Brück, B., \& Harnois-D\'{e}raps, J. 2021, Astronomy \&
  Astrophysics, 648, A74

\bibitem[{{Jones}(2017)}]{jones2017precision}
{Jones}, B.~J.~T. 2017, {Precision Cosmology: The First Half Million Years}
  (Cambridge University Press)

\bibitem[{{Kerscher} {et~al.}(1996){Kerscher}, {Schmalzing}, \&
  {Buchert}}]{schmalzing1996}
{Kerscher}, M., {Schmalzing}, J., \& {Buchert}, T. 1996, {Mapping, measuring
  and modelling the universe}, ed. P.~{Coles}, V.~{Martinez}, \& M.~J.~P.
  {Borderia}, 247--252

\bibitem[{Kono {et~al.}(2020)Kono, Takeuchi, Cooray, Nishizawa, \&
  Murakami}]{kono2020}
Kono, K.~T., Takeuchi, T.~T., Cooray, S., Nishizawa, A.~J., \& Murakami, K.
  2020, arXiv e-prints, arXiv:2006.02905

\bibitem[{Mahalanobis(1936)}]{mahalanobis}
Mahalanobis, P.~C. 1936, in Proceedings National Institute of Science, India,
  Vol.~2, 49--55

\bibitem[{{Masi}(2002)}]{boomerang}
{Masi}, S. 2002, Progress in Particle and Nuclear Physics, 48, 243

\bibitem[{{Matsubara}(2010)}]{matsubara2010}
{Matsubara}, T. 2010, \prd, 81, 083505

\bibitem[{{Mecke} {et~al.}(1994){Mecke}, {Buchert}, \& {Wagner}}]{mecke94}
{Mecke}, K.~R., {Buchert}, T., \& {Wagner}, H. 1994, Astronomy \& Astrophysics,
  288, 697

\bibitem[{Moraleda {et~al.}(2019)Moraleda, Valous, Xiong, \&
  Halama}]{moraleda2019}
Moraleda, R., Valous, N., Xiong, W., \& Halama, N. 2019, Computational Topology
  for Biomedical Image and Data Analysis: Theory and Applications, Focus Series
  in Medical Physics and Biomedical Engineering (CRC Press)

\bibitem[{Morozov(2005)}]{morozov2005}
Morozov, D. 2005, BioGeometry News, Dept. Comput. Sci., Duke Univ

\bibitem[{Munkres(1984)}]{munkres1984}
Munkres, J. 1984, Elements of Algebraic Topology, Advanced book classics
  (Perseus Books)

\bibitem[{{Park} {et~al.}(2013){Park}, {Pranav}, {Chingangbam}, {van de
  Weygaert}, {Jones}, {Vegter}, {Kim}, {Hidding}, \& {Hellwing}}]{ppc13}
{Park}, C., {Pranav}, P., {Chingangbam}, P., {et~al.} 2013, Journal of Korean
  Astronomical Society, 46, 125

\bibitem[{Park(2004)}]{park2004}
Park, C.-G. 2004, \mnras, 349, 313

\bibitem[{{Peebles} \& {Yu}(1970)}]{peebles1970}
{Peebles}, P.~J.~E. \& {Yu}, J.~T. 1970, \apj, 162, 815

\bibitem[{{Planck Collaboration} {et~al.}(2016{\natexlab{a}}){Planck
  Collaboration}, {Ade}, {Aghanim}, {Akrami}, {Aluri}, {Arnaud}, {Ashdown},
  {Aumont}, {Baccigalupi}, {Banday}, {Barreiro}, {Bartolo}, {Basak},
  {Battaner}, {Benabed}, {Beno{\^\i}t}, {Benoit-L{\'e}vy}, {Bernard},
  {Bersanelli}, {Bielewicz}, {Bock}, {Bonaldi}, {Bonavera}, {Bond}, {Borrill},
  {Bouchet}, {Boulanger}, {Bucher}, {Burigana}, {Butler}, {Calabrese},
  {Cardoso}, {Casaponsa}, {Catalano}, {Challinor}, {Chamballu}, {Chiang},
  {Christensen}, {Church}, {Clements}, {Colombi}, {Colombo}, {Combet},
  {Contreras}, {Couchot}, {Coulais}, {Crill}, {Cruz}, {Curto}, {Cuttaia},
  {Danese}, {Davies}, {Davis}, {de Bernardis}, {de Rosa}, {de Zotti},
  {Delabrouille}, {D{\'e}sert}, {Diego}, {Dole}, {Donzelli}, {Dor{\'e}},
  {Douspis}, {Ducout}, {Dupac}, {Efstathiou}, {Elsner}, {En{\ss}lin},
  {Eriksen}, {Fantaye}, {Fergusson}, {Fernandez-Cobos}, {Finelli}, {Forni},
  {Frailis}, {Fraisse}, {Franceschi}, {Frejsel}, {Frolov}, {Galeotta}, {Galli},
  {Ganga}, {Gauthier}, {Ghosh}, {Giard}, {Giraud-H{\'e}raud}, {Gjerl{\o}w},
  {Gonz{\'a}lez-Nuevo}, {G{\'o}rski}, {Gratton}, {Gregorio}, {Gruppuso},
  {Gudmundsson}, {Hansen}, {Hanson}, {Harrison}, {Henrot-Versill{\'e}},
  {Hern{\'a}ndez-Monteagudo}, {Herranz}, {Hildebrandt}, {Hivon}, {Hobson},
  {Holmes}, {Hornstrup}, {Hovest}, {Huang}, {Huffenberger}, {Hurier}, {Jaffe},
  {Jaffe}, {Jones}, {Juvela}, {Keih{\"a}nen}, {Keskitalo}, {Kim}, {Kisner},
  {Knoche}, {Kunz}, {Kurki-Suonio}, {Lagache}, {L{\"a}hteenm{\"a}ki},
  {Lamarre}, {Lasenby}, {Lattanzi}, {Lawrence}, {Leonardi}, {Lesgourgues},
  {Levrier}, {Liguori}, {Lilje}, {Linden-V{\o}rnle}, {Liu},
  {L{\'o}pez-Caniego}, {Lubin}, {Mac{\'\i}as-P{\'e}rez}, {Maggio}, {Maino},
  {Mandolesi}, {Mangilli}, {Marinucci}, {Maris}, {Martin},
  {Mart{\'\i}nez-Gonz{\'a}lez}, {Masi}, {Matarrese}, {McGehee}, {Meinhold},
  {Melchiorri}, {Mendes}, {Mennella}, {Migliaccio}, {Mikkelsen}, {Mitra},
  {Miville-Desch{\^e}nes}, {Molinari}, {Moneti}, {Montier}, {Morgante},
  {Mortlock}, {Moss}, {Munshi}, {Murphy}, {Naselsky}, {Nati}, {Natoli},
  {Netterfield}, {N{\o}rgaard-Nielsen}, {Noviello}, {Novikov}, {Novikov},
  {Oxborrow}, {Paci}, {Pagano}, {Pajot}, \& et.al.}]{planckIsotropy2015}
{Planck Collaboration}, {Ade}, P.~A.~R., {Aghanim}, N., {et~al.}
  2016{\natexlab{a}}, \aap, 594, A16

\bibitem[{{Planck Collaboration} {et~al.}(2016{\natexlab{b}}){Planck
  Collaboration}, {Ade}, {Aghanim}, {Arnaud}, {Arroja}, {Ashdown}, {Aumont},
  {Baccigalupi}, {Ballardini}, {Banday}, \& et~al.}]{planckcollaboration2016a}
{Planck Collaboration}, {Ade}, P.~A.~R., {Aghanim}, N., {et~al.}
  2016{\natexlab{b}}, \aap, 594, A17

\bibitem[{{Planck Collaboration} {et~al.}(2016{\natexlab{c}}){Planck
  Collaboration}, {Ade}, {Aghanim}, {Arnaud}, {Ashdown}, {Aumont},
  {Baccigalupi}, {Banday}, {Barreiro}, {Bartlett}, \& et~al.}]{plancksims}
{Planck Collaboration}, {Ade}, P.~A.~R., {Aghanim}, N., {et~al.}
  2016{\natexlab{c}}, \aap, 594, A12

\bibitem[{{Planck Collaboration} {et~al.}(2020{\natexlab{a}}){Planck
  Collaboration}, {Aghanim}, {Akrami}, {Arroja}, {Ashdown}, {Aumont},
  {Baccigalupi}, {Ballardini}, {Banday}, {Barreiro}, {Bartolo}, {Basak},
  {Battye}, {Benabed}, {Bernard}, {Bersanelli}, {Bielewicz}, {Bock}, {Bond},
  {Borrill}, {Bouchet}, {Boulanger}, {Bucher}, {Burigana}, {Butler},
  {Calabrese}, {Cardoso}, {Carron}, {Casaponsa}, {Challinor}, {Chiang},
  {Colombo}, {Combet}, {Contreras}, {Crill}, {Cuttaia}, {de Bernardis}, {de
  Zotti}, {Delabrouille}, {Delouis}, {D{\'e}sert}, {Di Valentino}, {Dickinson},
  {Diego}, {Donzelli}, {Dor{\'e}}, {Douspis}, {Ducout}, {Dupac}, {Efstathiou},
  {Elsner}, {En{\ss}lin}, {Eriksen}, {Falgarone}, {Fantaye}, {Fergusson},
  {Fernandez-Cobos}, {Finelli}, {Forastieri}, {Frailis}, {Franceschi},
  {Frolov}, {Galeotta}, {Galli}, {Ganga}, {G{\'e}nova-Santos}, {Gerbino},
  {Ghosh}, {Gonz{\'a}lez-Nuevo}, {G{\'o}rski}, {Gratton}, {Gruppuso},
  {Gudmundsson}, {Hamann}, {Handley}, {Hansen}, {Helou}, {Herranz},
  {Hildebrandt}, {Hivon}, {Huang}, {Jaffe}, {Jones}, {Karakci}, {Keih{\"a}nen},
  {Keskitalo}, {Kiiveri}, {Kim}, {Kisner}, {Knox}, {Krachmalnicoff}, {Kunz},
  {Kurki-Suonio}, {Lagache}, {Lamarre}, {Langer}, {Lasenby}, {Lattanzi},
  {Lawrence}, {Le Jeune}, {Leahy}, {Lesgourgues}, {Levrier}, {Lewis},
  {Liguori}, {Lilje}, {Lilley}, {Lindholm}, {L{\'o}pez-Caniego}, {Lubin}, {Ma},
  {Mac{\'\i}as-P{\'e}rez}, {Maggio}, {Maino}, {Mandolesi}, {Mangilli},
  {Marcos-Caballero}, {Maris}, {Martin}, {Martinelli},
  {Mart{\'\i}nez-Gonz{\'a}lez}, {Matarrese}, {Mauri}, {McEwen}, {Meerburg},
  {Meinhold}, {Melchiorri}, {Mennella}, {Migliaccio}, {Millea}, {Mitra},
  {Miville-Desch{\^e}nes}, {Molinari}, {Moneti}, {Montier}, {Morgante}, {Moss},
  {Mottet}, {M{\"u}nchmeyer}, {Natoli}, {N{\o}rgaard-Nielsen}, {Oxborrow},
  {Pagano}, {Paoletti}, {Partridge}, {Patanchon}, {Pearson}, {Peel}, {Peiris},
  {Perrotta}, {Pettorino}, {Piacentini}, {Polastri}, {Polenta}, {Puget},
  {Rachen}, {Reinecke}, {Remazeilles}, {Renault}, {Renzi}, {Rocha}, {Rosset},
  {Roudier}, {Rubi{\~n}o-Mart{\'\i}n}, {Ruiz-Granados}, {Salvati}, {Sandri},
  {Savelainen}, {Scott}, {Shellard}, {Shiraishi}, {Sirignano}, {Sirri},
  {Spencer}, {Sunyaev}, {Suur-Uski}, {Tauber}, {Tavagnacco}, {Tenti},
  {Terenzi}, {Toffolatti}, {Tomasi}, {Trombetti}, {Valiviita}, {Van Tent},
  {Vibert}, {Vielva}, {Villa}, {Vittorio}, {Wandelt}, {Wehus}, {White},
  {White}, {Zacchei}, \& {Zonca}}]{planckOverview2018}
{Planck Collaboration}, {Aghanim}, N., {Akrami}, Y., {et~al.}
  2020{\natexlab{a}}, \aap, 641, A1

\bibitem[{{Planck Collaboration} {et~al.}(2020{\natexlab{b}}){Planck
  Collaboration}, {Akrami}, {Andersen}, {Ashdown}, {Baccigalupi}, {Ballardini},
  {Banday}, {Barreiro}, {Bartolo}, {Basak}, {Benabed}, {Bernard}, {Bersanelli},
  {Bielewicz}, {Bond}, {Borrill}, {Burigana}, {Butler}, {Calabrese},
  {Casaponsa}, {Chiang}, {Colombo}, {Combet}, {Crill}, {Cuttaia}, {de
  Bernardis}, {de Rosa}, {de Zotti}, {Delabrouille}, {Di Valentino}, {Diego},
  {Dor{\'e}}, {Douspis}, {Dupac}, {Eriksen}, {Fernandez-Cobos}, {Finelli},
  {Frailis}, {Fraisse}, {Franceschi}, {Frolov}, {Galeotta}, {Galli}, {Ganga},
  {Gerbino}, {Ghosh}, {Gonz{\'a}lez-Nuevo}, {G{\'o}rski}, {Gruppuso},
  {Gudmundsson}, {Handley}, {Helou}, {Herranz}, {Hildebrandt}, {Hivon},
  {Huang}, {Jaffe}, {Jones}, {Keih{\"a}nen}, {Keskitalo}, {Kiiveri}, {Kim},
  {Kisner}, {Krachmalnicoff}, {Kunz}, {Kurki-Suonio}, {Lasenby}, {Lattanzi},
  {Lawrence}, {Le Jeune}, {Levrier}, {Liguori}, {Lilje}, {Lilley}, {Lindholm},
  {L{\'o}pez-Caniego}, {Lubin}, {Mac{\'\i}as-P{\'e}rez}, {Maino}, {Mandolesi},
  {Marcos-Caballero}, {Maris}, {Martin}, {Mart{\'\i}nez-Gonz{\'a}lez},
  {Matarrese}, {Mauri}, {McEwen}, {Meinhold}, {Mennella}, {Migliaccio},
  {Mitra}, {Molinari}, {Montier}, {Morgante}, {Moss}, {Natoli}, {Paoletti},
  {Partridge}, {Patanchon}, {Pearson}, {Pearson}, {Perrotta}, {Piacentini},
  {Polenta}, {Rachen}, {Reinecke}, {Remazeilles}, {Renzi}, {Rocha}, {Rosset},
  {Roudier}, {Rubi{\~n}o-Mart{\'\i}n}, {Ruiz-Granados}, {Salvati},
  {Savelainen}, {Scott}, {Sirignano}, {Sirri}, {Spencer}, {Suur-Uski},
  {Svalheim}, {Tauber}, {Tavagnacco}, {Tenti}, {Terenzi}, {Thommesen},
  {Toffolatti}, {Tomasi}, {Tristram}, {Trombetti}, {Valiviita}, {Van Tent},
  {Vielva}, {Villa}, {Vittorio}, {Wandelt}, {Wehus}, {Zacchei}, \&
  {Zonca}}]{npipe}
{Planck Collaboration}, {Akrami}, Y., {Andersen}, K.~J., {et~al.}
  2020{\natexlab{b}}, \aap, 643, A42

\bibitem[{{Planck Collaboration} {et~al.}(2020{\natexlab{c}}){Planck
  Collaboration}, {Akrami}, {Ashdown}, {Aumont}, {Baccigalupi}, {Ballardini},
  {Banday}, {Barreiro}, {Bartolo}, {Basak}, {Benabed}, {Bersanelli},
  {Bielewicz}, {Bock}, {Bond}, {Borrill}, {Bouchet}, {Boulanger}, {Bucher},
  {Burigana}, {Butler}, {Calabrese}, {Cardoso}, {Casaponsa}, {Chiang},
  {Colombo}, {Combet}, {Contreras}, {Crill}, {de Bernardis}, {de Zotti},
  {Delabrouille}, {Delouis}, {Di Valentino}, {Diego}, {Dor{\'e}}, {Douspis},
  {Ducout}, {Dupac}, {Efstathiou}, {Elsner}, {En{\ss}lin}, {Eriksen},
  {Fantaye}, {Fernandez-Cobos}, {Finelli}, {Frailis}, {Fraisse}, {Franceschi},
  {Frolov}, {Galeotta}, {Galli}, {Ganga}, {G{\'e}nova-Santos}, {Gerbino},
  {Ghosh}, {Gonz{\'a}lez-Nuevo}, {G{\'o}rski}, {Gruppuso}, {Gudmundsson},
  {Hamann}, {Handley}, {Hansen}, {Herranz}, {Hivon}, {Huang}, {Jaffe}, {Jones},
  {Keih{\"a}nen}, {Keskitalo}, {Kiiveri}, {Kim}, {Krachmalnicoff}, {Kunz},
  {Kurki-Suonio}, {Lagache}, {Lamarre}, {Lasenby}, {Lattanzi}, {Lawrence}, {Le
  Jeune}, {Levrier}, {Liguori}, {Lilje}, {Lindholm}, {L{\'o}pez-Caniego}, {Ma},
  {Mac{\'\i}as-P{\'e}rez}, {Maggio}, {Maino}, {Mandolesi}, {Mangilli},
  {Marcos-Caballero}, {Maris}, {Martin}, {Mart{\'\i}nez-Gonz{\'a}lez},
  {Matarrese}, {Mauri}, {McEwen}, {Meinhold}, {Mennella}, {Migliaccio},
  {Miville-Desch{\^e}nes}, {Molinari}, {Moneti}, {Montier}, {Morgante}, {Moss},
  {Natoli}, {Pagano}, {Paoletti}, {Partridge}, {Perrotta}, {Pettorino},
  {Piacentini}, {Polenta}, {Puget}, {Rachen}, {Reinecke}, {Remazeilles},
  {Renzi}, {Rocha}, {Rosset}, {Roudier}, {Rubi{\~n}o-Mart{\'\i}n},
  {Ruiz-Granados}, {Salvati}, {Savelainen}, {Scott}, {Shellard}, {Sirignano},
  {Sunyaev}, {Suur-Uski}, {Tauber}, {Tavagnacco}, {Tenti}, {Toffolatti},
  {Tomasi}, {Trombetti}, {Valenziano}, {Valiviita}, {Van Tent}, {Vielva},
  {Villa}, {Vittorio}, {Wandelt}, {Wehus}, {Zacchei}, {Zibin}, \&
  {Zonca}}]{planckIsotropy2018}
{Planck Collaboration}, {Akrami}, Y., {Ashdown}, M., {et~al.}
  2020{\natexlab{c}}, \aap, 641, A7

\bibitem[{{Planck Collaboration} {et~al.}(2020{\natexlab{d}}){Planck
  Collaboration}, {Akrami}, {Ashdown}, {Aumont}, {Baccigalupi}, {Ballardini},
  {Banday}, {Barreiro}, {Bartolo}, {Basak}, {Benabed}, {Bersanelli},
  {Bielewicz}, {Bond}, {Borrill}, {Bouchet}, {Boulanger}, {Bucher}, {Burigana},
  {Calabrese}, {Cardoso}, {Carron}, {Casaponsa}, {Challinor}, {Colombo},
  {Combet}, {Crill}, {Cuttaia}, {de Bernardis}, {de Rosa}, {de Zotti},
  {Delabrouille}, {Delouis}, {Di Valentino}, {Dickinson}, {Diego}, {Donzelli},
  {Dor{\'e}}, {Ducout}, {Dupac}, {Efstathiou}, {Elsner}, {En{\ss}lin},
  {Eriksen}, {Falgarone}, {Fernandez-Cobos}, {Finelli}, {Forastieri},
  {Frailis}, {Fraisse}, {Franceschi}, {Frolov}, {Galeotta}, {Galli}, {Ganga},
  {G{\'e}nova-Santos}, {Gerbino}, {Ghosh}, {Gonz{\'a}lez-Nuevo}, {G{\'o}rski},
  {Gratton}, {Gruppuso}, {Gudmundsson}, {Handley}, {Hansen}, {Helou},
  {Herranz}, {Hildebrandt}, {Huang}, {Jaffe}, {Karakci}, {Keih{\"a}nen},
  {Keskitalo}, {Kiiveri}, {Kim}, {Kisner}, {Krachmalnicoff}, {Kunz},
  {Kurki-Suonio}, {Lagache}, {Lamarre}, {Lasenby}, {Lattanzi}, {Lawrence}, {Le
  Jeune}, {Levrier}, {Liguori}, {Lilje}, {Lindholm}, {L{\'o}pez-Caniego},
  {Lubin}, {Ma}, {Mac{\'\i}as-P{\'e}rez}, {Maggio}, {Maino}, {Mandolesi},
  {Mangilli}, {Marcos-Caballero}, {Maris}, {Martin},
  {Mart{\'\i}nez-Gonz{\'a}lez}, {Matarrese}, {Mauri}, {McEwen}, {Meinhold},
  {Melchiorri}, {Mennella}, {Migliaccio}, {Miville-Desch{\^e}nes}, {Molinari},
  {Moneti}, {Montier}, {Morgante}, {Natoli}, {Oppizzi}, {Pagano}, {Paoletti},
  {Partridge}, {Peel}, {Pettorino}, {Piacentini}, {Polenta}, {Puget}, {Rachen},
  {Reinecke}, {Remazeilles}, {Renzi}, {Rocha}, {Roudier},
  {Rubi{\~n}o-Mart{\'\i}n}, {Ruiz-Granados}, {Salvati}, {Sandri}, {Savelainen},
  {Scott}, {Seljebotn}, {Sirignano}, {Spencer}, {Suur-Uski}, {Tauber},
  {Tavagnacco}, {Tenti}, {Thommesen}, {Toffolatti}, {Tomasi}, {Trombetti},
  {Valiviita}, {Van Tent}, {Vielva}, {Villa}, {Vittorio}, {Wandelt}, {Wehus},
  {Zacchei}, \& {Zonca}}]{ffp10}
{Planck Collaboration}, {Akrami}, Y., {Ashdown}, M., {et~al.}
  2020{\natexlab{d}}, \aap, 641, A4

\bibitem[{{Pogosyan} {et~al.}(2009){Pogosyan}, {Gay}, \&
  {Pichon}}]{pogosyan2009}
{Pogosyan}, D., {Gay}, C., \& {Pichon}, C. 2009, \prd, 80, 081301

\bibitem[{Pranav(2015)}]{pranavthesis}
Pranav, P. 2015, PhD thesis, University of Groningen

\bibitem[{Pranav(2021)}]{pranavReview2021}
Pranav, P. 2021, IEEE Signal Processing Magazine, 38, 130

\bibitem[{{Pranav}(2021)}]{pranav2021topology2}
{Pranav}, P. 2021, arXiv e-prints, arXiv:2109.08721

\bibitem[{{Pranav} {et~al.}(2019{\natexlab{a}}){Pranav}, {Adler}, {Buchert},
  {Edelsbrunner}, {Jones}, {Schwartzman}, {Wagner}, \& {van de
  Weygaert}}]{pranav2019b}
{Pranav}, P., {Adler}, R.~J., {Buchert}, T., {et~al.} 2019{\natexlab{a}}, \aap,
  627, A163

\bibitem[{{Pranav} {et~al.}(2017){Pranav}, {Edelsbrunner}, {van de Weygaert},
  {Vegter}, {Kerber}, {Jones}, \& {Wintraecken}}]{pranav2017}
{Pranav}, P., {Edelsbrunner}, H., {van de Weygaert}, R., {et~al.} 2017, \mnras,
  465, 4281

\bibitem[{{Pranav} {et~al.}(2019{\natexlab{b}}){Pranav}, {van de Weygaert},
  {Vegter}, {Jones}, {Adler}, {Feldbrugge}, {Park}, {Buchert}, \&
  {Kerber}}]{pranav2019a}
{Pranav}, P., {van de Weygaert}, R., {Vegter}, G., {et~al.} 2019{\natexlab{b}},
  \mnras, 485, 4167

\bibitem[{Ryden(2003)}]{ryden2003}
Ryden, B. 2003, Introduction to Cosmology (Addison-Wesley)

\bibitem[{Sahni {et~al.}(1998)Sahni, Sathyprakash, \& Shandarin}]{sahni1998}
Sahni, V., Sathyprakash, B., \& Shandarin, S. 1998, {\apj}, 507, L109

\bibitem[{{Schmalzing} \& {Gorski}(1998)}]{schmalzinggorski}
{Schmalzing}, J. \& {Gorski}, K.~M. 1998, \mnras, 297, 355

\bibitem[{{Schwarz} {et~al.}(2016){Schwarz}, {Copi}, {Huterer}, \&
  {Starkman}}]{cmbanomaliesstarkman}
{Schwarz}, D.~J., {Copi}, C.~J., {Huterer}, D., \& {Starkman}, G.~D. 2016,
  Classical and Quantum Gravity, 33, 184001

\bibitem[{Shivashankar {et~al.}(2016)Shivashankar, Pranav, Natarajan, van~de
  Weygaert, Bos, \& Rieder}]{shivashankar2015}
Shivashankar, N., Pranav, P., Natarajan, V., {et~al.} 2016, {IEEE} Trans. Vis.
  Comput. Graph., 22, 1745

\bibitem[{{Starobinsky}(1982)}]{starobinsky1982}
{Starobinsky}, A.~A. 1982, Physics Letters B, 117, 175

\bibitem[{{Telschow} {et~al.}(2019){Telschow}, {Schwartzman}, {Cheng}, \&
  {Pranav}}]{eecestimate}
{Telschow}, F., {Schwartzman}, A., {Cheng}, D., \& {Pranav}, P. 2019, arXiv
  e-prints, arXiv:1908.02493

\bibitem[{{The CGAL Project}(2021)}]{cgal}
{The CGAL Project}. 2021, {CGAL} User and Reference Manual, {5.3} edn. ({CGAL
  Editorial Board})

\bibitem[{Tukey(1975)}]{depth}
Tukey, J.~W. 1975, in Proceedings of the 1974 international congress of
  mathematicians, Vol.~2, 523--531

\bibitem[{van~de Weygaert {et~al.}(2011)van~de Weygaert, Vegter, Edelsbrunner,
  Jones, Pranav, Park, Hellwing, Eldering, Kruithof, Bos, Hidding, Feldbrugge,
  ten Have, van Engelen, Caroli, \& Teillaud}]{isvd10}
van~de Weygaert, R., Vegter, G., Edelsbrunner, H., {et~al.} 2011, Transactions
  on Computational Science, 14, 60

\bibitem[{Wilding {et~al.}(2021)Wilding, Nevenzeel, van de Weygaert, Vegter,
  Pranav, Jones, Efstathiou, \& Feldbrugge}]{wilding2020}
Wilding, G., Nevenzeel, K., van de Weygaert, R., {et~al.} 2021, Monthly
  Notices of the Royal Astronomical Society, 507, 2968–2990

\bibitem[{Xu {et~al.}(2019)Xu, Cisewski-Kehe, Green, \& Nagai}]{xu2019}
Xu, X., Cisewski-Kehe, J., Green, S., \& Nagai, D. 2019, Astronomy and
  Computing, 27, 34–52

\end{thebibliography}

\begin{appendix}

\section{Data and methods}
\label{sec:methods}

In this section, we briefly describe the dataset we used in the experiments. We also present a short account of the computational pipeline, referring to \cite{pranav2019b} for details.

\subsection{Datasets}

Over a period of time, the Planck team has invested significant effort in understanding and calibrating the source of noise as well as systematics that affect observations. This has led to the release of a series of datasets over a period of time, with successive data releases achieving better calibration and accuracy. In \cite{pranav2019b}, we performed our experiments on DR2, which is one of the intermediate data releases. In this paper, we concentrate on two data releases simultaneously, which we briefly describe below. 

\medskip\noindent\textbf{\textup{Planck 2020} Data Release 4 \texttt{NPIPE} dataset.} The primary results are based on the observational maps and simulations based on the \texttt{NPIPE} \citep{npipe} analysis pipeline (hereafter just \texttt{NPIPE}), which employs the \texttt{SEVEM} component separation technique. The \texttt{NPIPE} dataset is the final data release from the Planck team, and the pipeline is the most evolved and sophisticated of all  data-generation pipelines. It combines some of the most powerful features of the separate LFI and HFI analysis piplelines, resulting in frequency and component maps that have lower levels of noise and systematics at essentially all angular scales (cf. \cite{npipe}). The observational maps are accompanied by a suite of $600$ simulations modeled as isotropic, homogeneous Gaussian random fields.

\medskip\noindent\textbf{\textup{Planck 2018} Data Release 3 \texttt{FFP10} dataset.} For comparison, we also analyzed the temperature maps from \textup{Planck 2018} full-mission data release (\texttt{DR3} hereafter), which is the third data release by the Planck team, and employs the \texttt{SMICA} component separation technique. The noise and systematics in this release are better constrained than in the previous data releases. The component-separated observational maps are accompanied by a  suite of  $300$ simulations generated using the \textup{Full Focal Plane Plane} pipeline, hereafter referred to as the \texttt{FFP10} simulations \citep{ffp10}, also based on the standard model.

\subsection{Computational pipeline}

\begin{figure}
        \centering
        \subfloat{\includegraphics[width=0.5\textwidth]{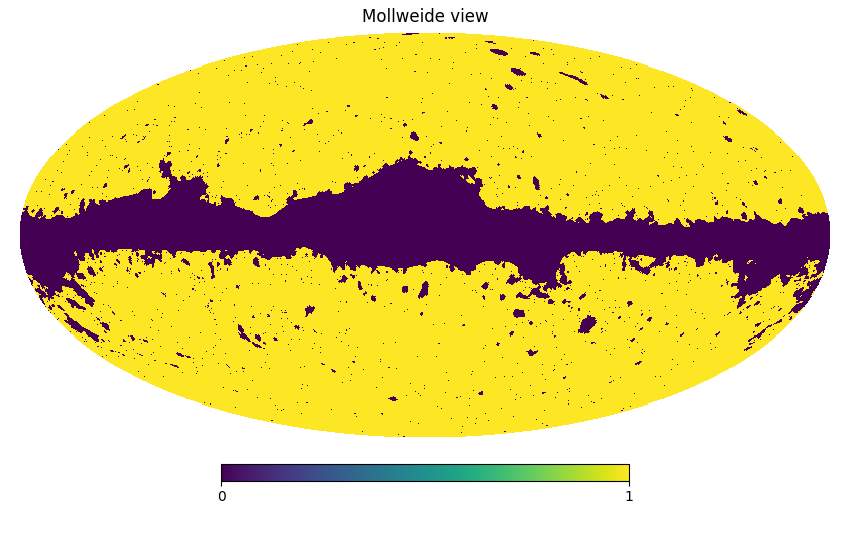} }\\
        \caption{Visualization of the mask employed for the analyses in this paper. It is the common mask released in Data Release 3 (2018) and masks our galaxy as well as other bright foreground sources, including point sources.}
        \label{fig:mask}
\end{figure}

\begin{figure*}
        \centering
        \subfloat{\includegraphics[width=0.4\textwidth]{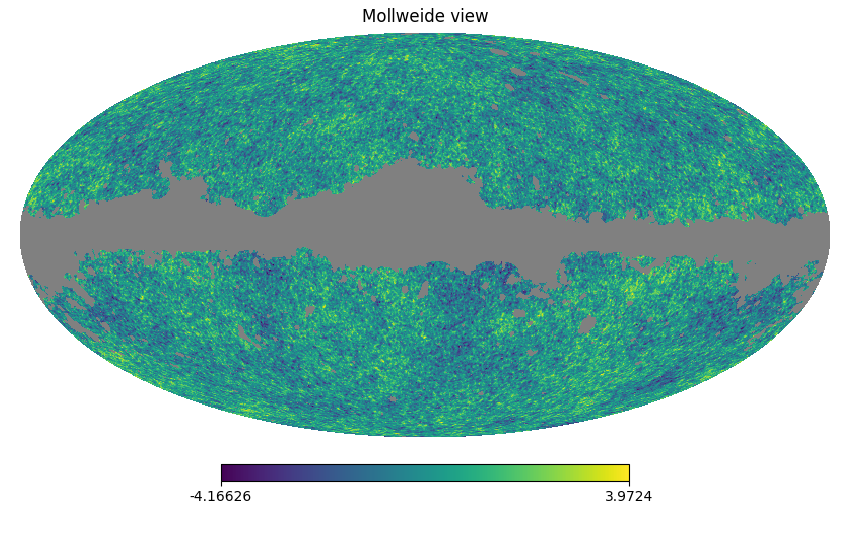} }
        \subfloat{\includegraphics[width=0.4\textwidth]{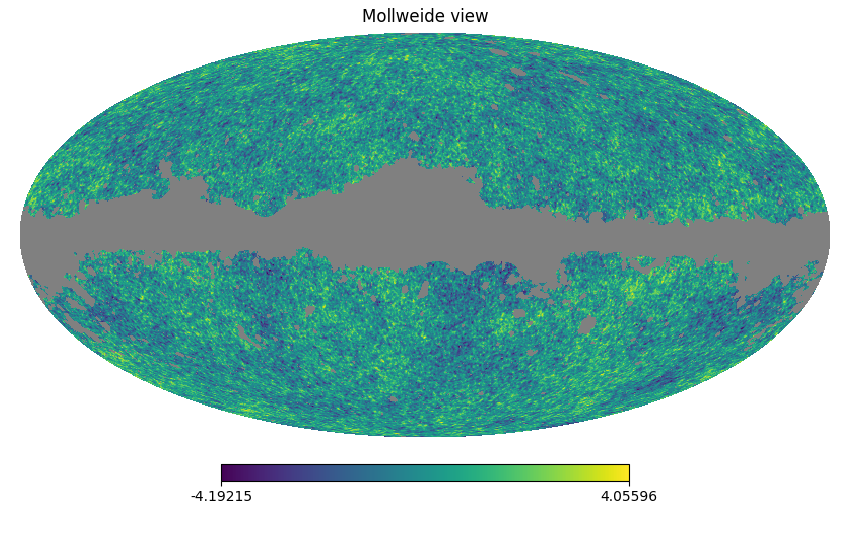} }\\
        \subfloat{\includegraphics[width=0.4\textwidth]{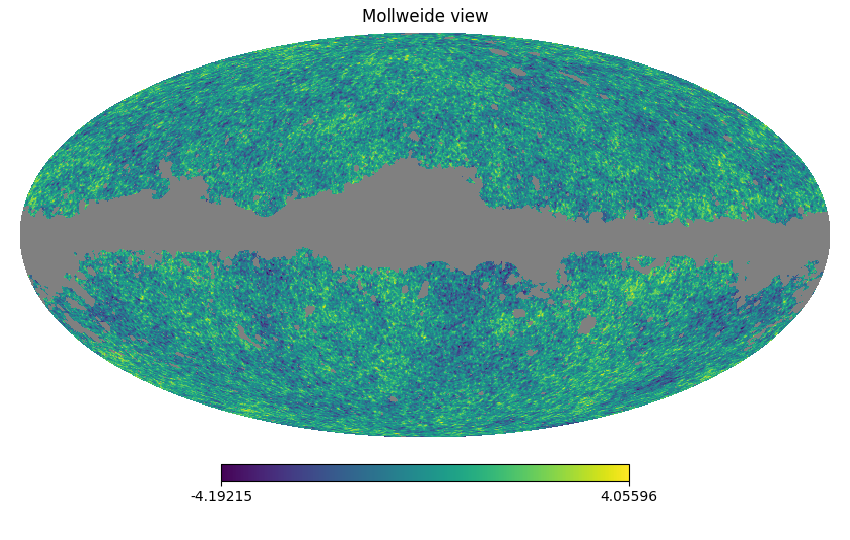} }
        \subfloat{\includegraphics[width=0.4\textwidth]{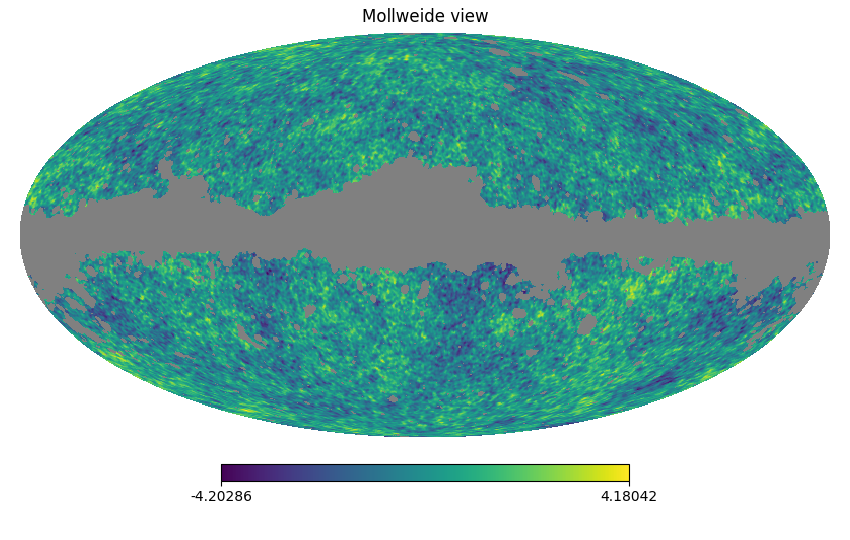} }\\
        \subfloat{\includegraphics[width=0.4\textwidth]{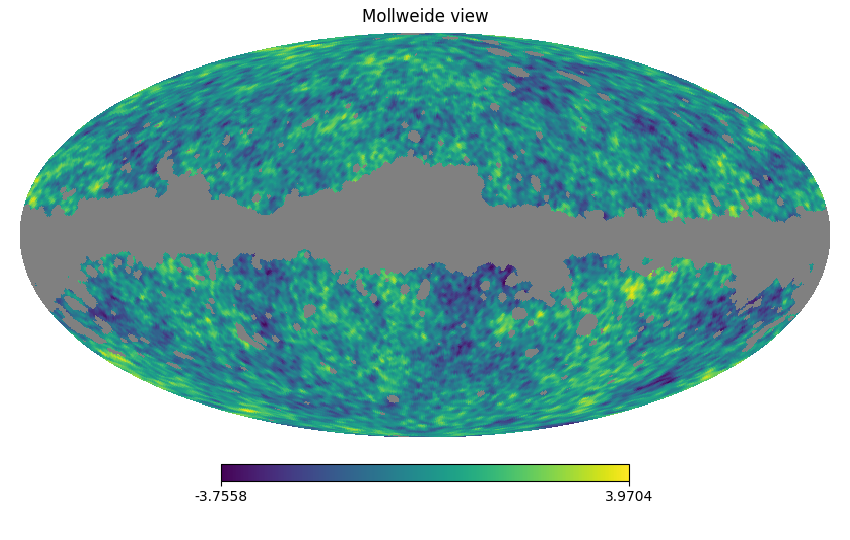} }
        \subfloat{\includegraphics[width=0.4\textwidth]{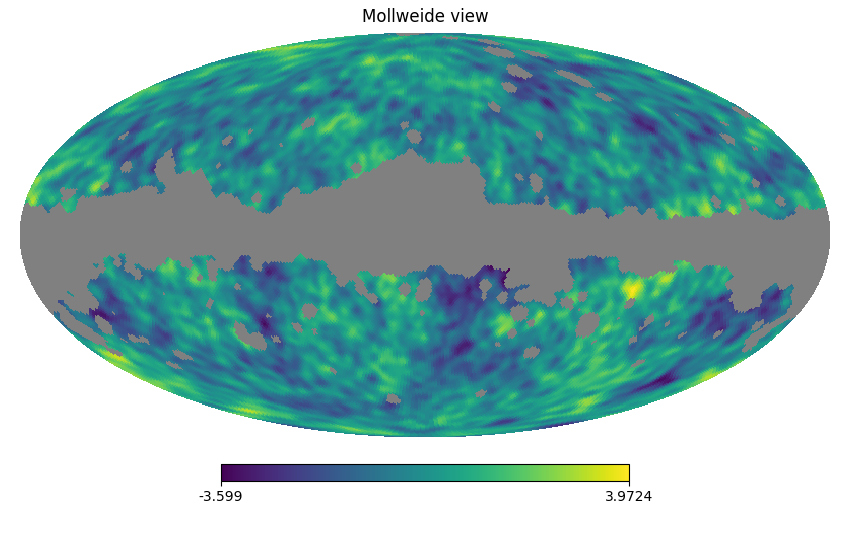} }\\
        \subfloat{\includegraphics[width=0.4\textwidth]{figs/npipe_obs_sevem_32_masked.png} }
        \subfloat{\includegraphics[width=0.4\textwidth]{figs/npipe_obs_sevem_16_masked.png} }\\
        \caption{Visualization of the degraded, smoothed, and masked observational maps employed in this paper. The maps are degraded at $\Res = 2048,1024,512,256,128,64,32,\text{and }16$, and smoothed at  $FWHM = 5', 10', 20', 40', 80', 160', 320',\text{ and } 640'$ , respectively. The mask is degraded and smoothed at the same scales as the map, and subsequently thresholded at $0.9$ for rebinarization.}
        \label{fig:masked_maps}
\end{figure*}

\begin{table} 
	\begin{center} 
		\caption{Percentage of sky area covered by the unmasked regions in the sky.} 
		\label{tab:unmasked_area}
		\begin{tabular}{|r|r|r|} 
			\hline
			Resolution   &   FWHM(arcmin) &  \% unmasked  \\ \hline \hline
			2048			&	5 						&	77.9			\\ \hline 	
			1024			&	10 						&	76.9			\\ \hline 	
			512				&	20						&	75.6			\\ \hline 
			256				&	40						&	74.7			\\ \hline 
			128				&	80						&	73.6		\\ \hline 
			64				&	160						&	71.3		\\ \hline 
			32				&	320						&	68.8			\\ \hline 
			16				&	640						&	64.5			\\ \hline 
		\end{tabular} 
	\end{center} 
	\tablefoot{The values are presented for the various degraded resolutions and smoothing scales. analyzed in this paper. The re-binarization cut-off for masking is set at $0.9$.} 
\end{table} 

The computational pipeline is composed of the steps that we briefly describe below. See \cite{pranav2019b} for a substantially detailed description.

\medskip\noindent\textbf{Preprocessing.} We performed a range of preprocessing steps to the maps using the \texttt{HealPix} package \citep{healpix1}, which is also the format of the initial data. First, the CMB and the noise maps were added pixel-wise for each realization of the simulations. Subsequently, the observational and noise-added simulation maps were degraded and smoothed at a range of scales for the values presented in Table~\ref{tab:unmasked_area}.  The mask presented in Figure~\ref{fig:mask}, which is a binary map, was also degraded and smoothed at the same resolution as the CMB maps, and subsequently thresholded at a value of $0.9$ to make it binary again. Table~\ref{tab:unmasked_area} presents the percentage of sky covered by the unmasked region for the degraded resolution and associated smoothing scales analyzed in this paper. Figure~\ref{fig:masked_maps} presents a visualization of the degraded and smoothed maps after applying the mask. Next, we computed the mean and standard deviation from the unmaksed pixels for each realization and resolution and rescaled the maps pixel-wise by the standard deviation after subtracting the mean. As a final step, we assigned the value infinity to the masked pixels (numerically simulated by a very large number). 

\medskip\noindent\textbf{Triangulation.} As a first step, we projected pixels of the maps  projected onto $\Sspace^2$, and we triangulated this set of points in $\Rspace^3$. Taking the convex hull of this triangulation produces a triangulation of the point-set on $\Sspace^2$. This triangulation consists of $V = 12 \Res^2$ vertices, $3V - 6$ edges, and $2V - 4$ triangles, where $\Res$ is the resolution parameter in \texttt{HealPix} format. This was the input to all downstream computations and represents the temperature field, $f \colon \Sspace^2 \to \Rspace$, by storing the temperature value at each vertex; see \cite{pranav2019b}. We assumed a piece-wise linear interpolation along higher dimensional simplices. 

\medskip\noindent\textbf{Upper-star filtration.} Given the triangulation $K$ constructed in the previous step, we ordered its simplices such that $\ssx$ precedes $\tsx$ if (i) $f(\ssx) > f(\tsx)$ or (ii) $f(\ssx) = f(\tsx)$ and $\dime{\ssx} < \dime{\tsx}$, in which $f(\ssx)$ is the minimum temperature value of the vertices of $\ssx$. Any ordering that satisfies (i) and (ii) is called an
\textup{upper-star filter} of $K$ and $f$. The corresponding \textup{upper-star filtration} consists of all
prefixes of the filter, each representing an excursion set of $f$. 

\medskip\noindent\textbf{Persistence computation.} We constructed a boundary matrix from the upper-star filtration of $K$. Writing $\ssx_1, \ssx_2, \ldots, \ssx_n$ for the sorted simplices of the upper-star filtration, the  boundary matrix $\partial [1..n, 1..n]$ is defined by  $\partial [i,j] = 1$, if $\ssx_i$ is a face of $\ssx_j$ and $\dime{\ssx_i} = \dime{\ssx_j} - 1$, and $\partial [i, j] = 0$, otherwise. Computing the persistence birth death pair from this ordered boundary matrix involves reduction of the columns to the \textup{lowest-j} form. We resorted to reduction from left to right, and a column of the matrix was reduced if it was zero or its lowest $1$ had only $\text{zero}$s in the same row to its left.  Each column with a unique \textup{lowest-j} contributes to the birth death pair of the persistence diagram, where the index of the birth and death simplices are precisely the row and column indices of the \textup{lowest-j}. Ranks of homology groups relative to the mask were inferred from the persistence diagrams by setting the vertices belonging to the mask at $+\infty; \text{see}$ \cite{pranav2019b}:

\begin{align}
\relBetti{0}  &= \# \{[b,d) \in Dgm_0(\Excursion \cup \Mask) \mid  +\infty > b \geq \nu > d \} ; \\ \nonumber
\relBetti{1}  &= \# \{[b,d) \in Dgm_0(\Excursion \cup \Mask) \mid  +\infty = b > d \geq \nu \} \\ \nonumber
&+ \# \{[b,d) \in Dgm_1(\Excursion \cup \Mask) \mid  +\infty > b \geq \nu > d \} ;\\ \nonumber
\relBetti{2} &= \# \{[b,d) \in Dgm_1(\Excursion \cup \Mask) \mid  +\infty = b > d \geq \nu \} \\ \nonumber
&+ \# \{[b,d) \in Dgm_2(\Excursion \cup \Mask) \mid  +\infty > b \geq \nu > d \} .
\end{align}

\medskip\noindent\textbf{Tests for determining statistical significance.} 
\label{sec:stat}
Our main aim was comparing the observational CMB maps with the null-hypothesis simulation maps, which assume isotropy, homogeneity, and Gaussianity. Let $\x_i \in \R^m$, $i=1,\ldots,n$, be a sample of i.i.d.\ $m$-dimensional  vectors,  drawn from a distribution $G$. Let $\y \in \R^m$ be another sample point, assumed to be drawn from a distribution $F$. We wish to test the (null) hypothesis that $F=G$, and give the test results in terms of \textup{$p$-values}, which compute the probability that $\y$ is `consistent' with this hypothesis. We employed two different statistical tests that we briefly describe below. 

\medskip\noindent\textup{Mahalanobis distance.} The first is the parametric \textup{Mahalanobis distance}, or the familiar $\chi^2$ test \citep{mahalanobis}. If $G$ is assumed to be Gaussian and $n$ is large,  then under the hypothesis that $G=F,$ the squared Mahalanobis distance is approximately distributed as a   $\chi^2$ distribution with $m$ degrees of freedom. Thus the corresponding  $p$-value is

\begin{equation}\label{eq:Mahal-p}
p_{\rm Mahal}(\y) = P[\chi^2_m > d^2_{\rm Mahal}(\y)].
\end{equation}

\medskip\noindent\textup{Tukey depth.} The second method is a nonparametric test based on the \textup{Tukey depth} \citep{depth}. It \pp{is} a general metric  for identifying outliers  in a flexible manner and in a nonparametric setting, making no assumptions on the structure of $F$ and $G$. Let $\mathbf{z}$ be any point in $\mathbb R^m$.  Then the half-space depth $d_{\rm dep}(\mathbf{z}; \x_1,\ldots,\x_n)$ of $\mathbf{z}$ within the sample of the $\x_i$ is the smallest fraction of the $n$ points $\x_1,\ldots,\x_n$ to either side of any hyperplane passing through $\mathbf{z}$.  Points that have the same depth constitute a nonparametric estimate of the  isolevel contour of the distribution $F$. We first compute d$d_j = d_{\rm dep}(\x_j; \x_1,\ldots,\x_n)$ for every point $\x_j$, $j=1,\ldots,n$, yielding an empirical distribution of depth. The $p$-value of  $\y$ was computed as the proportion of points whose depth is lower than that of $\y$, 

\begin{equation}\label{eq:hsd-p}
p_{\rm dep} (\y)  =  \#\{j \mid d_j > d_{dep}(y) \} / n
.\end{equation}

\section{Code description} 
\label{sec:codes}
All the analyses in this paper have been performed by deploying \texttt{TopoS2}.  \texttt{TopoS2} is a specialized software written for computing the topology of scalar functions on S2, especially adapted to the HealPix data format. It is written in the context of the topological analysis of Planck CMB maps, however useful for other scalar functions on $\mathbb{S}^2$. \texttt{TopoS2} is tested on Debian Linux 9, with the following dependencies:

\begin{table}
	\begin{center}
		\begin{tabular}{lcl}
	* [boost]& ---  &   general purpose C++ libraries\\
	* [CGAL]-4.7 &---&   triangulation and filtration building\\
	* [CMake] &---&     build process\\
	* [Dionysus] &---&  data structure and topology computation\\
	* [HealPiX] &---&    data-processing operations on the sphere\\
	* [PHAT] &---&      data structure and topology computation\\
\end{tabular}
	\end{center}
\end{table}

\begin{table}
	\begin{center}
		\begin{tabular}{ll}
* [boost]:       \url{ http://www.boost.org}\\
* [CGAL]:         \url{http://www.cgal.org}\\
* [CMake]:        \url{http://www.cmake.org}\\
* [Dionysus]:     \url{https://www.mrzv.org/software/dionysus/}\\
* [HealPix]:      \url{https://healpix.jpl.nasa.gov/}\\
* [PHAT]:        \url{ https://github.com/blazs/phat}\\

\end{tabular}
\end{center}
\end{table}

\subsection{Description}

\texttt{TopoS2} has three executables that work in tandem to produce the persistence diagrams 
of scalar fields on $\mathbb{S}^2$. For Planck CMB files supplied in HealPix format the code
\texttt{helper\_func/fits\_RDS\_maskRDS\_nrm.cpp} reads the CMB and mask fits files, degrades and 
smooths them to specified values, and outputs an ascii masked file that serves as input 
to the next code in sequence \url{filtration/streaming_filtration_builder_superlevelset.cpp}.
The output of this operation is a filtration of superlevel sets to the standard output 
stream. This stream is fed to the final part of the code \texttt{persistent\_homology/diagram\_generator.cpp},
which outputs the persistence diagram to the standard output stream. 
The code can be run in isolation for each data file, or batch-processed on large clusters. Higher 
resolutions are memory intensive:

\begin{itemize}
		\item 2048  $\sim$120 GB\\
		\item 1024  $\sim$  35 GB\\
		\item 512    $\sim$   8 GB\\
\end{itemize}

\subsection{Building}

We assume we are in TopoS2 root directory. 

1.) For the fits operation handling, the code is compiled against healpix C++ library. This amounts to modifying
the \texttt{helper\_func/compile} file to point to the healpix directories, and executing it.

\begin{verbatim}
	cd helper_func/
	./compile
	cd ..
\end{verbatim}

2.) Building the triangulation and filtration invokes the CGAL library \citep{cgal}. It is best compiled by invoking the 
\texttt{cgal\_create\_cmake\_script} scriptfile to generate the cmake components. Dependency problems, 
if arising, are resolved by interactively editing the cmake configuration file, and setting the right paths. 
If the CGAL root directory is \texttt{\$CGAL\_DIR}, this amounts to executing the following commands in sequence:

\begin{verbatim}
cd filtration/
$CGAL_DIR/scripts/cgal_create_cmake_script
cmake . (or  ccmake  .  for interactive)
make
cd ..
\end{verbatim}

3.) Computing persistence diagrams is based on Dionysus \citep{morozov2005} and PHAT \citep{phat} library. The code is compiled by executing the following commands in sequence:

\begin{verbatim}
cd persistent_homology
mkdir build
cd build
cmake ..
make 
cd ../..
\end{verbatim}

%
%

\subsection{Analyzing planck dataset}

The core code and associated helper scripts have evolved over a period of time keeping in mind the central requirement 
to analyze the full Planck dataset as fast as possible. Given the memory requirements as stated above, the best strategy
is to engage full  nodes with multiple cores on large clusters for each run, especially at higher resolutions.
The codes are written such that they default on memory overflow. As memory consumption on the higher resolutions is the 
bottleneck, the codes are written with a capability to run on a single processor, as well as multiple processors, and crash 
if memory overflows.

\subsubsection{Generating persistence diagrams}

Given all of this, and a successful compilation of the codes, and an assumption that we are in the directory containing 
the NPIPE observation file \texttt{npipe6v20\_sevem\_cmb\_005a\_2048.fits}, and simulations with CMB and Noise added (numbering from 200
to 799, SEVEM kind), we provide two scripts that run on SunGrid Engine cluster environment and produce a multi-scale analysis
in an automated fashion. The helper script \texttt{sge\_job\_script\_npipe\_DS\_obs.py} in the "example"
directory serves as an example to process multiple resolutions for the observation files on a cluster running the SGE grid
engine. The script file \texttt{sge\_job\_script\_npipe\_DS.py} processes the simulations. In the script, the variables "binary1", 
"binary2" and "binary3" set the paths to fits processing script, filtration builder code, and persistent homology calculator code. 
At this moment we have the persistence diagrams from the observations and simulations, from which the Betti numbers may be 
calculated.

\subsubsection{Computing Betti numbers}

To compute the Betti numbers from the persistence diagrams, we need to insert a heading in each \texttt{.dgm} file which is the 
number of rows in it. The script \texttt{heading\_fixer.py} does 
this for the simulations, similarly  \texttt{heading\_fixer\_obs.py} for observations. Thereafter the Betti numbers are calculated by the 
C++ code \texttt{betti\_superlevel\_relative} (to be compiled like a normal c++ code). For batch processing two script files 
\texttt{betti\_calculator.py} and \texttt{betti\_calculator\_obs.py} do this. The scripts in this paragraph need to be placed in the \texttt{outdir} directory of \texttt{sge\_job\_script\_npipe\_DS.py}.

\section{Validation of the statistical tests} 
\label{sec:stat_validity}

\begin{figure*}
        \centering
        \subfloat{\includegraphics[width=0.8\textwidth]{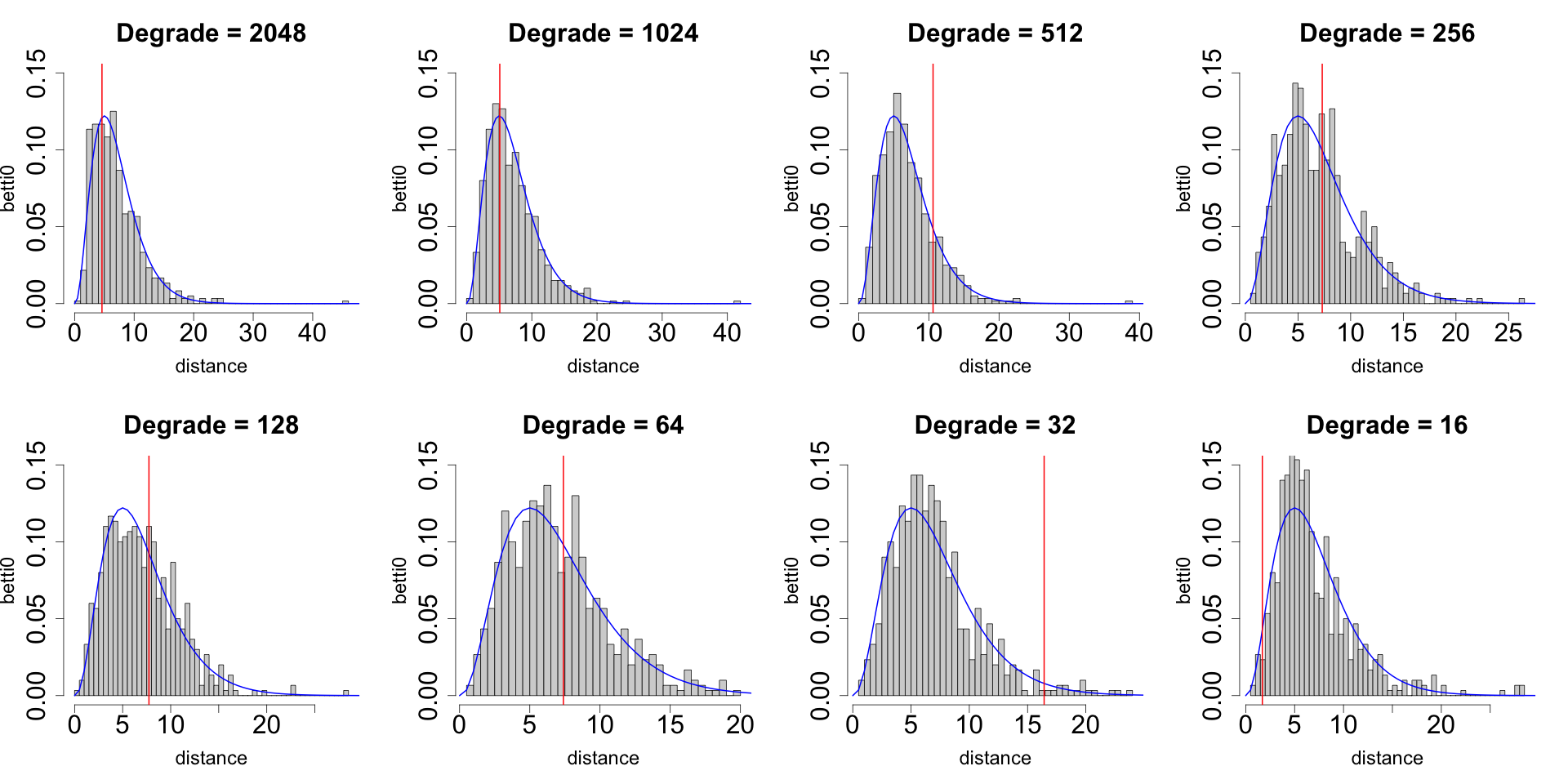} }\\
        \subfloat{\includegraphics[width=0.8\textwidth]{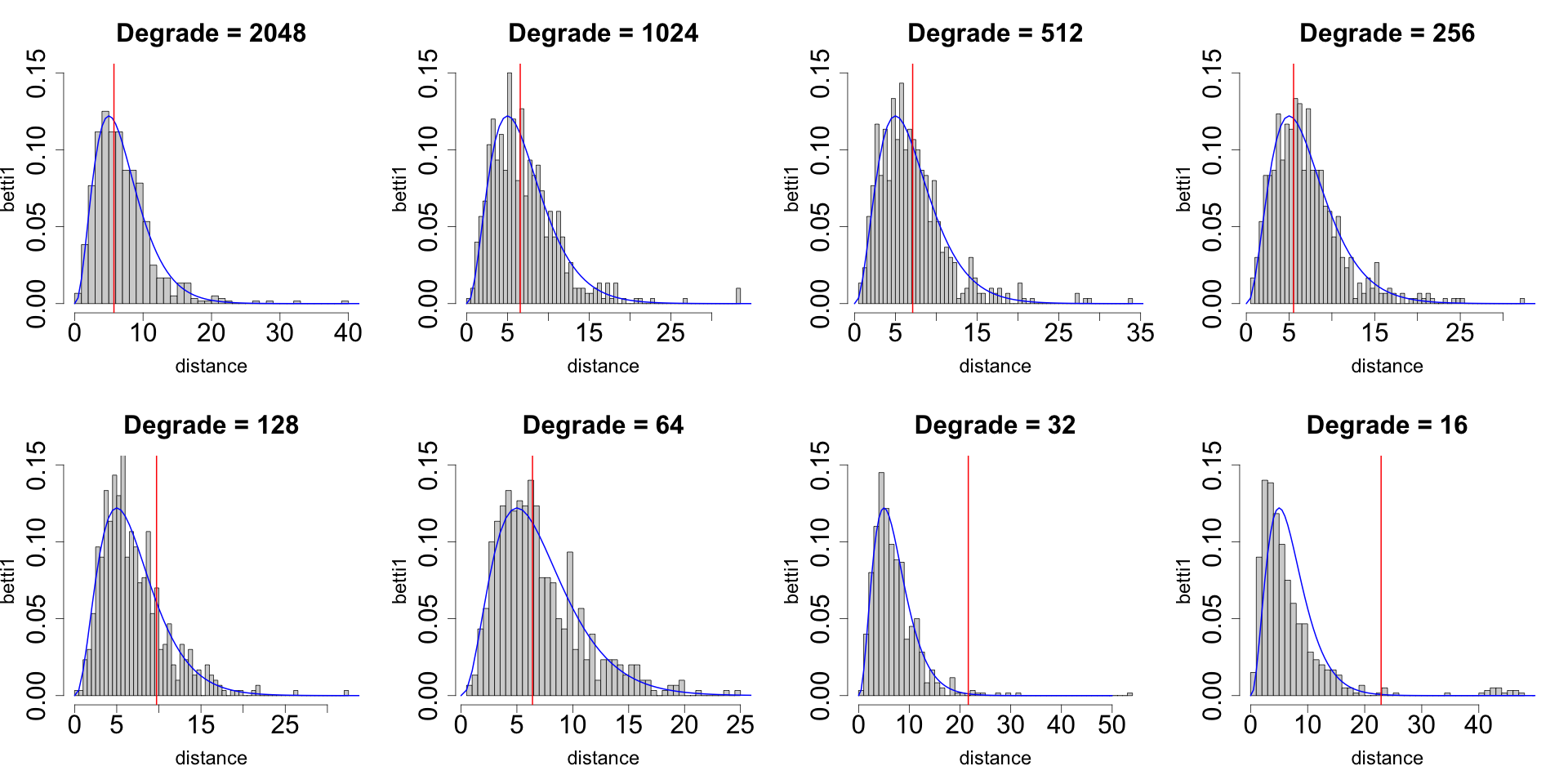} }\\
        \subfloat{\includegraphics[width=0.8\textwidth]{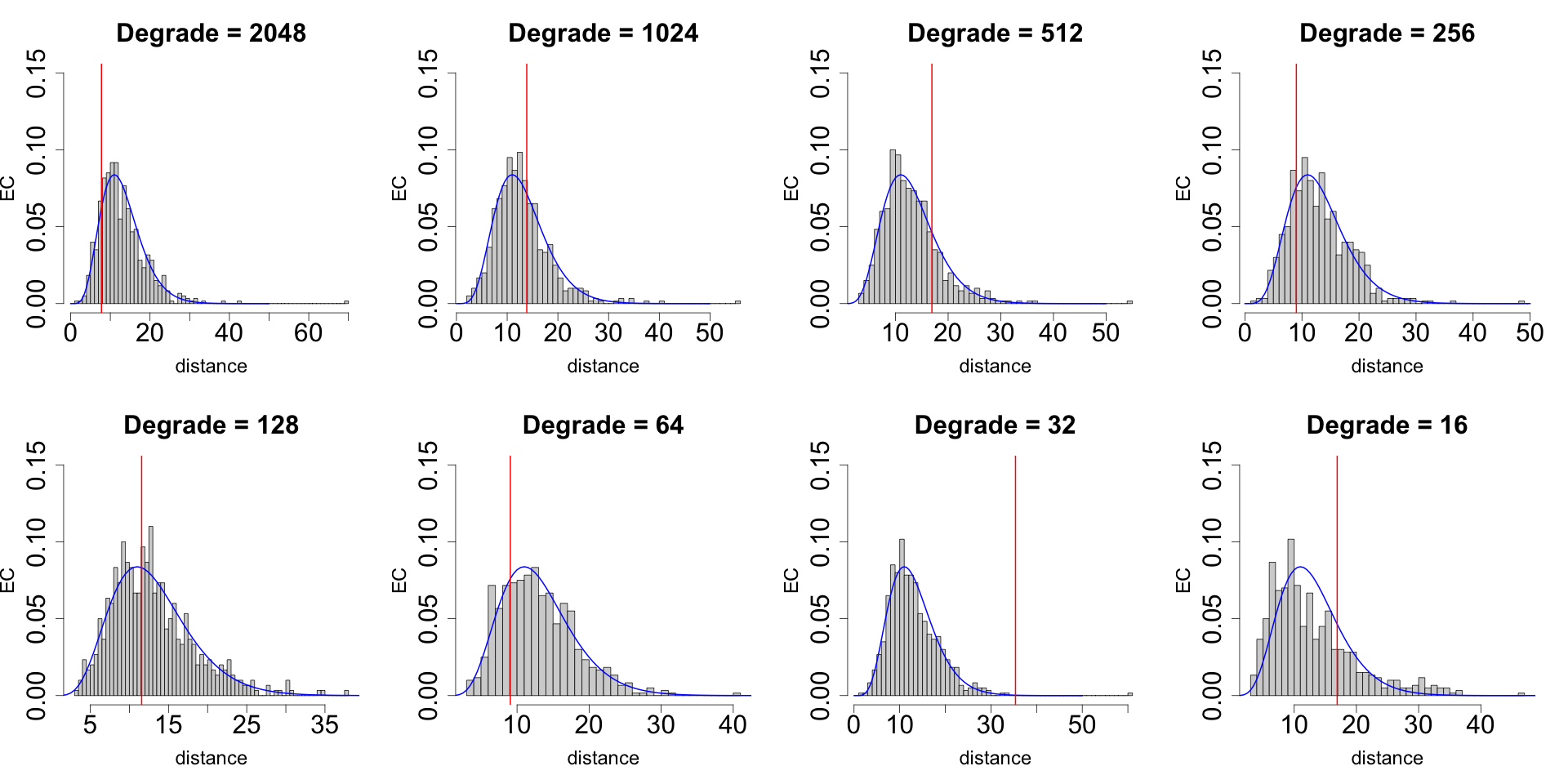} }\\
        \caption{Distribution of the Mahalanobis distance. The box histograms present the distribution of the Mahalanobis distance of the simulations with respect to the mean. The blue curve plots the theoretical $\chi^2$ distribution for the given degrees of freedom, and the red vertical line presents the Mahalanobis distance of the observation with respect to mean. The generally good fit of the theoretical curve with the histogram of the simulations for all resolutions lends credibility to the validity of the statistical tests in the chosen regimes.}
        \label{fig:hist_maha}
\end{figure*}

\begin{figure*}
        \centering
        \subfloat{\includegraphics[width=0.8\textwidth]{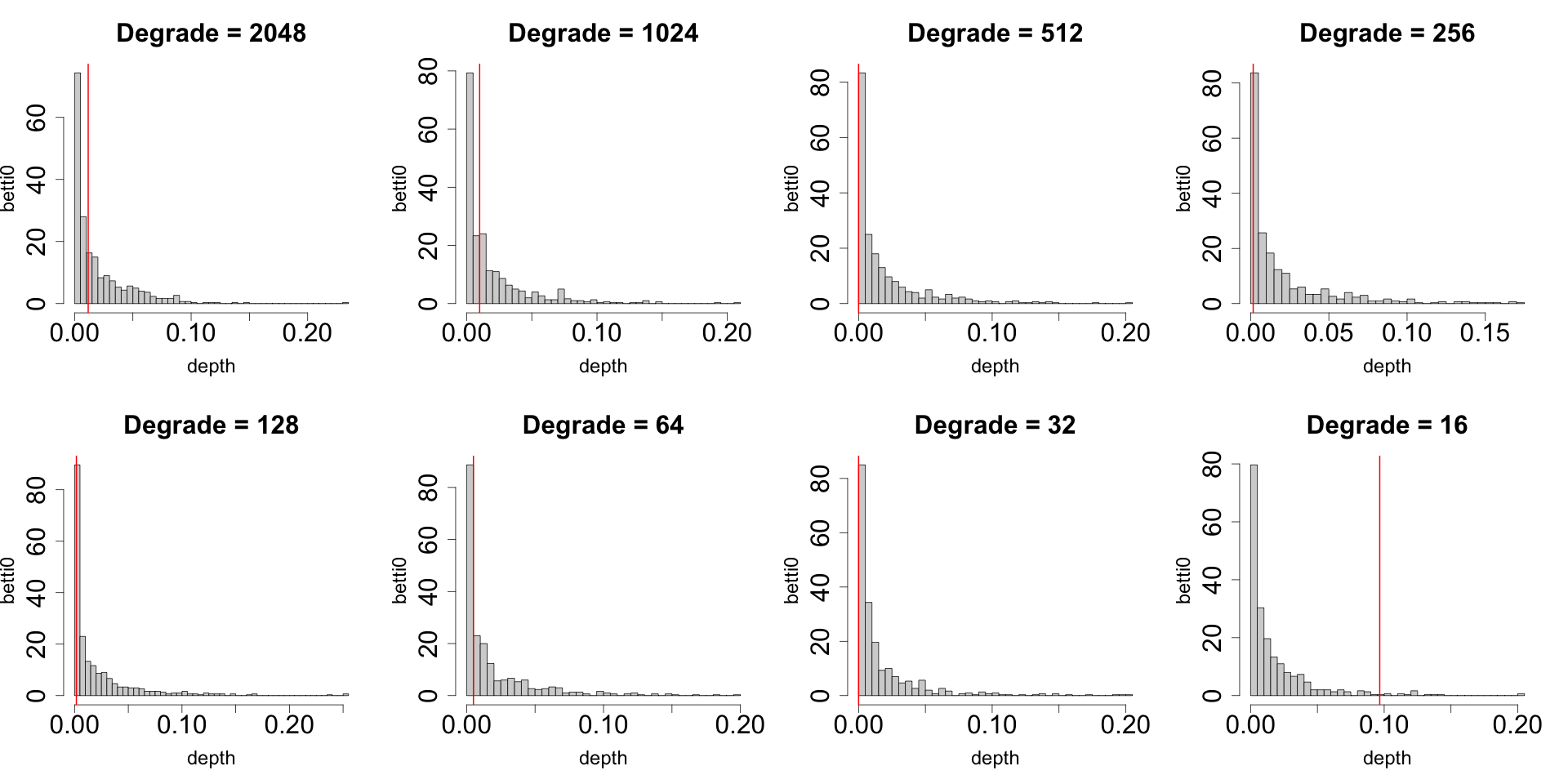} }\\
        \subfloat{\includegraphics[width=0.8\textwidth]{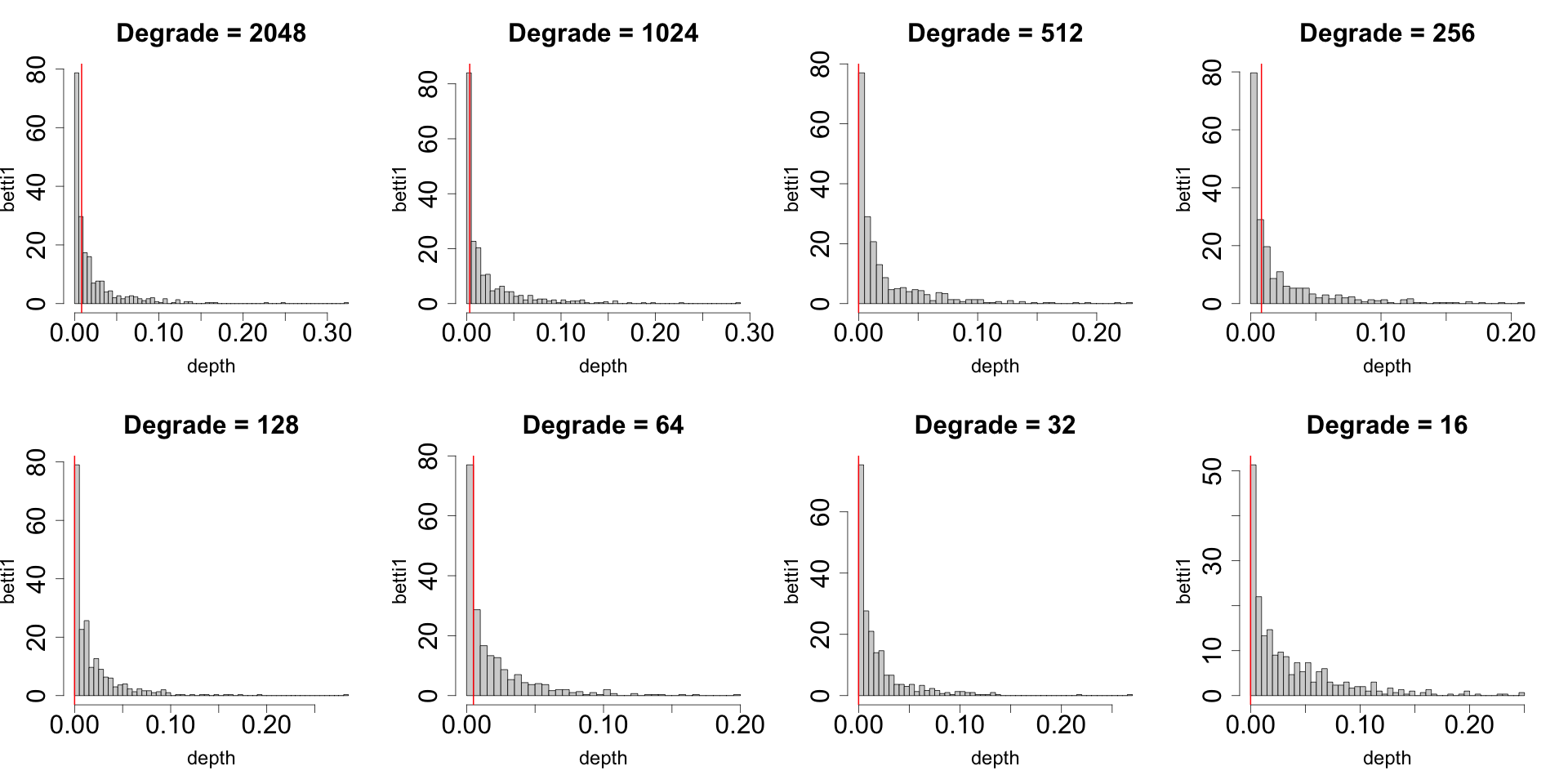} }\\
        \subfloat{\includegraphics[width=0.8\textwidth]{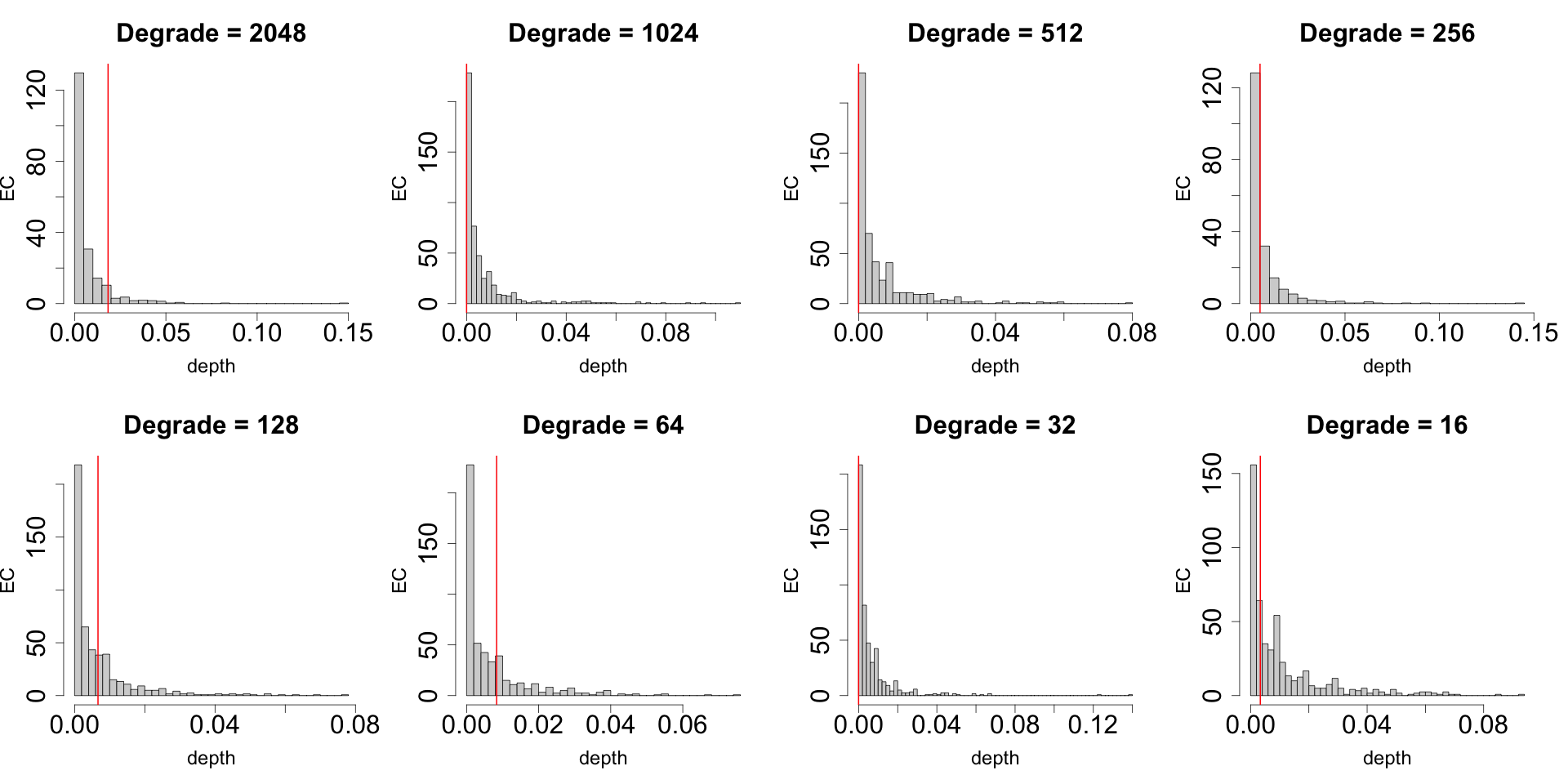} }\\
        \caption{Distribution of the Tukey depth. The box histograms present the distribution of the depth of the simulations, and the red vertical line presents the depth of the observation.}
        \label{fig:hist_tukey}
\end{figure*}

In this section, we examine the validity of the regime of our statistical tests. To this end, we examined the empirical distribution of the $\chi^2$ and Tukey depth values from the simulations, and compared it with the observations.

Figure~\ref{fig:hist_maha} presents the case for the $\chi^2$ distribution. The top two rows present the values for $b_0$, the middle two rows for $b_1$, and the bottom two rows for the Euler characteristic. The distribution of the values from the $600$ \texttt{NPIPE} simulations is denoted by the gray box histograms. The blue curve denotes the theoretical curve for the given degrees of freedom, and the red vertical line shows the value of the observational map with respect to the distribution for the simulations. In general, we find a good agreement between the theoretical curves and the histograms from the simulations, which lends additional support to the validity of the $\chi^2$ test in our regimes of testing.

Figure~\ref{fig:hist_tukey} presents the case for the Tukey depth, with the values from the simulations presented in the gray box, and the observational value presented by the red vertical line. The top two rows present the values for $b_0$, the middle two rows for $b_1$, and the bottom two rows for the Euler characteristic. We note the well-behaved nature of the distribution from the histograms. To help interpret these diagrams, we note that depth values increase from left to right on the horizontal axis, and the $p$-values are computed as the fraction of points that have lower depth than the candidate. The extreme $p$-values of $0.0$ in the main paper then simply means that the observational points are true outliers, and no simulated points are shallower than the observation. The $p$-values are consistent with the distribution characteristics, where in these cases, the observed red vertical line is to the left of all simulated box histograms.

\end{appendix}

\end{document}